\documentclass[aps,prd,showkeys,superscriptaddress,singlecolumn,nofootinbib,floatfix]{revtex4-2}

\usepackage{subfiles}
\usepackage{amsmath}
\usepackage{amssymb}
\usepackage{amsthm}
\usepackage{mathrsfs}
\usepackage{graphicx}
\usepackage{epstopdf}
\usepackage{fancyhdr}
\usepackage{array}
\usepackage[all]{xy}
\usepackage{eufrak}
\usepackage{euscript}
\usepackage{enumerate}
\usepackage{slashed}
\usepackage{physics}
\usepackage{hyperref}
\usepackage{xcolor}
\usepackage[normalem]{ulem} 
\usepackage{subfigure} 
\usepackage{epstopdf} 

\hypersetup{colorlinks=true,linkcolor=blue,citecolor=blue,menucolor=black,urlcolor=blue,filecolor=blue}

\begin{document}

\title{Homogeneous isotropization dynamics and entropy production in a hot and dense strongly interacting fluid}

\author{Gustavo de Oliveira}
\email{gustav.o.liveira@discente.ufg.br}
\affiliation{Instituto de F\'{i}sica, Universidade Federal de Goi\'{a}s, Av. Esperan\c{c}a - Campus Samambaia, CEP 74690-900, Goi\^{a}nia, Goi\'{a}s, Brazil}

\author{Willians Barreto}
\email{willians.barreto@ufabc.edu.br}
\affiliation{Centro de Ci\^{e}ncias Naturais e Humanas, Universidade Federal do ABC, Av. dos Estados 5001, 09210-580 Santo Andr\'{e}, S\~{a}o Paulo, Brazil}
\affiliation{Centro de F\'{i}sica Fundamental, Universidad de Los Andes, M\'{e}rida 5101, Venezuela}

\author{Romulo Rougemont}
\email{rougemont@ufg.br}
\affiliation{Instituto de F\'{i}sica, Universidade Federal de Goi\'{a}s, Av. Esperan\c{c}a - Campus Samambaia, CEP 74690-900, Goi\^{a}nia, Goi\'{a}s, Brazil}

\begin{abstract}
We numerically investigate the time evolution of the non-equilibrium entropy during the homogeneous isotropization dynamics of the 2 R-Charge Black Hole (2RCBH) model, corresponding to a top-down holographic fluid defined at finite temperature and R-charge density. In addition to the entropy production, we also analyze the time evolution of the pressure anisotropy and the scalar condensate of the medium. When the system is far from equilibrium the dominant and weak energy conditions can be transiently violated. For all initial conditions considered, we observe the emergence of a periodic sequence of several close plateaus forming a stairway for the entropy as the system approaches thermodynamic equilibrium. The entropy stairway allows for the entropy to encode a periodic structure without violating the second law of thermodynamics. In fact, the complex frequency of the lowest quasinormal mode (QNM) of the system is directly tied to the periodic structure of the entropy stairway, which provides another explicit numerical confirmation of a quite general connection between entropy production and QNMs previously discovered in the literature. Furthermore, when the chemical potential of the 2RCBH fluid exceeds a certain threshold, the pressure anisotropy exhibits a late-time decay governed by a purely imaginary QNM, and as the system is doped with increasing values of R-charge chemical potential the late-time equilibration pattern of the pressure anisotropy gets increasingly deformed, eventually losing the oscillatory behavior observed at lower values of chemical potential.
\end{abstract}

\maketitle
\tableofcontents


\section{Introduction}
\label{sec:intro}

The holographic gauge-gravity duality~\cite{Maldacena:1997re,Gubser:1998bc,Witten:1998qj,Witten:1998zw} is a mathematical dictionary that relates physical observables of strongly interacting quantum field theories (QFTs) to semi-classical gravity calculations in higher dimensional asymptotically Anti-de Sitter (AdS) spacetime manifolds. It provides an interesting framework for investigating the dynamical evolution of strongly coupled quantum media initially prepared in far-from-equilibrium states, allowing for a non-perturbative understanding of thermalization processes. In fact, many different far-from-equilibrium dynamics have been investigated in the literature across several different holographic models. For instance, the computation of many physical observables in the homogeneous isotropization dynamics of different holographic systems, some of them encompassing effects due to a chemical potential or due to external electromagnetic fields, has been explored e.g. in~\cite{Chesler:2008hg,Fuini:2015hba,Critelli:2017euk,Cartwright:2019opv,Cartwright:2020qov,Rougemont:2024hpf}. Several analyses of the boost invariant holographic Bjorken flow, including discussions on the role played by different hydrodynamic attractors in the hydrodynamization of strongly coupled quantum fluids with gravity duals, some of them encompassing effects due to a chemical potential or due to higher order derivative corrections in the bulk, have been presented e.g. in~\cite{Chesler:2009cy,Heller:2011ju,Heller:2012je,Jankowski:2014lna,Pedraza:2014moa,Bellantuono:2015hxa,Romatschke:2017vte,Spalinski:2017mel,DiNunno:2017obv,Casalderrey-Solana:2017zyh,Critelli:2018osu,Kurkela:2019set,Rougemont:2021qyk,Rougemont:2021gjm,Cartwright:2021maz,Rougemont:2022piu,Cartwright:2022hlg}. Several other far-from-equilibrium dynamics which have also been explored in the literature include the holographic radial flow~\cite{vanderSchee:2012qj,vanderSchee:2014qwa}, the conformal Gubser flow~\cite{Banerjee:2023djb,Mitra:2024zfy}, gravitational shockwave collisions~\cite{Chesler:2010bi,Casalderrey-Solana:2013aba,Chesler:2015bba,Casalderrey-Solana:2016xfq,Grozdanov:2016zjj,Attems:2016tby,Attems:2017zam,Attems:2018gou,Waeber:2019nqd,Ecker:2021ukv}, and holographic turbulence~\cite{Adams:2012pj,Adams:2013vsa,Waeber:2021xba}. See also~\cite{Chesler:2013lia} for an excellent review on some of the above topics, with a focus on numerical techniques of general relativity adapted to asymptotically AdS spacetimes, as usually required for holographic applications.

In the present work, we numerically investigate the homogeneous isotropization dynamics of a top-down holographic construction called the 2 R-Charge Black Hole (2RCBH) model~\cite{DeWolfe:2011ts,DeWolfe:2012uv}. The 2RCBH plasma corresponds to a strongly interacting fluid defined at finite temperature and R-charge density, which appears as a particular case of the more general STU model~\cite{Behrndt:1998jd,Cvetic:1999ne}. In fact, the STU model describes five-dimensional black branes charged under the $U(1)\times U(1)\times U(1)$ Cartan subgroup of the global $SU(4)$ R-symmetry of strongly coupled $\mathcal{N}=4$ Supersymmetric Yang-Mills (SYM) theory living at the conformally flat four-dimensional boundary of the bulk manifold. In its most general form, the STU model has three conserved Abelian $U(1)$ R-charges, therefore featuring three different Maxwell fields in the five-dimensional gravity action, in addition to the metric field and two scalar fields. Different particular cases of the STU model can be obtained by turning off some of the associated R-charges. The 2RCBH model is obtained by setting one of the R-charges to zero and then identifying the remaining two R-charges: in this case, the effective bulk action comprises, besides the metric field, only one Maxwell field and one scalar field.

Another interesting particular case of the STU model, called 1 R-Charge Black Hole (1RCBH) model, is obtained by setting two of the three R-charges to zero. In this case, the gravity dual action also features a metric field, a single Maxwell field, and a single scalar field, with the 1RCBH plasma also describing a strongly coupled fluid at finite temperature and R-charge density, but with a rather different physics from the 2RCBH model. In fact, while the 1RCBH model features two branches of black hole solutions, one thermodynamically stable and another one unstable, with both branches merging at the critical point of the model, the 2RCBH model has no critical point and features a single branch of stable black hole solutions~\cite{DeWolfe:2011ts,DeWolfe:2012uv}.

Several different physical observables have been previously considered in both models of strongly interacting R-charged plasmas. In fact, their thermodynamics have been analyzed in~\cite{DeWolfe:2011ts,DeWolfe:2012uv,Finazzo:2016psx,deOliveira:2024bgh}, the structure of Fermi surfaces was addressed in~\cite{DeWolfe:2012uv}, some hydrodynamic transport coefficients were discussed in~\cite{DeWolfe:2011ts,Asadi:2021hds}, the spectra of homogeneous quasinormal modes (QNMs) were obtained in~\cite{Finazzo:2016psx,Critelli:2017euk,deOliveira:2024bgh}, while some information-theoretic quantities were computed in~\cite{Ebrahim:2018uky,Ebrahim:2020qif,Amrahi:2021lgh,Karan:2023hfk,deOliveira:2025qwe}, and chaotic properties were investigated in~\cite{Karan:2023hfk,Amrahi:2023xso,Lilani:2025wnd}. In what concerns far-from-equilibrium dynamics for the 1RCBH model, holographic renormalization and homogeneous isotropization dynamics were discussed in~\cite{Critelli:2017euk,Rougemont:2024hpf}, while the boost-invariant Bjorken flow dynamics was investigated in~\cite{Critelli:2018osu,Rougemont:2022piu}. The 2RCBH model has not yet had its properties investigated under different out-of-equilibrium dynamics, and in the present work, we provide the first analysis in this direction by considering the calculation of the pressure anisotropy, the scalar condensate, and the entropy production in the 2RCBH plasma undergoing homogeneous isotropization dynamics.

In fact, in a recent work on the homogeneous isotropization of the 1RCBH model~\cite{Rougemont:2024hpf}, it was observed a stairway structure for the entropy density comprising ever-increasing and progressively closer steps with transient periods of zero entropy production near thermodynamic equilibrium. Furthermore, the frequency of formation of those near-equilibrium transient plateaus for the entropy density was observed to be twice the real part of the lowest quasinormal eigenfrequency of the system. This constitutes a particular instance of a quite general relation between entropy production and QNMs discovered in~\cite{Jansen:2016zai,Jansen:2020ign}. On the other hand, very recently in~\cite{deOliveira:2024bgh}, the analysis of the QNM spectra of the 2RCBH model unveiled new purely imaginary QNMs which become the dominant, lowest QNMs of the system for sufficiently high values of R-charge chemical potential. Both results motivated us to investigate the homogeneous isotropization dynamics of the 2RCBH model to check whether also in this case an entropy stairway is observed near thermodynamic equilibrium, and if its relationship with the lowest QNM of the system holds the same. Another motivation was to investigate the influence of the purely imaginary QNMs on the late-time isotropization and thermalization of the 2RCBH plasma. In the present work, we address these questions in detail and provide another explicit numerical confirmation of the general connection between entropy production and QNMs originally found in~\cite{Jansen:2016zai,Jansen:2020ign}. Moreover, we also observe that when the 2RCBH plasma is progressively doped with values of R-charge chemical potential above the threshold of dominance of a purely imaginary QNM in the quintuplet channel of the $SO(3)$ rotation symmetry group of the homogeneous fluid, the late-time equilibration pattern of the pressure anisotropy gets increasingly deformed, eventually losing the oscillatory behavior observed at lower values of chemical potential, in consonance with the prediction made in~\cite{deOliveira:2024bgh}.

The manuscript is organized as follows. In section~\ref{sec:1stedoSystem}, we review the gravity actions for the holographic 2RCBH and 1RCBH models, the ansatze for bulk fields associated with the homogeneous isotropization dynamics, and outline the general algorithm used to numerically solve the field equations of motion. In section~\ref{sec:nearboundexp}, we discuss the ultraviolet near-boundary expansions of the bulk fields. In section~\ref{sec:renorm}, we discuss the renormalized one-point functions corresponding to the expectation values of the boundary energy-momentum tensor, the boundary vector current associated with the conserved R-charge, and the scalar condensate dual to the bulk scalar field, besides the holographic formula for the non-equilibrium dynamical entropy associated with the apparent horizon of bulk black brane solutions. In section~\ref{sec:far}, we discuss different topics relevant for the numerical solutions of the far-from-equilibrium homogeneous isotropization dynamics. Our numerical results are then presented in section~\ref{sec:results} and in section~\ref{sec:conc} we conclude with a summary of our main findings.

In the present work, we use natural units with $\hbar = c = k_B = 1$ and a mostly plus metric signature.

\section{Homogeneous isotropization for conformal Einstein-Maxwell-Dilaton models}
\label{sec:1stedoSystem}

Both the 2RCBH and 1RCBH models lie within the class of general holographic Einstein-Maxwell-Dilaton (EMD) constructions with five-dimensional holographically renormalized actions of the form~\cite{DeWolfe:2011ts,DeWolfe:2012uv,Critelli:2017euk,Rougemont:2024hpf,deOliveira:2024bgh,deOliveira:2025qwe},
\begin{equation}
\label{EMDaction}
    S=\frac{1}{2\kappa_5^2}\int_{\mathcal{M}_5} \dd ^5x~\sqrt{-g}\left[R-\frac{f(\phi)}{4}F_{\mu\nu}^2-\frac{1}{2}(\partial_\mu\phi)^2-V(\phi)\right]+S_{\text{GHY}}+S_{\text{ct}},
\end{equation}
where $\kappa_5^2\equiv 8\pi G_5$ is the five-dimensional gravitational constant, $V(\phi)$ is the dilaton potential and $f(\phi)$ is the Maxwell-dilaton coupling function. The two boundary terms in~\eqref{EMDaction} correspond respectively to: (a) the universal Gibbons-Hawking-York action~\cite{York:1972sj,Gibbons:1976ue} required for the specification of a well-posed variational problem with Dirichlet boundary conditions in spacetime manifolds with boundaries~\cite{Poisson:2009pwt}, and (b) a particular counterterm action constructed through the holographic renormalization procedure \cite{Bianchi:2001kw,Skenderis:2002wp,deHaro:2000vlm,Papadimitriou:2011qb,Lindgren:2015lia,Elvang:2016tzz}, which is needed for the systematic elimination of the divergences of the full on-shell boundary action, rendering finite predictions for the renormalized physical observables of the dual QFT at the boundary.

Within the class of general EMD actions, the 2RCBH and 1RCBH models are specified by the following particular dilaton functions~\cite{DeWolfe:2011ts,DeWolfe:2012uv,deOliveira:2024bgh},
\begin{align}
&\text{1RCBH model:} && \text{2RCBH model:}\\
&V(\phi) = -\frac{1}{L^2} \left(8 e^{\phi/\sqrt{6}} + 4 e^{-\sqrt{2/3}\,\phi} \right),&&V(\phi) = -\frac{1}{L^2} \left(8 e^{\phi/\sqrt{6}} + 4 e^{-\sqrt{2/3}\,\phi} \right),\label{eq:V}\\ 
&f(\phi) = e^{- 2\sqrt{2/3}\,\phi}, && f(\phi) = e^{\sqrt{2/3}\,\phi},\label{eq:f}
\end{align}
where $L$ is the radius of the asymptotically AdS$_5$ background geometry $\mathcal{M}_5$. The ultraviolet, near-boundary expansion of the dilaton potential~\eqref{eq:V}, $V(\phi\to 0)=[-12-2\phi^2+\mathcal{O}(\phi^4)]/L^2$, reveals a tachyonic dilaton mass, $m_\phi^2=\partial_\phi^2 V(\phi=0)=-4/L^2$, satisfying the Breitenlohner-Freedman (BF) bound, $m_\phi^2\ge -d^2/4L^2$, for stable asymptotically AdS$_{d+1}$ spacetimes \cite{Breitenlohner:1982jf,Breitenlohner:1982bm} (see also \cite{Ramallo:2013bua}), where $d$ is the spacetime dimension of the boundary ($d=4$ for the models considered in the present work). The holographic dictionary relates the scaling dimension $\Delta_\phi$ of the boundary QFT operator $\mathcal{O}_\phi$ dual to the bulk dilaton field $\phi$ with its mass according to, $m_\phi^2 L^2 = \Delta_\phi(\Delta_\phi-4)$, which then yields $\Delta_\phi=2$ for both the 2RCBH and 1RCBH models (for simplicity, we set $L=1$ from now on). Therefore, the counterterms for both models have the same general form of the counterterm action for the Coulomb branch of the holographic renormalization group flow of the SYM theory~\cite{Bianchi:2001kw}. Indeed, since from~\eqref{eq:f}, $f(\phi=0)=1$ is the same for both the 2RCBH and 1RCBH models, the counterterm action for the 2RCBH model is the same as for the 1RCBH model, which was discussed in detail in~\cite{Critelli:2017euk}.

The main goal of this work is to explore how a hot and dense strongly coupled fluid described holographically by the 2RCBH model, if prepared initially in a homogeneous but anisotropic state, equilibrates over time to become isotropic, while conserving both internal energy density and R-charge density. Additionally, we compare the obtained results with those found for the 1RCBH model in Refs.~\cite{Critelli:2017euk,Rougemont:2024hpf}. The far-from-equilibrium dynamics involved is known as \textit{homogeneous isotropization}~\cite{Chesler:2008hg,Chesler:2013lia,vanderSchee:2014qwa} and it can be implemented in our holographic settings by employing the following ansatze for the bulk fields in the generalized infalling Eddington-Finkelstein (EF) coordinates~\cite{Critelli:2017euk},
\begin{subequations}
\label{eq:ansatze}
\begin{align}
ds^2&= 2dv\left[dr-A(v,r) dv \right]+\Sigma(v,r)^2\left[e^{B(v,r)}(dx^2+dy^2)+e^{-2B(v,r)}dz^2\right],\label{eq:ansatzeds2}\\
A_\mu dx^\mu &= \Phi(v,r)dv,\\
\phi &= \phi(v,r).
\end{align}
\end{subequations}
In these ansatze, the bulk EF-time coordinate $v$ is related to the time coordinate $t$ at the boundary QFT according to,
\begin{equation}
dv = dt +\sqrt{-\frac{g_{rr}}{g_{tt}}}dr,
\label{eq:EFtime}
\end{equation}
a relation that ensures that ingoing radial null geodesics have a constant $v$, thereby removing coordinate singularities at the event horizon of bulk black brane solutions, while also guaranteeing outgoing null geodesics to obey $dr/dv=A(v,r)$~\cite{Chesler:2008hg}.\footnote{Note that $A(v,r)$ here and in~\cite{Critelli:2017euk,Rougemont:2024hpf} is half the corresponding metric function in the convention adopted in~\cite{Chesler:2008hg}.} In particular, because $g_{rr}$ and $g_{tt}$ are the radial and temporal components of the metric field of an asymptotically AdS$_{5}$ spacetime, the bulk EF time $v$ approaches the boundary time coordinate $t$ as the holographic radial coordinate tends to the boundary of the bulk manifold, $r\to\infty$. This follows from the Poincare form of the metric, where $\sqrt{-g_{rr}/g_{tt}} \to 1/r^2$ vanishes in the aforementioned limit. Moreover, the metric \eqref{eq:ansatzeds2} also has a residual diffeomorphism invariance under radial shifts, $r \to r + \lambda(v)$, where $\lambda(v)$ is an arbitrary function of the EF time $v$ \cite{Chesler:2013lia}.

Extremizing the general EMD action~\eqref{EMDaction} with respect to the bulk fields yields the EMD field equations~\cite{Critelli:2017euk,Rougemont:2024hpf},
\begin{subequations}
\label{EqsofMotion}
    \begin{align}
\label{eq:Einstein} R_{\mu\nu}-\frac{g_{\mu\nu}}{3}\left[V(\phi)-\frac{f(\phi)}{4}F_{\alpha\beta}F^{\alpha\beta}\right]-\frac{1}{2}\partial_\mu \phi \partial_\nu \phi-\frac{f(\phi)}{2} F_{\mu\rho}F_\nu\,^\rho &=0,\\
   \label{eq:Maxwell} \partial_\mu\left(\sqrt{-g}f(\phi) F^{\mu\nu}\right)&=0,\\
   \label{eq:Dilaton} \frac{1}{\sqrt{-g}} \partial_\mu\left(\sqrt{-g} g^{\mu\nu} \partial_\nu\phi\right) - \frac{\partial_\phi f(\phi)}{4}F_{\mu\nu}F^{\mu\nu}-\partial_\phi V(\phi)&=0.
\end{align}
\end{subequations}
By substituting the specific ansatze~\eqref{eq:ansatze} into the general EMD field equations \eqref{EqsofMotion}, one obtains the following set of coupled $1+1$ partial differential equations of motion for the homogeneous isotropization dynamics~\cite{Critelli:2017euk,Rougemont:2024hpf},
\begin{subequations}
\label{eq:EOMs}
\begin{align}
\partial_v\mathcal{E}+A\mathcal{E}'+\left(3\frac{d_+\Sigma}{\Sigma}+\frac{\partial_\phi f}{f}d_+\phi\right)\mathcal{E} &=0,\label{eq:pdea}\\
\frac{3 \Sigma '}{\Sigma } + \frac{\partial_\phi f}{f}\phi ' + \frac{\mathcal{E}'}{\mathcal{E}} &=0,\label{eq:pdeb}\\
4 \Sigma (d_+\phi)'+6 \phi'd_+\Sigma + 6 \Sigma'd_+\phi+\Sigma\mathcal{E}^2 \partial_\phi f - 2 \Sigma  \partial_\phi V &=0,\label{eq:pdec}\\
(d_+\Sigma)' +\frac{2 \Sigma '}{\Sigma }d_+\Sigma+\frac{\Sigma}{12} \left(2V + f \mathcal{E}^2\right) &=0,\label{eq:pded}\\
\Sigma \, (d_{+}B)'+\frac{3}{2}(B' d_+\Sigma + \Sigma'd_+ B) &=0,\label{eq:pdee}\\
A'' + \frac{1}{12} \left(18 B'd_{+}B-\frac{72 \Sigma'd_{+}\Sigma}{\Sigma^2} + 6 \phi'd_{+}\phi-7 f \mathcal{E}^2-2 V\right) &=0,\label{eq:pdef}\\
\Sigma'' + \frac{\Sigma}{6} \left(3 \left(B'\right)^2+\left(\phi '\right)^2\right) &=0,\label{eq:pdeg}\\
d_+(d_+\Sigma)+\frac{\Sigma}{2}(d_+B)^2-A'd_+\Sigma+\frac{\Sigma}{6}(d_+\phi)^2 &=0,\label{eq:pdeh}
\end{align}
\end{subequations}
where the prime denotes the radial derivative $\partial_r$, the special symbol defined as,
\begin{equation}
\label{eq:dirder}
    d_+\equiv\partial_v+A(v,r)\partial_r,
\end{equation} 
corresponds to the directional derivative along outgoing radial null geodesics, and, finally, 
\begin{equation}
    \mathcal{E}\equiv-\Phi',
\end{equation}
is the bulk ``electric field''~\cite{Fuini:2015hba}. Eqs.~\eqref{eq:pdea} and~\eqref{eq:pdeb} are the non-trivial components of Maxwell's equations, Eq.~\eqref{eq:pdec} is the dilaton equation, and Eqs.~\eqref{eq:pded} ---~\eqref{eq:pdeh} are the non-trivial components of Einstein's equations. Notice there are five unknown functions, $\{A(v,r),\Sigma(v,r),B(v,r),\phi(v,r),\mathcal{E}(v,r)\}$, which are to be determined by the five dynamical equations of motion~\eqref{eq:pdeb} ---~\eqref{eq:pdef}, besides three constraints given by Eqs.~\eqref{eq:pdea},~\eqref{eq:pdeg}, and~\eqref{eq:pdeh}. Eqs.~\eqref{eq:pdeb} ---~\eqref{eq:pdeg} correspond to a nested set of differential equations which can be numerically solved in succession, while the constraint Eqs.~\eqref{eq:pdea} and~\eqref{eq:pdeh} may be used to check the accuracy of the corresponding numerical solutions~\cite{Rougemont:2024hpf}.

It follows from Eq.~\eqref{eq:pdeb} the following relation between $\phi$, $\Sigma$, and $\mathcal{E}$,
    \begin{align}
    &\text{1RCBH model:} && \text{2RCBH model:}\nonumber\\
&\ln\left(\Sigma^3\mathcal{E}\right)-2\sqrt{\frac{2}{3}}\phi=\text{constant}, &&
        \ln\left(\Sigma^3\mathcal{E}\right)+\sqrt{\frac{2}{3}}\phi=\text{constant},\label{eq:maxwellrel}
    \end{align}
and in the next section, we shall connect the still unspecified constants in~\eqref{eq:maxwellrel} to the $U(1)$ R-charge density $\rho_c$ for each model.

The approach followed here is called the \textit{characteristic formulation of numerical general relativity}~\cite{Bondi:1962px,Sachs:1962wk}, as adequately adapted by Chesler and Yaffe in~\cite{Chesler:2008hg,Chesler:2013lia} to deal with asymptotically AdS spacetimes, which are the kind of manifolds featured in holographic models. In the characteristic formulation, one foliates the bulk spacetime manifold in slices of constant $v$, which in the EF coordinates correspond to null hypersurfaces, and then evolves the equations of motion in the EF time. In this formulation, the set of coupled partial differential equations of motion for the bulk fields is organized into a nested structure of equations which can be systematically solved in succession. Another typical feature of the characteristic formulation is that the dynamical equations of motion to be solved are of first order in time derivatives. Then, the algorithm used here to solve the nested system of $1+1$ partial differential equations~\eqref{eq:EOMs} comprises the following general steps~\cite{Critelli:2017euk,Rougemont:2024hpf}:
\begin{enumerate}
    \item Choose initial profiles for the metric anisotropy function $B(v_0,r)$ and the dilaton field $\phi(v_0,t)$ together with a numerical value for the dimensionless ratio of R-charge chemical potential over temperature, $\mu/T$, where we take here $v_0=0$ as the initial time slice. If one works with a non-trivial radial shift function $\lambda(v)$, its initial value must also be specified, and we choose here $\lambda(v_0=0)=0$ (more about the radial shift function will be discussed in section~\ref{sec:AH});
    \item Using $B(v_0,r)$ and $\phi(v_0,r)$, radially solve Eq.~\eqref{eq:pdeg} to obtain $\Sigma(v_0,r)$, allowing thus to determine the value of $\mathcal{E}(v_0,r)$ through the relation~\eqref{eq:maxwellrel}, which follows from~\eqref{eq:pdeb};
    \item Given the above determined profiles for $B(v_0,r)$, $\phi(v_0,r)$, $\Sigma(v_0,r)$, and $\mathcal{E}(v_0,r)$, radially solve~\eqref{eq:pded} to obtain $\dd_+\Sigma(v_0,r)$;
    \item Next radially solve~\eqref{eq:pdee} to obtain $\dd_+ B(v_0,t)$;
    \item Proceed by radially solving~\eqref{eq:pdec} to obtain $\dd_+ \phi(v_0,t)$;
    \item Next radially solve~\eqref{eq:pdef} to obtain $A(v_0,t)$;
    \item With $B(v_0,r)$, $\dd_+ B(v_0,r)$, $\phi(v_0,r)$, and $\dd_+ \phi(v_0,r)$ at hand, it follows from the definition given in~\eqref{eq:dirder} the expressions for $\partial_v B(v_0,r)$ and $\partial_v\phi(v_0,r)$. Using $\{B(v_0,r),\partial_vB(v_0,r)\}$ and $\{\phi(v_0,r),\partial_v \phi(v_0,r)\}$ so obtained, one has the necessary initial conditions required to evolve $B(v_0,r)$ and $\phi(v_0,r)$ in time from $v_0$ to $v_0+\Delta v$, where $\Delta v$ is the chosen numerical time step;
    \item Repeat the above process to obtain the fields for all the subsequent time slices by using discrete numerical integration techniques (here we use the pseudospectral method \cite{boyd01} to deal with integration in the radial direction, while integration in the time direction is done with the fourth-order Adams-Bashforth method). The procedure is iterated until reaching any desired end time $v_\textrm{end}$ for the numerical simulations.
    \end{enumerate}

The accuracy of the numerical solutions can then be checked through the constraint Eqs.~\eqref{eq:pdea} and~\eqref{eq:pdeh}.

\section{Near-boundary expansions of the bulk fields}
\label{sec:nearboundexp}

In this section, we consider the ultraviolet (UV) near-boundary expansions for the bulk fields in the 2RCBH and 1RCBH models undergoing homogeneous isotropization dynamics. For both models, the UV expansions of the EMD fields can be cast into the following form~\cite{Critelli:2017euk,Rougemont:2024hpf},
\begin{subequations}
\label{eq:nearexpansion}
\begin{align}
A(v,r) & = \frac{\left[r+\lambda(v)\right]^2}{2}-\partial_v\lambda(v) + \sum_{n=1}^{\infty}\frac{A_n(v)}{r^n}, \label{eq:expA}\\
\Sigma(v,r) & = r+\lambda(v) + \sum_{n=1}^{\infty}\frac{\Sigma_n(v)}{r^n}, \label{eq:expSig}\\
B(v,r) & = \sum_{n=1}^{\infty}\frac{B_n(v)}{r^n}, \label{eq:expB}\\
\phi(v,r) & = \sum_{n=2}^{\infty}\frac{\phi_n(v)}{r^n}, \label{eq:expphi}\\
\Phi(v,r) & = \Phi_0(v) + \sum_{n=2}^{\infty}\frac{\Phi_n(v)}{r^n}.\label{eq:expPhi}
\end{align}
\end{subequations}

If one substitutes the near-boundary expansions~\eqref{eq:nearexpansion} into the equations of motion~\eqref{eq:EOMs} and solves for all possible UV coefficients in terms of the others, the UV asymptotic behavior of the EMD fields reads as follows,

\begin{subequations}
\label{eq:UVexp}
    \begin{align}
       A(v,r)&=\frac{r^2}{2}+r \lambda (v)+\frac{\lambda (v)^2}{2} -\dot{\lambda}(v)+\frac{H-\frac{1}{18} \phi _2(v){}^2}{r^2}+\frac{\lambda (v) \left(\frac{1}{9} \phi _2(v){}^2-2 H\right)-\frac{1}{18} \phi _2(v) \dot{\phi_2}(v)}{r^3}+\mathcal{O}(r^{-4}),\label{eq:UVexpA}\\
       \Sigma(v,r)&=r+\lambda (v)-\frac{\phi _2(v){}^2}{18 r^3}+\frac{\phi _2(v) \left(5 \lambda (v) \phi _2(v)-3 \dot{\phi} _2(v)\right)}{30 r^4}+\mathcal{O}(r^{-5}),\label{eq:UVexpS}\\
       \dd_+\Sigma(v,r)&=\frac{r^2}{2}+r \lambda (v)+\frac{\lambda (v)^2}{2}+\frac{H+\frac{1}{36} \phi _2(v){}^2}{r^2}+\mathcal{O}(r^{-3}),\label{eq:UVexpdS}\\
       B(v,r)&=\frac{B_4(v)}{r^4}+\frac{\dot{B_4}(v)-4 B_4(v) \lambda (v)}{r^5}+\mathcal{O}(r^{-6}),\label{eq:UVexpB}\\
       \dd_+ B(v,r)&=-\frac{2 B_4(v)}{r^3}+\frac{6 B_4(v) \lambda (v)-\frac{3}{2}\dot{B}_4(v)}{r^4}+\mathcal{O}(r^{-6}),\label{eq:UVexpdB}\\
       \phi(v,r)&=\frac{\phi _2(v)}{r^2}+\frac{\dot{\phi_2}(v)-2 \lambda (v) \phi _2(v)}{r^3}+\frac{-3 \lambda (v) \dot{\phi_2}(v)+3 \lambda (v)^2 \phi _2(v)+\frac{1}{12} \left(9 \ddot{\phi _2}(v)+\sqrt{6} \phi _2(v){}^2\right)}{r^4}+\mathcal{O}(r^{-5}),\label{eq:UVexpphi}\\
       \dd_+\phi(v,r)&=-\frac{\phi_2(v)}{r}+\frac{\lambda (v) \phi_2(v)-\frac{1}{2} \dot{\phi_2}(v)}{r^2}+\mathcal{O}(r^{-3}),\label{eq:UVexpdphi}
    \end{align}
\end{subequations}
where the dot designates differentiation with respect to the EF time $v$, and all the above UV expansions are identical for both the 2RCBH and 1RCBH models up to order $\mathcal{O}(r^{-5})$. At this order, only the UV expansion of the bulk field $\Phi(v,r)$ differs explicitly in both models,

\begin{align}
    &\text{1RCBH model:} &&\text{2RCBH model:}\nonumber\\
    &\Phi(v,r)=\Phi _0(v)+\frac{\Phi _2(v)}{r^2}-\frac{2 \lambda (v) \Phi _2(v)}{r^3} &&\Phi(v,r)=\Phi _0(v)+\frac{\Phi _2(v)}{r^2}-\frac{2 \lambda (v) \Phi _2(v)}{r^3}\\
    &+\frac{ 9 \lambda (v)^2+\sqrt{6} \phi _2(v)}{3 r^4}\Phi_2(v)+\mathcal{O}(r^{-5}), &&+\frac{ 18 \lambda (v)^2+\sqrt{6} \phi _2(v)}{6 r^4}\Phi_2(v)+\mathcal{O}(r^{-5}).\nonumber
\end{align}

By analyzing the imbalance between the number of equations and independent variables one concludes that this near-boundary treatment cannot uniquely determine five UV coefficients: $A_2(v)$ (or, interchangeably, the coefficient $H$ defined below), $B_4(v)$, $\phi_2(v)$, $\Phi_0(v)$, and $\Phi_2(v)$. The conclusion we draw from this fact is that, with the sole exception of $\Phi_0$, these coefficients are of a \textit{dynamical} nature, i.e.~besides the near-boundary expansions, they also require the full numerical bulk solutions to be determined. On the other hand, the coefficient $\Phi_0(v)$ is fixed by the Dirichlet boundary condition for the bulk Maxwell field, which imposes that its boundary value corresponds to the R-charge chemical potential of the dual strongly coupled fluid in thermodynamic equilibrium (which is attained only at asymptotically large times),
\begin{equation}
    \mu = \lim_{v\to\infty}\lim_{r\to\infty}\Phi(v,r) = \lim_{v\to\infty}\Phi_0(v).
\end{equation}

In the case of the UV coefficient $H$ defined as,
\begin{equation}
\label{eq:Hdef}
    H\equiv A_2(v)+\frac{\phi_2(v)^2}{18},
\end{equation}
by manipulating the near-boundary expansions of the equations of motion to order $\mathcal{O}(r^{-3})$, we find that $H$ is a constant of motion. In fact, we will later show in Eq.~\eqref{eq:varepsilon} that this coefficient gives (up to a numerical factor) the conserved internal energy density of the homogeneous medium.
 
Another constant of motion is the UV coefficient $\Phi_2(v)$, a fact that can be demonstrated by manipulating the near-boundary expansions of the equations of motion to order $\mathcal{O}(r^{-5})$. Additionally, using the relation~\eqref{eq:maxwellrel} together with Eqs.~\eqref{eq:UVexp} one ends up with the following expressions,
\begin{align}
&\text{1RCBH model:} &&\text{2RCBH model:}\nonumber\\&\mathcal{E}(v,r)= 2\Phi_{2}\Sigma(v,r)^{-3}e^{2\sqrt{\frac{2}{3}}\phi(v,r)},
&& \mathcal{E}(v,r) = 2\Phi_{2}\Sigma(v,r)^{-3}e^{-\sqrt{\frac{2}{3}}\phi(v,r)}.
\label{eq:Erhofield2R2}
\end{align}
Indeed, as we will later show the fact that the UV coefficient $\Phi_2(v)$ is constant implies that the R-charge density is conserved in the homogeneous medium.

Thus, one concludes that only the dynamical UV coefficients $B_4(v)$ and $\phi_2(v)$ are time-dependent quantities, which, as we shall see, are related, respectively, to the pressure anisotropy and the scalar condensate of the strongly interacting fluid at the dual boundary QFT.

\section{Renormalized one-point functions}
\label{sec:renorm}

To investigate the homogeneous isotropization and thermalization dynamics of the 2RCBH and 1RCBH models, we shall compute the renormalized one-point functions of different boundary QFT operators. We consider here the expectation value of the energy-momentum tensor $\langle T_{\mu\nu}\rangle$, holographically dual to the bulk metric field $g_{\mu\nu}$; the expectation value of the $U(1)$ R-charge vector current $\langle J^\mu\rangle$, holographically dual to the bulk Maxwell field $A_{\mu}$; and the scalar condensate $\langle \mathcal{O}_\phi\rangle$, holographically dual to the bulk dilaton field $\phi$. We refer to isotropization as the stabilization of the pressure anisotropy in the zero value within some numerical tolerance, while the latter thermalization refers to the equilibration of the observable, which takes longer to attain thermodynamic equilibrium in the system, which is generally the scalar condensate.

The holographic renormalization procedure is usually implemented in the Fefferman-Graham (FG) coordinates, where the bulk spacetime metric takes the following form~\cite{Critelli:2017euk,Bianchi:2001kw},
\begin{equation}
\label{eq:ansatzequil}
    \dd s^2_{\text{FG}}=\frac{\dd \rho^2}{4\rho^2}+\gamma_{\mu\nu}\,\dd x^\mu \dd x^{\nu},
\end{equation}
where $\rho$ is the FG radial coordinate, in terms of which the boundary is located at $\rho=0$, and in~\eqref{eq:ansatzequil} the Greek indices $(\mu,\nu)$ run over the coordinates of the dual QFT; here $\gamma_{\mu\nu}$ corresponds to the induced metric on a constant-$\rho$ hypersurface, being related to the bulk metric by $g_{\mu\nu}(x)=\rho\,\gamma_{\mu\nu}(\rho,x)$.

In terms of the FG coordinates, the EMD field expansions near the boundary take the following form (where the coefficients of the logarithmic terms below vanish for the conformal 2RCBH and 1RCBH plasmas)~\cite{Critelli:2017euk,Bianchi:2001kw},
\begin{subequations}
\label{eq:NearBoundExpOPF}
    \begin{align}
    \gamma_{\mu\nu}(\rho,x)&=\frac{1}{\rho}\gamma_{(0)\mu\nu}(x)+\gamma_{(2)\mu\nu}(x)+\gamma_{(2,1)\mu\nu}(x)\ln\rho\nonumber\\
    &+\rho\left[\gamma_{(4)\mu\nu}(x)+\gamma_{(4,1)\mu\nu}\ln \rho+\gamma_{(4,2)\mu\nu}\ln^2 \rho\right]+\mathcal{O}(\rho²),\\
    A_\mu(\rho,x)&=A_{(0)\mu}(x)+\rho\left[A_{(2)\mu}(x)+A_{(2,1)\mu}(x)\ln\rho\right]+\mathcal{O}(\rho^2),\\
    \phi(\rho,x)&=\rho\left[\phi_{(0)}(x)+\phi_{(0,1)}(x)\ln\rho\right]+\mathcal{O}(\rho^2).
\end{align}
\end{subequations}

As discussed in detail in Appendix A of~\cite{Critelli:2017euk}, for the 1RCBH model, and also for the 2RCBH model, the aforementioned renormalized one-point functions can be expressed in terms of the UV coefficients of the FG near-boundary expansions~\eqref{eq:NearBoundExpOPF} as follows,
\begin{subequations}
\label{eq:opf}
    \begin{align}
        \langle T_{\mu\nu} \rangle &= \frac{1}{\kappa_5^2}\left(2\gamma_{(4)\mu\nu}+\gamma_{(0)\mu\nu}\frac{\phi_{(0)}^2}{6}\right),\label{eq:optmn}\\
        \langle J_{\mu} \rangle&=\frac{1}{\kappa_5^2}A_{(2)\mu},\label{eq:opJm}\\
         \langle \mathcal{O}_{\phi} \rangle&=-\frac{1}{\kappa_5^2}\phi_{(0)}.\label{eq:opphi}
    \end{align}
\end{subequations}
To relate the FG UV coefficients appearing in Eqs.~\eqref{eq:opf} to the UV coefficients in the EF coordinates, one needs to obtain a map between the different coordinate charts, as we are going to discuss in a moment.

\subsection{Equilibrium solutions}

Equilibrium solutions for both the 2RCBH and the 1RCBH models are represented by static, spatially homogeneous and isotropic charged EMD black holes, as encoded in the following ansatze written in modified EF coordinates denoted with a tilde~\cite{Critelli:2017euk},
\begin{align}
\label{eq:AnsatzEqs}
\dd s^2=\dd v\left[2e^{ a(\tilde{r})+b(\tilde{r})}\dd\tilde{r}-e^{ 2a(\tilde{r})}h(\tilde{r})\dd v\right]+e^{2 a(\tilde{r})}\dd \mathbf{x}^2,\qquad A_\mu = \Phi(\tilde{r})\delta_\mu^0,\qquad \phi=\phi(\tilde{r}).
\end{align}
In thermodynamic equilibrium, analytical solutions can be found for the EMD field equations for the 2RCBH and 1RCBH models, which are specified in terms of the black hole charge $Q$ and mass $M$ as follows (in units of $L=1$)~\cite{DeWolfe:2011ts,DeWolfe:2012uv,Finazzo:2016psx,Critelli:2017euk,deOliveira:2024bgh},
\begin{subequations}
\label{eq:AnsatzAll}
    \begin{align}
    &\text{1RCBH model:}\qquad &&\text{2RCBH model:}\nonumber \\
   &a(\tilde{r})=\ln \tilde{r} +\frac{1}{6}\ln\left(1+\frac{Q^2}{\tilde{r}^2}\right),\qquad & &a(\tilde{r})=\ln \tilde{r} +\frac{1}{3}\ln\left(1+\frac{Q^2}{\tilde{r}^2}\right),\\ 
   &b(\tilde{r})=-\ln \tilde{r} -\frac{1}{3}\ln\left(1+\frac{Q^2}{\tilde{r}^2}\right),\qquad & &b(\tilde{r})=-\ln \tilde{r} -\frac{2}{3}\ln\left(1+\frac{Q^2}{\tilde{r}^2}\right),\\
   &h(\tilde{r})=1-\frac{M^2}{\tilde{r}^2(\tilde{r}^2+Q^2)}, \qquad & &h(\tilde{r})=1-\frac{M^2}{(\tilde{r}^2+Q^2)^2},\\
   &\phi(\tilde{r})=-\sqrt{\frac{2}{3}}\ln\left(1+\frac{Q^2}{\tilde{r}^2}\right),& &\phi(\tilde{r})=\sqrt{\frac{2}{3}}\ln\left(1+\frac{Q^2}{\tilde{r}^2}\right),\\
   &\Phi(\tilde{r}) = -\frac{MQ}{\tilde{r}^2+Q^2}+\frac{MQ}{\tilde{r}_{H}^2+Q^2} ,&&\Phi(\tilde{r}) = -\frac{\sqrt{2}MQ}{\tilde{r}^2+Q^2}+\frac{\sqrt{2}MQ}{\tilde{r}_{H}^2+Q^2},
\end{align}
\end{subequations}
where the radial position of the black brane event horizon is given by the largest real zero of the blackening function $h(r)$,
\begin{align}
    &\text{1RCBH model:}&&\text{2RCBH model:}\nonumber\\
    &\tilde{r}_{H}=\sqrt{\frac{1}{2}\left(\sqrt{Q^4+4M^2}-Q^2\right)},  
    &&\tilde{r}_{H}=\sqrt{M-Q^2}.\label{eq:bbrp}
\end{align}

\subsection{One-point functions in thermodynamic equilibrium}
\label{sec:equilibriumOPf}

To calculate the equilibrium one-point functions $\langle T_{\mu\nu}\rangle_{\text{eq}}$, $\langle J^\mu\rangle_{\text{eq}}$ and $\langle \mathcal{O}_\phi\rangle_{\text{eq}}$ from Eqs.~\eqref{eq:opf}, one needs first to find a map relating the FG radial coordinate $\rho$, in terms of which the holographic renormalization program was carried out, and the modified EF radial coordinate $\tilde{r}$, in terms of which the above equilibrium solutions are written. For this sake, one first rewrites in diagonal form the equilibrium line element given in~\eqref{eq:AnsatzEqs},
\begin{align}
\label{eq:AnsatzEqs2}
\dd s^2=e^{2a(\tilde{r})}\left[-h(\tilde{r})\dd t^2 +\dd \mathbf{x}^2\right]+\frac{e^{2 b(\tilde{r})}}{h(\tilde{r})}\dd\tilde{r}^2,
\end{align}
and then imposes that $(g_{\tilde{r}\tilde{r}}\,\dd \tilde{r}^2)_\textrm{EF}=(g_{\rho\rho}\,\dd \rho^2)_\textrm{FG}$, leading to the following integral relation,
\begin{equation}
    \int \frac{e^{b(\tilde{r})}}{\sqrt{h(\tilde{r})}}\dd \tilde{r}=- \int \frac{\dd\rho}{2\rho}=-\frac{1}{2}\ln\rho,
\end{equation}
where the minus sign follows from the fact that $\tilde{r}\to\infty\Rightarrow\rho\to 0$. Solving the above integral perturbatively in $\tilde{r}$ close to the boundary using the equilibrium solutions, it follows that,
\begin{align}
    &\text{1RCBH model:}&&\tilde{r}(\rho)=\frac{1}{\sqrt{\rho }}-\frac{\sqrt{\rho } Q^2}{6}+\frac{1}{72} \rho ^{3/2} \left(9 M^2+Q^4\right)+\mathcal{O}(\rho^{5/2}),\\
    &\text{2RCBH model:}&&\tilde{r}(\rho)=\frac{1}{\sqrt{\rho }}-\frac{\sqrt{\rho } Q^2}{3}+\frac{1}{72} \rho ^{3/2} \left(9 M^2-2Q^4\right)+\mathcal{O}(\rho^{5/2}).
\end{align}

By expanding the equilibrium solutions~\eqref{eq:AnsatzAll} close to the boundary using the diagonal coordinates~\eqref{eq:AnsatzEqs2} and then using the above expressions for $\tilde{r}(\rho)$ to compare with the form of the FG UV expansions~\eqref{eq:ansatzequil}, one finds that,
\begin{subequations}
\label{eq:coeff}
    \begin{align}
        &\text{1RCBH model:}&&\gamma_{tt}(\rho)=-\frac{1}{\rho }+\rho  \left(\frac{3 M^2}{4}+\frac{Q^4}{18}\right)+\rho ^2 \left(-\frac{7 M^2 Q^2}{24}-\frac{13 Q^6}{648}\right)+\mathcal{O}(\rho^3),\\
        & &&\gamma_{xx}(\rho)=\frac{1}{\rho }+\rho  \left(\frac{ M^2}{4}-\frac{Q^4}{18}\right)+\rho ^2 \left(-\frac{ M^2 Q^2}{24}+\frac{13 Q^6}{648}\right)+\mathcal{O}(\rho^3),\\
        & && \phi(\rho)=-\sqrt{\frac{2}{3}} \rho  Q^2+\frac{\rho ^2 Q^4}{3 \sqrt{6}}+\mathcal{O}(\rho^3),\\
         & &&\Phi(\rho)=\frac{M Q}{Q^2+\tilde{r}_H^2}-M \rho  Q+\frac{2}{3} M \rho ^2 Q^3+\mathcal{O}(\rho^3),\\
         & &&\nonumber\\
         &\text{2RCBH model:}&&\gamma_{tt}(\rho)=-\frac{1}{\rho }+\rho  \left(\frac{3 M^2}{4}+\frac{Q^4}{18}\right)+\rho ^2 \left(\frac{Q^6}{162}-\frac{7 M^2 Q^2}{12}\right)+\mathcal{O}(\rho^3),\\
         & &&\gamma_{xx}(\rho)=\frac{1}{\rho }+\rho  \left(\frac{M^2}{4}-\frac{Q^4}{18}\right)+\rho ^2 \left(-\frac{M^2 Q^2}{12}-\frac{Q^6}{162}\right)+\mathcal{O}(\rho^3),\\
         & && \phi(\rho)=\sqrt{\frac{2}{3}} \rho  Q^2+\frac{\rho ^2 Q^4}{3 \sqrt{6}}+\mathcal{O}(\rho^3),\\
         & &&\Phi(\rho)=\frac{\sqrt{2} M Q}{Q^2+r_H^2}-\sqrt{2} M \rho  Q+\frac{1}{3} \sqrt{2} M \rho ^2 Q^3+\mathcal{O}(\rho^3).
    \end{align}
\end{subequations}

By comparing Eqs.~\eqref{eq:coeff} with Eqs.~\eqref{eq:NearBoundExpOPF},\footnote{As discussed in~\cite{Critelli:2017euk}, the coefficients of the logarithmic terms in the FG UV expansions~\eqref{eq:NearBoundExpOPF} vanish for the 1RCBH model, and also for the 2RCBH model, due to the conformal symmetry of both models, the conformal flatness of the boundary, and the scaling dimension $\Delta_\phi=2$ of the scalar QFT operator $\mathcal{O}_\phi$ dual to the bulk dilaton field $\phi$ in both models.} one identifies the FG UV expansion coefficients in thermodynamic equilibrium, and by substituting them into Eqs.~\eqref{eq:opf} for the renormalized one-point functions, it follows that,\footnote{We use the holographic relation, $\kappa_5^2=8\pi G_5 = 4\pi^2L^3/N_c^2=4\pi^2/N_c^2$ (in units of $L=1$), which is valid for the SYM theory~\cite{Gubser:1996de,Natsuume:2014sfa}, where $N_c$ is the number of color charges at the boundary gauge theory.}
\begin{subequations}
    \begin{align}
    &\text{1RCBH model:} &&\text{2RCBH model:}\nonumber\\
    \varepsilon_{\text{eq}}&\equiv\langle T_{tt}\rangle_{\text{eq}} = \frac{1}{\kappa_5^2}\frac{3M^2}{2} =\frac{N_c^2}{4\pi^2}\frac{3M^2}{2}, && \varepsilon_{\text{eq}}\equiv\langle T_{tt}\rangle_{\text{eq}}= \frac{1}{\kappa_5^2}\frac{3M^2}{2} =\frac{N_c^2}{4\pi^2}\frac{3M^2}{2},\label{eq:eeq}\\
    p_{\text{eq}}&\equiv\langle T_{xx}\rangle_{\text{eq}} = \frac{1}{\kappa_5^2}\frac{M^2}{2} =\frac{N_c^2}{4\pi^2}\frac{M^2}{2}, && p_{\text{eq}}\equiv\langle T_{xx}\rangle_{\text{eq}} = \frac{1}{\kappa_5^2}\frac{M^2}{2} =\frac{N_c^2}{4\pi^2}\frac{M^2}{2},\label{eq:peq}\\
    \rho_{c,{\text{eq}}}&\equiv\langle J^t \rangle_{\text{eq}} = \frac{1}{\kappa_5^2}MQ =\frac{N_c^2}{4\pi^2}MQ, &&  \rho_{c,{\text{eq}}}\equiv\langle J^t \rangle_{\text{eq}} = \frac{1}{\kappa_5^2}\sqrt{2}MQ =\frac{N_c^2}{4\pi^2}\sqrt{2}MQ,\label{eq:rhoeq}\\
    \langle \mathcal{O}_\phi \rangle_{\text{eq}} &= \frac{1}{\kappa_5^2}\sqrt{\frac{2}{3}}Q^2 = \frac{N_c^2}{4\pi^2}\sqrt{\frac{2}{3}}Q^2, && 
    \langle \mathcal{O}_\phi \rangle_{\text{eq}} = -\frac{1}{\kappa_5^2}\sqrt{\frac{2}{3}}Q^2 = -\frac{N_c^2}{4\pi^2}\sqrt{\frac{2}{3}}Q^2,\label{eq:ScalCondEq}
\end{align}
\end{subequations}
where $\varepsilon$ is the energy density, $p$ is the isotropic equilibrium pressure, and $\rho_c$ is the $U(1)$ R-charge density at the boundary QFT in thermodynamic equilibrium.

By inspecting the expressions of the equilibrium solutions in Eqs.~\eqref{eq:AnsatzAll} and the radial position of the black brane event horizon in Eqs.~\eqref{eq:bbrp}, one notices that both 2RCBH and 1RCBH equilibrium backgrounds are described in terms of two non-negative parameters, namely $(Q,M)$ or, equivalently, $(Q,\tilde{r}_H)$. The expressions for the Hawking temperature of the black brane and the $U(1)$ R-charge chemical potential are given by (in units of $L=1$)~\cite{DeWolfe:2011ts,DeWolfe:2012uv,Finazzo:2016psx,Critelli:2017euk,deOliveira:2024bgh},
\begin{subequations}
\begin{align}
&\text{1RCBH model:} &&\text{2RCBH model:}\nonumber\\
&T=\frac{\sqrt{-(g_{tt})'(g^{\tilde{r}\tilde{r}})'}}{4\pi}\Bigg|_{\tilde{r}=\tilde{r}_{H}}=\frac{Q^2+2\tilde{r}_{H}^2}{2\pi\sqrt{Q^2+\tilde{r}_{H}^2}}, &&T=\frac{\tilde{r}_{H}}{\pi},\label{eq:TThermDef}\\
&\mu=\lim_{\tilde{r}\to\infty}\Phi(\tilde{r})=\frac{\tilde{r}_{H}Q}{\sqrt{Q^2+\tilde{r}_{H}^2}}, &&\mu=\sqrt{2}Q.
\end{align}
\end{subequations}

\begin{figure}
\centering  
\subfigure[Entropy density]{\includegraphics[width=0.425\linewidth]{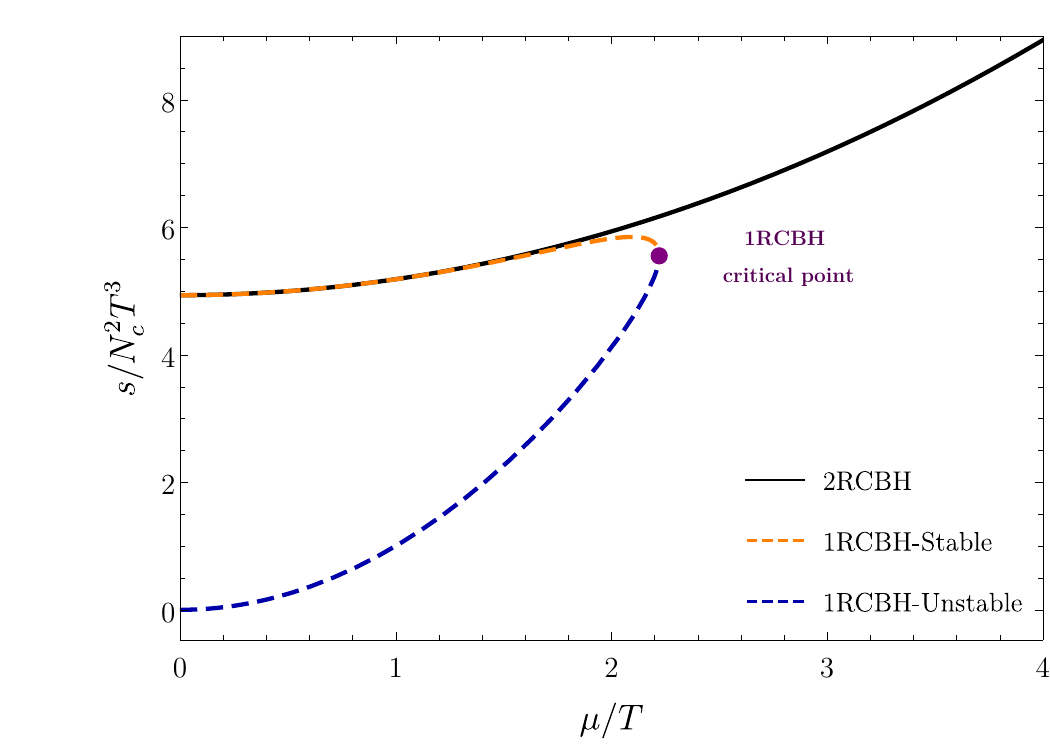}\label{fig:sThermo}}
\subfigure[R-charge density]{\includegraphics[width=0.425\linewidth]{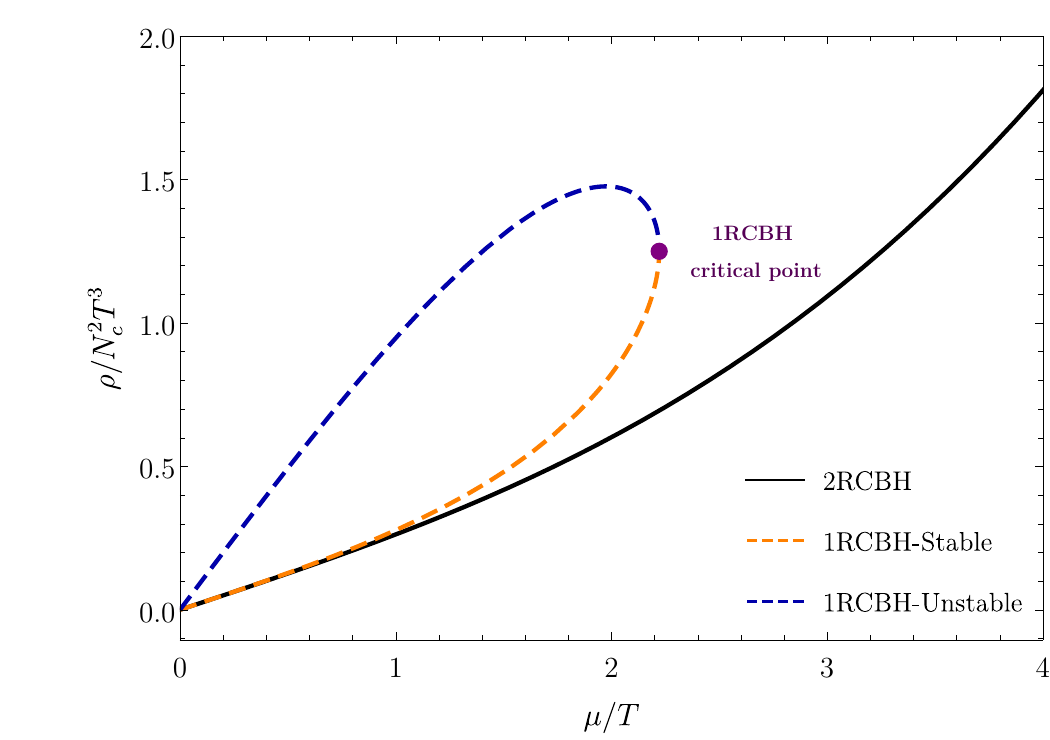}\label{fig:rhoThermo}}
\subfigure[Pressure]{\includegraphics[width=0.425\linewidth]{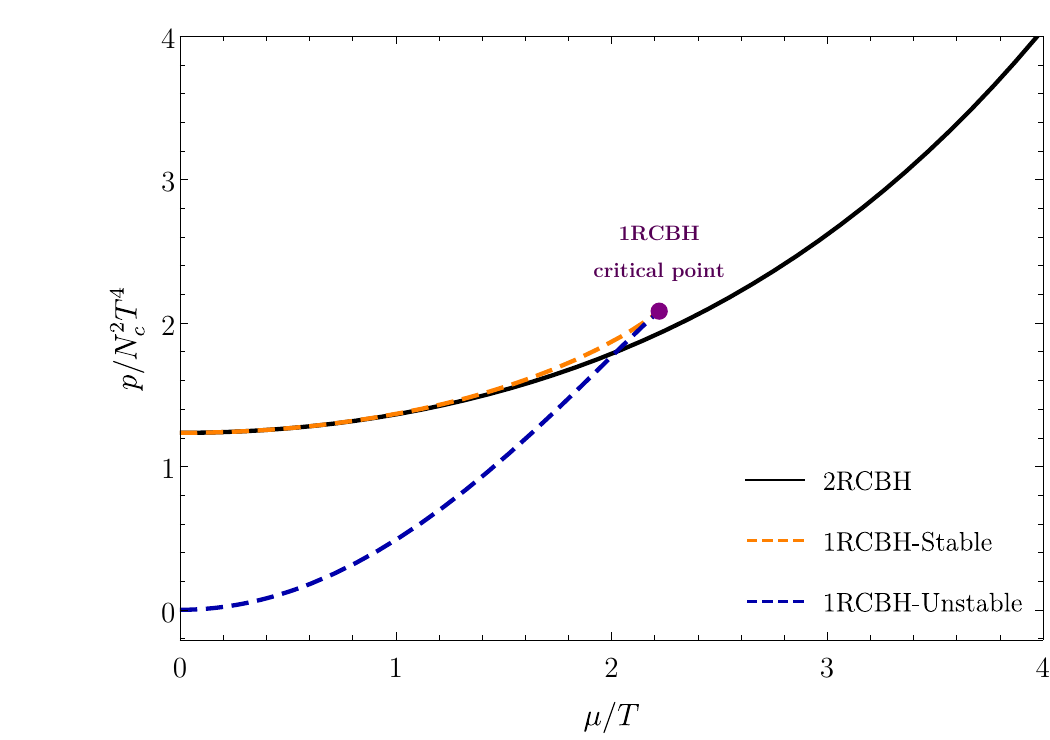}\label{fig:pThermo}}
\subfigure[Scalar condensate]{\includegraphics[width=0.425\linewidth]{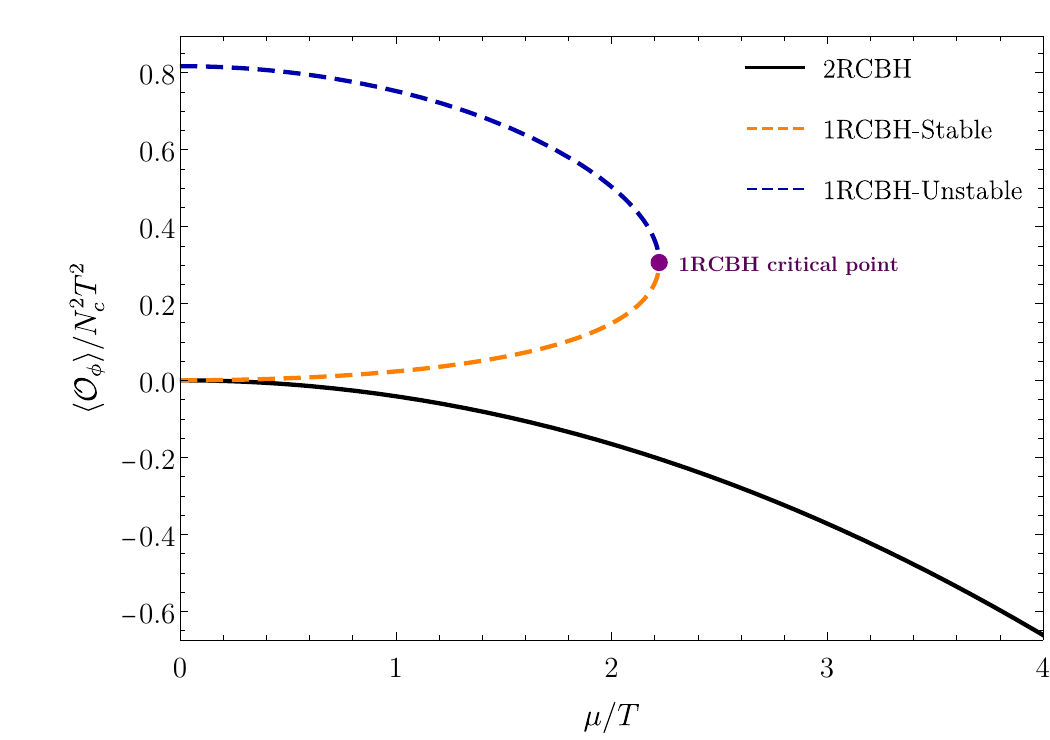}\label{fig:OThermo}}
\subfigure[Specific heat]{\includegraphics[width=0.425\linewidth]{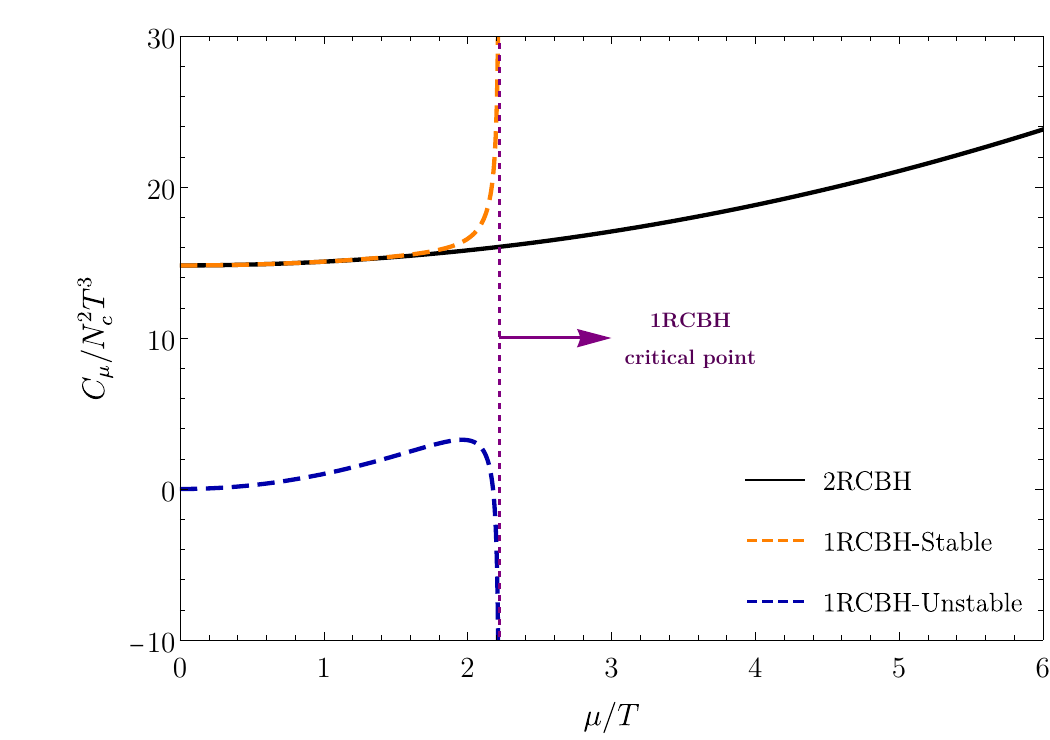}\label{fig:CThermo}}
\subfigure[R-charge susceptibility]{\includegraphics[width=0.425\linewidth]{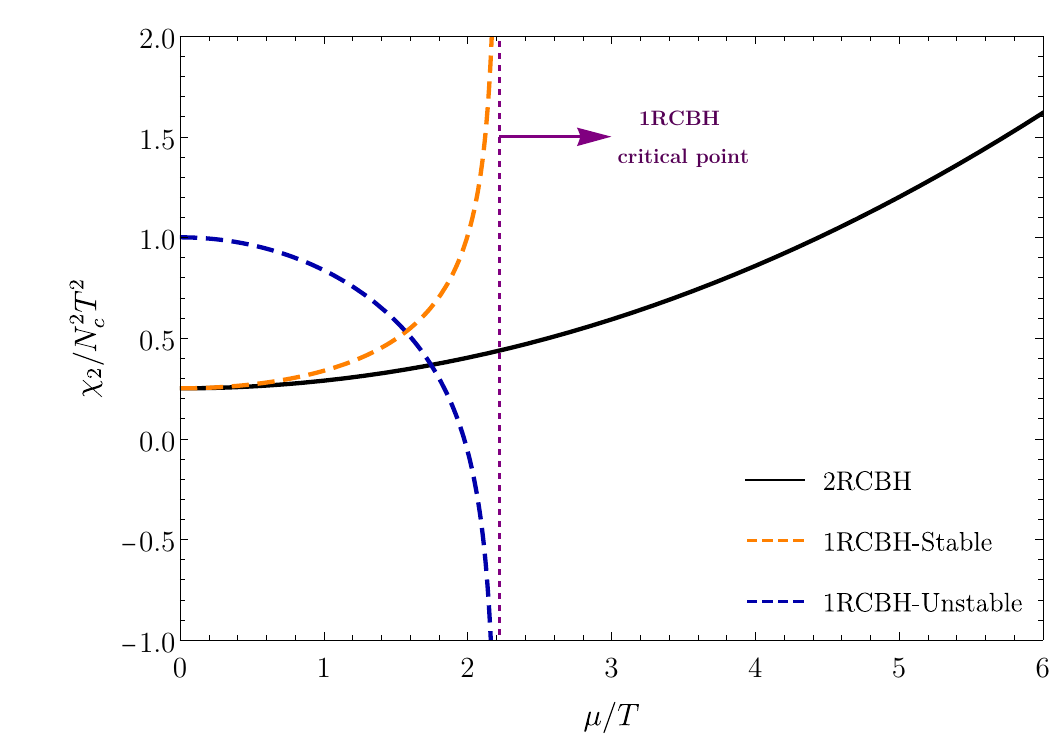}\label{fig:chi2Thermo}}
\caption{Equation of state, scalar condensate, specific heat, and R-charge susceptibility for the 2RCBH and 1RCBH models.}
\label{fig:Thermo}
\end{figure}

The bulk dimensionless ratio $Q/\tilde{r}_H$ specifying the different equilibrium black hole solutions in the gravity theory, and the boundary dimensionless ratio $\mu/T$ specifying the different thermodynamic equilibrium states at the dual boundary QFT, can then be related as follows,

\begin{align}
&\text{1RCBH model:} &&\text{2RCBH model:}\nonumber\\
&\frac{Q}{\tilde{r}_H}=\sqrt{2}\left(\frac{1\pm\sqrt{1-\left(\frac{\mu/T}{\pi/\sqrt{2}}\right)^2}}{\frac{\mu/T}{\pi/\sqrt{2}}}\right), &&\frac{Q}{\tilde{r}_H}=\frac{1}{\pi\sqrt{2}}\frac{\mu}{T }\label{eq:Qoverr}.
\end{align}

Since $Q/\tilde{r}_H$ is non-negative, Eqs.~\eqref{eq:Qoverr} imply that $\mu/T\in [0,\pi/\sqrt2]$ in the 1RCBH model, whereas $\mu/T\in[0,\infty)$ in the 2RCBH model. Moreover, for every value of $\mu/T$ in the 1RCBH model, there are two distinct values of $Q/\tilde{r}_H$ corresponding to two different branches of equilibrium black hole solutions: a thermodynamically stable branch with $Q/\tilde{r}_{H} \in [0,\sqrt{2}]$, corresponding to the lower sign, and an unstable branch with $Q/\tilde{r}_{H} \in [\sqrt{2},\infty)$, corresponding to the upper sign, while in the 2RCBH model there is only one branch of equilibrium black hole solutions~\cite{DeWolfe:2011ts,DeWolfe:2012uv,Finazzo:2016psx,Critelli:2017euk,deOliveira:2024bgh,deOliveira:2025qwe}. The value $Q/\tilde{r}_H=\sqrt{2}\Rightarrow\mu/T=\pi/\sqrt{2}$ where both 1RCBH branches merge corresponds to the critical point of the 1RCBH model, while for $Q/\tilde{r}_H=0\Rightarrow\mu/T = 0$ one recovers from both the 1RCBH and 2RCBH models the purely thermal SYM plasma with zero R-charge density.

One can holographically compute the thermodynamic equilibrium entropy of the strongly coupled fluid through the well-known Bekenstein-Hawking's relation~\cite{Bekenstein:1973ur,Hawking:1975vcx},
\begin{align}
S_H=\frac{A_H}{4G_5}= \frac{N_c^2}{2\pi}\int_{\mathcal{M}_H}d^3x \sqrt{|\gamma_H|} = \frac{N_c^2}{2\pi} g_{xx}^{3/2}(\tilde{r}_H)\int_{\mathcal{M}_H}d^3x = \frac{N_c^2}{2\pi} g_{xx}^{3/2}(\tilde{r}_H) V_H,
\label{eq:sBH}
\end{align}
where $A_H$ is the hyperarea of the black brane event horizon in the gravity side of the holographic duality. Then, thermodynamic equilibrium entropy densities, $s\equiv S_H/V_H$, for the 2RCBH and 1RCBH plasmas are given by~\cite{DeWolfe:2011ts,DeWolfe:2012uv,Finazzo:2016psx,Critelli:2017euk,deOliveira:2024bgh},
\begin{align}
&\text{1RCBH model:} &&\text{2RCBH model:}\nonumber\\
&\frac{s_{\text{eq}}}{N_c^2 T^3}=\frac{\pi^2}{16}\left[3\pm \sqrt{1-\left(\frac{\mu/T}{\pi/\sqrt{2}}\right)^2}\right]^2\left[1\mp \sqrt{1-\left(\frac{\mu/T}{\pi/\sqrt{2}}\right)^2}\right], &\qquad &\frac{s_{\text{eq}}}{N_c^2 T^3}=\frac{\pi^2}{2}\left[1+\frac{(\mu/T)^2}{2\pi^2}\right].
\label{eq:seq}
\end{align}

The $U(1)$ R-charge density can be evaluated through the holographic relation $\rho_c =\lim_{\tilde{r}\to \infty}\delta S/\delta \Phi'$, such that~\cite{DeWolfe:2011ts,DeWolfe:2012uv,Finazzo:2016psx,Critelli:2017euk,deOliveira:2024bgh},
\begin{align}
    &\text{1RCBH model:} &&\text{2RCBH model:}\nonumber\\
&\frac{\rho_{c,{\text{eq}}}}{N_c^2 T^3}=\frac{\mu/T}{16}\left[3\pm \sqrt{1-\left(\frac{\mu/T}{\pi/\sqrt{2}}\right)^2}\right]^2, &\qquad& \frac{\rho_{c,{\text{eq}}}}{N_c^2 T^3}=\frac{\mu/T}{4}\left[1+\frac{(\mu/T)^2}{2\pi^2}\right].\label{eq:rho12}
\end{align}
Notice that Eqs.~\eqref{eq:rho12} already express Eqs.~\eqref{eq:rhoeq} directly in terms of $\mu/T$.

Using the Gibbs-Duhem thermodynamic relation, $\dd p=s\dd T+\rho \dd \mu$, one may integrate Eqs. \eqref{eq:seq} and \eqref{eq:rho12} to obtain the isotropic plasma pressure in thermodynamic equilibrium~\cite{DeWolfe:2011ts,DeWolfe:2012uv,Finazzo:2016psx,Critelli:2017euk,deOliveira:2024bgh}, 
\begin{align}
    &\text{1RCBH model:} &&\text{2RCBH model:}\nonumber\\
& \frac{p_{\text{eq}}}{N_c^2 T^4}=\frac{\pi^2}{128}\left[3\pm \sqrt{1-\left(\frac{\mu/T}{\pi/\sqrt{2}}\right)^2}\right]^3\left[1\mp \sqrt{1-\left(\frac{\mu/T}{\pi/\sqrt{2}}\right)^2}\right],&\qquad& \frac{p_{\text{eq}}}{N_c^2 T^4}=\frac{\pi^2}{8}\left[1+\frac{(\mu/T)^2}{2\pi^2}\right]^2.
\label{eq:p12}
\end{align}
From Eq.~\eqref{eq:p12}, one can compute the specific heat at fixed chemical potential \cite{DeWolfe:2010he}, $C_\mu=T(\partial^2p/\partial T^2)_\mu=T(\partial s/\partial T)_\mu$, as well as the $n$th order R-charge susceptibility, $\chi_n=(\partial^n p/\partial \mu^n)_T=(\partial^{n-1}\rho/\partial \mu^{n-1})_T$.

Finally, the equilibrium value of the scalar condensate is given by Eqs.~\eqref{eq:ScalCondEq}, which when written directly in terms of $\mu/T$ reads as follows,
\begin{align}
    &\text{1RCBH model:} &&\text{2RCBH model:}\nonumber\\
   & \frac{\langle \mathcal{O}_{\phi}\rangle_{\text{eq}}}{N_c^2 T^2}=\frac{1}{4\sqrt{6}}\left[3\pm \sqrt{1-\left(\frac{\mu/T}{\pi/\sqrt{2}}\right)^2}\right]\left[1\pm \sqrt{1-\left(\frac{\mu/T}{\pi/\sqrt{2}}\right)^2}\right],&\qquad& \frac{\langle \mathcal{O}_{\phi}\rangle_{\text{eq}}}{N_c^2 T^2}=-\frac{(\mu/T)^2}{4\sqrt{6}\pi^2}.
\label{eq:O12}
\end{align}

In all the above equations, the upper/lower signs notation designates thermodynamically unstable/stable branches of black hole solutions in the 1RCBH model, while the 2RCBH model features only a single branch of black hole solutions in equilibrium.

In Fig.~\ref{fig:Thermo} we present the results for various thermodynamic observables of the 2RCBH and 1RCBH models in equilibrium. Since there are two branches of black hole solutions for the 1RCBH model,\footnote{At $\mu/T = 0$, there are also two solutions: the uncharged AdS$_5$-Schwarzschild black hole, which is dual to the purely thermal SYM plasma and belongs to the stable branch of the 1RCBH model, and a charged solution without an event horizon. This latter solution is not a black hole but rather a supersymmetric BPS solution, known as the ``superstar''~\cite{Myers:2001aq}, which lies in the unstable branch of solutions~\cite{DeWolfe:2011ts,Finazzo:2016psx}.} and global thermodynamic stability dictates that the preferred phase is the one that minimizes the free energy density $f$ or, equivalently, maximizes the pressure $p=-f$~\cite{DeWolfe:2010he}, one can clearly identify both branches in Fig.~\ref{fig:pThermo}. Furthermore, local thermodynamic stability under small thermal fluctuations requires the Jacobian of the matrix of susceptibilities to be positive~\cite{DeWolfe:2010he}, which in turn implies that $C_\mu$ and $\chi_2$ must be positive for thermodynamically stable solutions --- see Figs.~\ref{fig:CThermo} and \ref{fig:chi2Thermo}. Moreover, at a second-order phase transition such as at the critical point of the 1RCBH model, second and higher-order derivatives of the pressure diverge, which is precisely what happens with the specific heat and the R-charge susceptibility at the critical point $\mu/T=\pi/\sqrt{2}$ of the 1RCBH model --- see Figs.~\ref{fig:CThermo} and \ref{fig:chi2Thermo}. One can also see that the stable and unstable branches merge for the observables that remain finite at the critical point of the 1RCBH model, such as the entropy and R-charge densities, the pressure, and the scalar condensate. In the same figures, it can be observed that the 2RCBH model exhibits only a single branch of solutions and has no phase transition, with thermodynamic observables displaying a monotonic behavior as functions of $\mu/T$. One also notes that the stable branch of the 1RCBH model and the 2RCBH model coalesce at $\mu/T=0$, as expected, since both reduce to the purely thermal SYM plasma in this limit.

It is clear that the stable thermodynamic equilibrium states in both models are specified by the value of $\mu/T$, which is one of the initial data that must be specified in the homogeneous isotropization dynamics. The above equilibrium results are expected to be recovered in the long-time regime of the time evolution of initially out of equilibrium states undergoing homogeneous isotropization dynamics, constituting an important analytical consistency check of the late-time behavior of the numerical solutions to be discussed in the next sections.

\subsection{One-point functions out of equilibrium}

With the equilibrium solutions specified above, we now move on to the calculation of the out of equilibrium one-point functions, $\langle T_{\mu\nu}\rangle$, $\langle J^\mu\rangle$, and $\langle \mathcal{O}_\phi\rangle$. Once again, since the non-equilibrium metric ansatz in~\eqref{eq:ansatzeds2} is expressed in terms of the radial EF coordinate $r$, it is necessary to establish a map between $r$ and the radial FG coordinate $\rho$, in terms of which the renormalized one-point functions are expressed. By substituting the definition~\eqref{eq:EFtime} of the EF-time coordinate into the non-equilibrium metric~\eqref{eq:ansatzeds2}, and requiring the resulting $rt$-component to vanish, one obtains that, $g_{tt}=-2A$ and $g_{rr}=1/2A$, which diagonalizes the metric~\eqref{eq:ansatzeds2}. Thus, by imposing that $(g_{rr}\dd r^2)_{\text{EF}}=(g_{\rho\rho}\dd \rho^2)_{\text{FG}}$, it follows that,
\begin{equation}
    \int \frac{\dd r}{\sqrt{2 A(v,r)}}=- \int \frac{\dd\rho}{2\rho}=-\frac{1}{2}\ln\rho, \label{eq:RelIntRo}
\end{equation}
where the minus sign follows from the fact that $r\to\infty\Rightarrow\rho\to 0$. By perturbatively evaluating the above integral close to the boundary, one can connect $r$ and $\rho$ with the help of Eq.~\eqref{eq:UVexpA} through the following expression,
\begin{align}
    r(\rho)=&\frac{1}{\sqrt{\rho }}+\frac{1}{72} \rho ^{3/2} \left(\phi _2(v){}^2-18 H\right)+\frac{1}{90} \rho ^2 \phi _2(v) \dot{\phi}_2(v)+\mathcal{O}(\rho^{5/2}),
\end{align}
which is the same for both the 1RCBH and 2RCBH models. Substituting the above expression for $r(\rho)$ into Eq.~\eqref{eq:UVexp}, one obtains,
\begin{subequations}
    \begin{align}
        \gamma_{tt}(\rho)&=-\frac{1}{\rho }-\frac{2 \lambda (v)}{\sqrt{\rho }}+\left[2 \lambda '(v)-\lambda (v)^2\right]+\frac{1}{12} \rho  \left[\phi _2(v){}^2-18 H\right]+\mathcal{O}(\rho^{3/2}),\\
        \gamma_{xx}(\rho)&=\frac{1}{\rho }+\frac{2 \lambda (v)}{\sqrt{\rho }}+\lambda (v)^2+\rho  \left(B_4(v)-\frac{H}{2}-\frac{1}{12} \phi _2(v){}^2\right)+\mathcal{O}(\rho^{3/2}),\\
        \gamma_{zz}(\rho)&=\frac{1}{\rho }+\frac{2 \lambda (v)}{\sqrt{\rho }}+\lambda (v)^2+\rho  \left(-2 B_4(v)-\frac{H}{2}-\frac{1}{12} \phi _2(v){}^2\right)+\mathcal{O}(\rho^{3/2}),\\
        \phi(\rho)&=\phi_2(v)\rho+\mathcal{O}(\rho^{3/2}),\\
        \Phi(\rho)&=\Phi_0(v)+\rho\Phi_2(v)+\mathcal{O}(\rho^{3/2}).
    \end{align}
\end{subequations}
By extracting the relevant FG UV coefficients from the above set of equations, and substituting them into Eqs.~\eqref{eq:opf}, it follows that,
\begin{subequations}
\label{eq:observablesNE}
    \begin{align}
        \varepsilon&=\langle T_{tt}\rangle=- \frac{1}{\kappa_5^2}3H=-\frac{N_c^2}{4\pi^2}3H,\label{eq:varepsilon}\\
        p_T(v)&=\langle T_{xx}\rangle = \frac{1}{\kappa_5^2}\left(2B_4(v)-H\right),\label{eq:Tzz}\\
        p_L(v)&=\langle T_{zz}\rangle =- \frac{1}{\kappa_5^2}\left(4B_4(v)+H\right),\\
        \Delta p(v)&\equiv p_T - p_L=\frac{N_c^2}{4\pi^2}6B_4(v),\\
        \rho_c&=\langle J^t\rangle= -\frac{1}{\kappa_5^2}\Phi_2 =-\frac{N_c^2}{4\pi^2}\Phi_2,\\
        \langle\mathcal{O}_\phi\rangle(v)&= -\frac{1}{\kappa_5^2}\phi_2(v)=-\frac{N_c^2}{4\pi^2}\phi_2(v).
    \end{align}
\end{subequations}
where $p_T$ and $p_L$ are, respectively, the transverse and longitudinal pressures, and $\Delta p$ is the pressure anisotropy. Notice from above that the trace of the QFT energy-momentum tensor vanishes, $\langle T_\mu^\mu\rangle=-\varepsilon+2 p_T+p_L=0$, confirming that both the 2RCBH and 1RCBH models are conformal field theories (CFTs). In terms of the above R-charge density $\rho_c$, one can rewrite the bulk electric field given in Eq.~\eqref{eq:Erhofield2R2} as,
\begin{align}
&\text{1RCBH model:} &&\text{2RCBH model:}\nonumber\\
&\mathcal{E}(v,r)= -\frac{8\pi^2\rho_c}{N_c^2}\,\Sigma^{-3}(v,r)\,e^{2\sqrt{\frac{2}{3}}\phi(v,r)},
&& \mathcal{E}(v,r) = -\frac{8\pi^2\rho_c}{N_c^2}\,\Sigma^{-3}(v,r)\,e^{-\sqrt{\frac{2}{3}}\phi(v,r)}.
\label{eq:Efield2R2}
\end{align}

Another important observable to be numerically computed in this work is the non-equilibrium entropy density in the homogeneous isotropization dynamics. In the context of the gauge-gravity duality, the holographic dictionary relates thermodynamic entropy of a strongly coupled QFT in thermodynamic equilibrium to the Bekenstein-Hawking entropy for a dual black hole in equilibrium within the higher-dimensional bulk, as given in Eq.~\eqref{eq:sBH}. The Bekenstein-Hawking formula expresses the black hole entropy in terms of the hyperarea of its event horizon, which is a global property of the spacetime manifold, i.e.~it depends on the entire future evolution of the black hole geometry. The global character of a dynamic event horizon contrasts with the expected local behavior of the entropy production in a dissipative dynamical process that starts out of equilibrium. For this reason, Ref.~\cite{Figueras:2009iu} claimed that, in the case of strongly coupled systems with gravity duals, a consistent definition of the holographic non-equilibrium entropy should be given in terms of the hyperarea of the apparent horizon, which, contrary to the event horizon, can be determined locally in time. The same approach was also employed in several other works in the literature, e.g.~\cite{Muller:2020ziz,Chesler:2010bi,Heller:2011ju,Heller:2012je,vanderSchee:2014qwa,Jankowski:2014lna,Grozdanov:2016zjj,Buchel:2016cbj,Engelhardt:2017aux}. Finally, this reasoning is further strengthened by the fact that the apparent horizon, when present, is always located inside the black hole event horizon, and in the case of dissipative systems, such as the 2RCBH and 1RCBH models, at late-times both horizons coalesce. Therefore, the non-equilibrium entropy defined in terms of the hyperarea of the apparent horizon does converge, in the long-time regime, to the usual holographic thermodynamic entropy defined in terms of the event horizon of a black hole in equilibrium.

In terms of the infalling EF coordinates, one calculates the dynamic hypearea of the black hole apparent horizon as follows,
\begin{equation}
    A_{\text{AH}}(v)=\int_{\text{AH}}\dd^3x\sqrt{|\gamma_{\text{AH}}|}\,\Bigg|_{r=r_{\text{AH}}}=\int_{\text{AH}}\dd x \dd y \dd z \sqrt{g_{xx}g_{yy}g_{zz}}\,\Bigg|_{r=r_{\text{AH}}}=|\Sigma(v,r_{\text{AH}})|^3\, V_3,
\end{equation}
where $r_\textrm{AH}$ is the radial location of the apparent horizon, to be discussed in section~\ref{sec:AH}, and the volume of the planar black brane apparent horizon is given by  $V_3=\int_{\text{AH}}\dd x\dd y \dd z$. Furthermore, the holographic non-equilibrium entropy is defined by analogy with the Bekenstein-Hawking entropy formula, but exchanging the event horizon with the apparent horizon,
\begin{equation}
    s_{\text{AH}}(v)\equiv\frac{S_{\text{AH}}(v)}{V_3}=\frac{A_{\text{AH}}(v)/4G_5}{V_3}=\frac{2\pi}{\kappa_5^2} |\Sigma(v,r_{\text{AH}})|^3.
\end{equation}

By fixing the temperature scale as $T=1/\pi$, one can then define the following dimensionless ratio involving the non-equilibrium entropy density, 
\begin{equation}
    \frac{\hat{s}_{\text{AH}}(v)}{T^3}\equiv \frac{\kappa_5^2}{L^3} \frac{s_{\text{AH}}(v)}{T^3}=2\pi^4 |\Sigma(v,r_{\text{AH}})|^3.
\end{equation}
Notice that $\hat{s}_\textrm{AH}(v)/T^3=4\pi^2 s_\textrm{AH}(v)/N_c^2T^3$, which therefore, for any chosen value of $\mu/T$, should converge in the long-time regime, $v\to\infty$, to $4\pi^2$ times the analytical equilibrium results given in Eqs.~\eqref{eq:seq} and illustrated in Fig.~\ref{fig:sThermo}.

We furthermore define for any given observable $Y$ the hatted quantity: $\hat{Y}\equiv Y\, \kappa_5^2/L^3 = 4\pi^2 Y/N_c^2$.

\subsection{Relating equilibrium and non-equilibrium coefficients}

Finally, we finish this section by relating the black hole parameters $Q$, $M$, and $\tilde{r}_H$ of the equilibrium solutions~\eqref{eq:AnsatzAll} with the UV coefficients of the non-equilibrium solutions in the near-boundary expansions~\eqref{eq:UVexp}. Since the equilibrium solutions are attained only at asymptotically late times, we begin by defining the equilibrium UV coefficients of the non-equilibrium solutions as follows,
\begin{equation}
    X^{(\text{eq})}\equiv \lim_{v\to \infty}X(v), \quad \text{with} \quad \partial_v X^{(\text{eq})}=0,
\end{equation}
with $X(v)$ being any of the UV coefficients in the set $\{B_4(v),\phi_2(v),\Phi_0(v)\}$, where the UV coefficients $H$ and $\Phi_2$ are not included because they are constant, as stated before. By equating the metric component $(g_{\tilde{r}v})_\textrm{EF}$ of the modified EF coordinates used in the equilibrium ansatz~\eqref{eq:AnsatzEqs}, and the corresponding metric component in the original EF coordinates employed in the non-equilibrium ansatz~\eqref{eq:ansatzeds2}, one gets the following relation,
\begin{equation}
    e^{a(\tilde{r})+b(\tilde{r})}\dd \tilde{r}=\dd r,
\end{equation}
which can be analytically solved as below,
\begin{align}
    &\text{1RCBH model:}&&\text{2RCBH model:}\nonumber\\
    &r(\tilde{r})=\frac{3 \tilde{r}^{4/3} \, _2F_1\left(\frac{1}{6},\frac{2}{3};\frac{5}{3};-\frac{r^2}{Q^2}\right)}{4Q^{1/3} }+\frac{3 \sqrt{\pi } Q \Gamma \left(\frac{5}{3}\right)}{2 \Gamma \left(\frac{1}{6}\right)},&& r(\tilde{r})=\frac{3 r^{5/3} \, _2F_1\left(\frac{1}{3},\frac{5}{6};\frac{11}{6};-\frac{r^2}{Q^2}\right)}{5 Q^{2/3}}+\frac{6 \sqrt{\pi } Q \Gamma \left(\frac{11}{6}\right)}{5 \Gamma \left(\frac{1}{3}\right)},
\end{align}
where $_2 F_1$ is the hypergeometric function and the above integration constants were determined by imposing that $\lim_{\tilde{r}\to \infty}r(\tilde{r})-\tilde{r}=0$.

In the same way as we did before, we expand the above expressions in powers of $\tilde{r}$ close to the boundary, thus obtaining, 
    \begin{align}
      &\text{1RCBH model:}&&\text{2RCBH model:}\nonumber\\
      &r(\tilde{r})=\tilde{r}+\frac{Q^2}{6 \tilde{r}}-\frac{7 Q^4}{216 \tilde{r}^3}+\frac{91 Q^6}{6480 \tilde{r}^5}+\mathcal{O}(\tilde{r}^{-7}), &&r(\tilde{r})=\tilde{r}+\frac{Q^2}{3 \tilde{r}}-\frac{2 Q^4}{27 \tilde{r}^3}+\frac{14 Q^6}{405 r^5}+\mathcal{O}(\tilde{r}^{-7})\label{eq:rrexp}.
    \end{align}

By substituting relations~\eqref{eq:rrexp} into Eqs.~\eqref{eq:UVexp}, one obtains,
\begin{subequations}
\label{eq:nonequil}
    \begin{align}
        &\text{1RCBH model:}&&\text{2RCBH model:}\nonumber\\
        &g_{tt}(\tilde{r})=-\frac{Q^2}{3}-\tilde{r}^2&&g_{tt}(\tilde{r})=-\frac{2Q^2}{3}-\tilde{r}^2\nonumber\\
        &\qquad\qquad+\frac{-2 H+\frac{Q^4}{27}+\frac{1}{9} (\phi_2^{(\text{eq})})^2}{\tilde{r}^2}+\mathcal{O}(\tilde{r}^{-4}),&&\qquad\qquad+\frac{-2 H+\frac{Q^4}{27}+\frac{1}{9} (\phi_2^{(\text{eq})})^2}{\tilde{r}^2}+\mathcal{O}(\tilde{r}^{-4}),\\
        &g_{xx}(\tilde{r})=\frac{Q^2}{3}+\tilde{r}^2+\frac{-\frac{Q^4}{27}-\frac{1}{9} (\phi_2^{(\text{eq})})^2}{\tilde{r}^2}+\mathcal{O}(\tilde{r}^{-4}),&&g_{xx}(\tilde{r})=\frac{2Q^2}{3}+\tilde{r}^2+\frac{-\frac{Q^4}{27}-\frac{1}{9} (\phi_2^{(\text{eq})})^2}{\tilde{r}^2}+\mathcal{O}(\tilde{r}^{-4}),\\
        &\phi(\tilde{r})=\frac{\phi _2^{(\text{eq})}}{r^2}+\frac{\frac{(\phi _2^{(\text{eq})})^2}{2 \sqrt{6}}-\frac{1}{3} Q^2 \phi_2^{(\text{eq})}}{r^4}+\mathcal{O}(\tilde{r}^{-6}),&&\phi(\tilde{r})=\frac{\phi _2^{(\text{eq})}}{r^2}+\frac{\frac{(\phi _2^{(\text{eq})})^2}{2 \sqrt{6}}-\frac{2}{3} Q^2 \phi_2^{(\text{eq})}}{r^4}+\mathcal{O}(\tilde{r}^{-6}),\\
        &\Phi(\tilde{r})=\Phi_0^{(\text{eq})}+\frac{\Phi_2^{(\text{eq})}}{\tilde{r}^2}&&\Phi(\tilde{r})=\Phi_0^{(\text{eq})}+\frac{\Phi_2^{(\text{eq})}}{\tilde{r}^2}\nonumber\\
        &\quad\qquad+\frac{\sqrt{\frac{2}{3}} \Phi_2^{(\text{eq})} \phi_2^{(\text{eq})}-\frac{1}{3} Q^2 \Phi _2^{(\text{eq})}}{\tilde{r}^4}+\mathcal{O}(\tilde{r}^{-6}),&&\quad\qquad+\frac{\sqrt{\frac{2}{3}} \Phi_2^{(\text{eq})} \phi_2^{(\text{eq})}-\frac{2}{3} Q^2 \Phi _2^{(\text{eq})}}{\tilde{r}^4}+\mathcal{O}(\tilde{r}^{-6}).
    \end{align}
\end{subequations}

By expanding the analytical equilibrium solutions~\eqref{eq:AnsatzAll} close to the boundary, it follows that,
\begin{subequations}
\label{eq:equil}
    \begin{align}
        &\text{1RCBH model:}&&\text{2RCBH model:}\nonumber\\
        &g_{tt}(\tilde{r})=-\frac{Q^2}{3}-\tilde{r}^2+\frac{M^2+\frac{Q^4}{9}}{\tilde{r}^2}+\frac{-\frac{2 M^2 Q^2}{3}-\frac{5 Q^6}{81}}{\tilde{r}^4}&&g_{tt}(\tilde{r})=-\frac{2 Q^2}{3}-\tilde{r}^2+\frac{M^2+\frac{Q^4}{9}}{\tilde{r}^2}-\frac{4 \left(27 M^2 Q^2+Q^6\right)}{81 \tilde{r}^4}\nonumber\\
        &\quad+\frac{\frac{5 M^2 Q^4}{9}+\frac{10 Q^8}{243}}{\tilde{r}^6}+\mathcal{O}(\tilde{r}^{-8}),&&\quad+\frac{7 \left(54 M^2 Q^4+Q^8\right)}{243 \tilde{r}^6}+\mathcal{O}(\tilde{r}^{-8}),\\
        &g_{xx}(\tilde{r})=\frac{Q^2}{3}+\tilde{r}^2-\frac{Q^4}{9 \tilde{r}^2}+\frac{5 Q^6}{81 \tilde{r}^4}-\frac{10 Q^8}{243 \tilde{r}^6}+\mathcal{O}(\tilde{r}^{-8}),&&g_{xx}(\tilde{r})=\frac{2 Q^2}{3}+\tilde{r}^2-\frac{Q^4}{9 \tilde{r}^2}+\frac{4 Q^6}{81 \tilde{r}^4}-\frac{7 Q^8}{243 \tilde{r}^6}+\mathcal{O}(\tilde{r}^{-8}),\\
        &\phi(\tilde{r})=-\frac{\sqrt{\frac{2}{3}} Q^2}{\tilde{r}^2}+\frac{Q^4}{\sqrt{6} \tilde{r}^4}-\frac{\sqrt{\frac{2}{3}} Q^6}{3 \tilde{r}^6}+\mathcal{O}(\tilde{r}^{-8}),&&\phi(\tilde{r})=+\frac{\sqrt{\frac{2}{3}} Q^2}{\tilde{r}^2}-\frac{Q^4}{\sqrt{6} \tilde{r}^4}+\frac{\sqrt{\frac{2}{3}} Q^6}{3 \tilde{r}^6}+\mathcal{O}(\tilde{r}^{-6}),\\
        &\Phi(\tilde{r})=\frac{M Q}{Q^2+\tilde{r}_H^2}-\frac{M Q}{r^2}+\frac{M Q^3}{r^4}&&\Phi(\tilde{r})=-\frac{\sqrt{2} M Q}{\tilde{r}^2}+\frac{\sqrt{2} M Q}{Q^2+\tilde{r}_H^2}+\frac{\sqrt{2} M Q^3}{\tilde{r}^4}\nonumber\\
        &\qquad\quad\qquad-\frac{M Q^5}{\tilde{r}^6}+\mathcal{O}(\tilde{r}^{-8}),&&\qquad\quad\qquad-\frac{\sqrt{2} M Q^5}{\tilde{r}^6}+\mathcal{O}(\tilde{r}^{-8}).
    \end{align}
\end{subequations}

Finally, by comparing Eqs.~\eqref{eq:nonequil} and~\eqref{eq:equil}, one finds that,
\begin{subequations}
    \begin{align}
        &\text{1RCBH model:} && \text{2RCBH model: }\\
        &\phi_2^{(\text{eq})}=-\sqrt{\frac{2}{3}}Q^2,&&\phi_2^{(\text{eq})}=\sqrt{\frac{2}{3}}Q^2,\\
        &\Phi_2^{(\text{eq})}=-MQ,&&\Phi_2^{(\text{eq})}=-\sqrt{2}MQ,\\
        &H=-\frac{M^2}{2},&&H=-\frac{M^2}{2},\\
        &\Phi_0^{(\text{eq})}=\frac{ M Q}{Q^2+\tilde{r}_H^2}, &&\Phi_0^{(\text{eq})}=\frac{\sqrt{2} M Q}{Q^2+\tilde{r}_H^2}. 
    \end{align}
\end{subequations}

\section{Far-from-equilibrium solutions}
\label{sec:far}

\subsection{Field redefinitions and radial integration}
\label{sec:FieldRed}

As mentioned in section~\ref{sec:1stedoSystem}, the residual diffeomorphism invariance of the metric \eqref{eq:ansatzeds2} under radial shifts allows for the introduction of an arbitrary function of the Eddington-Finkelstein time $v$: the radial shift function $\lambda(v)$. Due to this redundancy, it follows that physical results must be independent of the specific choice of $\lambda(v)$, including the trivial choice $\lambda(v)=0$. However, depending on the circumstance, one may require the introduction of a nonzero radial shift function to keep the radial position of the apparent horizon fixed during the time evolution of the system~\cite{Chesler:2013lia,vanderSchee:2014qwa}. This may be needed for numerical stability, as in the case e.g.~of the holographic Bjorken flow dynamics~\cite{Critelli:2018osu,Rougemont:2021qyk,Rougemont:2021gjm,Rougemont:2022piu}. In the simpler case of the homogeneous isotropization dynamics considered here, the choice of using a non-trivial $\lambda(v)$ turns out to be optional. For completeness, we computed all the physical observables analyzed in this work both with and without the inclusion of a non-vanishing radial shift function. By doing so, we explicitly verified that all the results obtained are in fact invariant under such a choice, thus confirming the consistency of our calculations with the residual radial diffeomorphism invariance of the systems under consideration.

Having considered in the previous sections the introduction of a radial shift function $\lambda(v)$ in the UV expansions of the bulk fields, we now proceed with further considerations aimed to numerically solving the system of partial differential equations~\eqref{eq:EOMs}. For this sake, we define a compact radial coordinate, $u\equiv 1/r$, which is well-suited for numerical integration using the pseudospectral method~\cite{boyd01}. In this new radial coordinate, the boundary is located at $u=0$. Next, to facilitate the extraction of the relevant dynamical UV coefficients from the near-boundary expansions given in Eqs.~\eqref{eq:UVexp}, we consider some convenient rescalings and subtractions of the bulk fields, so that the redefined fields approach finite radial constants at the boundary. The subtracted bulk fields are defined through, $u^p \,Z_s(v,u)\equiv Z(v,u)-Z_\textrm{UV}(v,u)$, where $p$ is an integer and $Z_\textrm{UV}(v,u)$ is some UV truncation of a given bulk field $Z(v,u)$. The chosen truncation criterion is that $Z_s(v,u=0)$ should yield a finite radial constant. In the present case, the subtracted fields and their corresponding boundary conditions are specified as follows~\cite{Rougemont:2024hpf},
\begin{subequations}
\begin{align}
u^2 A_s(v,u) &\equiv A(v,u) - \frac{1}{2u^2} - \frac{\lambda(v)}{u} - \frac{\left[\lambda^2(v)-2\partial_v\lambda(v)\right]}{2}, \label{eq:Aa}\\
\Rightarrow A_s(v,u\to 0) &\to \frac{18 H - \phi_2^2(v)}{18} - \frac{\left[\phi_2(v)\partial_v\phi_2(v)+\lambda(v)\left(36 H-2\phi_2^2(v)\right)\right] u}{18} +\mathcal{O}(u^2), \label{eq:Ab}\\
\Rightarrow \partial_u A_s(v,u=0) &= -\frac{\phi_2(v) \left[\partial_u\phi_s(v,u=0)+2\lambda(v)\phi_2(v)\right]+\lambda(v)\left(36 H-2\phi_2^2(v)\right)}{18}, \label{eq:Ac}
\end{align}
\end{subequations}
where Eq.~\eqref{eq:Ac} was obtained from Eq.~\eqref{eq:Ab} via Eq.~\eqref{eq:fc},\footnote{Eq.~\eqref{eq:Ac} is employed as an extra boundary condition in the radial
integration of the subtracted function $A_s$.}

\begin{subequations}
\begin{align}
u^4 B_s(v,u) &\equiv B(v,u), \label{eq:Ba}\\
\Rightarrow B_s(v,u\to 0) &\to B_4(v) + \left[\partial_v B_4(v)-4\lambda(v) B_4(v)\right] u +\mathcal{O}\left(u^2\right), \label{eq:Bb}\\
\Rightarrow \partial_v B_4(v) &= \partial_u B_s(v,u=0) + 4\lambda(v) B_4(v), \label{eq:Bc}
\end{align}
\end{subequations}

\begin{subequations}
\begin{align}
u^2 \Sigma_s(v,u) &\equiv \Sigma(v,u) - \frac{1}{u} - \lambda(v), \label{eq:Sa}\\
\Rightarrow \Sigma_s(v,u\to 0) &\to -\frac{\phi_2^2(v) u}{18} - \frac{\left[3\phi_2(v)\partial_v\phi_2(v)-5\lambda(v)\phi_2^2(v)\right] u^2}{30} +\mathcal{O}\left(u^3\right), \label{eq:Sb}
\end{align}
\end{subequations}

\begin{subequations}
\begin{align}
u^2 \phi_s(v,u) &\equiv \phi(v,u), \label{eq:fa}\\
\Rightarrow \phi_s(v,u\to 0) &\to \phi_2(v) + \left[\partial_v \phi_2(v)-2\lambda(v) \phi_2(v)\right] u +\mathcal{O}\left(u^2\right), \label{eq:fb}\\
\Rightarrow \partial_v \phi_2(v) &= \partial_u \phi_s(v,u=0) + 2\lambda(v) \phi_2(v), \label{eq:fc}
\end{align}
\end{subequations}

\begin{align}
\mathcal{E}_s(v,u) \equiv \mathcal{E}(v,u), \label{eq:E}
\end{align}

\begin{subequations}
\begin{align}
u^2 \left(d_+\Sigma\right)_s(v,u) &\equiv \left(d_+\Sigma\right)(v,u) - \frac{1}{2u^2} - \frac{\lambda(v)}{u}-\frac{\lambda^2(v)}{2}, \label{eq:dSa}\\
\Rightarrow \left(d_+\Sigma\right)_s(v,u\to 0) &\to H + \frac{\phi_2^2(v)}{36} +\mathcal{O}\left(u\right), \label{eq:dSb}
\end{align}
\end{subequations}

\begin{subequations}
\begin{align}
u^3 \left(d_+B\right)_s(v,u) &\equiv \left(d_+B\right)(v,u), \label{eq:dBa}\\
\Rightarrow \left(d_+B\right)_s(v,u\to 0) &\to -2B_4(v) +\mathcal{O}\left(u\right), \label{eq:dBb}
\end{align}
\end{subequations}

\begin{subequations}
\begin{align}
u \left(d_+\phi\right)_s(v,u) &\equiv \left(d_+\phi\right)(v,u), \label{eq:dfa}\\
\Rightarrow \left(d_+\phi\right)_s(v,u\to 0) &\to -\phi_2(v) +\mathcal{O}\left(u^2\right). \label{eq:dfb}
\end{align}
\end{subequations}

Moreover, by changing the radial coordinate $r=1/u\Rightarrow \partial_r=-u^2\partial_u$ and introducing the subtracted field redefinitions from Eqs.~\eqref{eq:Aa}, \eqref{eq:Ba}, \eqref{eq:Sa}, \eqref{eq:fa}, \eqref{eq:dSa}, \eqref{eq:dBa}, \eqref{eq:dfa} into the original set of nested equations of motion~\eqref{eq:Efield2R2} and~\eqref{eq:pdec} ---~\eqref{eq:pdeg}, one obtains the following set of equations,
\begin{subequations}
\label{eqs:odefinalsolution}
    \begin{align}
 \text{1RCBH model:}\quad \mathcal{E}_s 
 =\, &\frac{2 \Phi_2 e^{2 \sqrt{\frac{2}{3}} u^2 \phi_s}}{(\lambda + u^2 \Sigma_s + \frac{1}{u})^3}, \hspace{1cm}\text{2RCBH model:}\quad \mathcal{E}_s 
 = \frac{2 \Phi_2 e^{- \sqrt{\frac{2}{3}} u^2 \phi_s}}{(\lambda + u^2 \Sigma_s + \frac{1}{u})^3}, \\
&\nonumber\\
4 u^3 \left(u^3 \Sigma_s+\lambda  u+1\right)(\dd_+\phi)_s'&+2 u^2  \left(3 u^4 \Sigma_s'+8 u^3 \Sigma_s+2 \lambda  u-1\right)(\dd_+\phi)_s+3 u^3  \left(2 u^4 (\dd_+\Sigma)_s+(\lambda  u+1)^2\right)\phi_s'\nonumber\\
+6 u^2  (2 u^4 (\dd_+\Sigma)_s&+(\lambda  u+1)^2)\phi_s-\left(u^3 \Sigma_s+\lambda  u+1\right) \left(\mathcal{E}_s^2 \partial_\phi f(u^2 \phi_s)-2 \partial_\phi V(u^2 \phi_s)\right)=0,\\
&\nonumber\\
\frac{1}{12}u^6\left(\mathcal{E}_s^2 f(u^2 \phi_s)+2 {V(u^2 \phi_s)}\right)\Sigma_s^2&-u^5  \left(u^3 \Sigma_s+\lambda  u+1\right)(\dd_+\Sigma)_s'-2 u^5  \left(\lambda +u^3 \Sigma_s'+3 u^2 \Sigma_s\right)(\dd_+\Sigma)_s\nonumber\\
&+\frac{1}{6} u^3  (\lambda  u+1) \left(\mathcal{E}_s^2 f(u^2 \phi_s)+2 V(u^2 \phi_s)-12 \lambda  u-6\right)\Sigma_s\nonumber\\
&+\frac{1}{12}(\lambda  u+1)^2 \left(\mathcal{E}_s^2 f(u^2 \phi_s)-12 u^4 \Sigma_s'+2V(u^2 \phi_s)+24\right)=0,\\
&\nonumber\\
u\left(u^3 \Sigma_s+\lambda  u+1\right)(\dd_+B)_s'&+\frac{3}{2} \left(u^4 \Sigma_s'+4 u^3 \Sigma_s+2 \lambda  u+1\right)(\dd_+B)_s\nonumber\\
&+\frac{3}{4} \left(u B_s'+4 B_s\right) \left(2 u^4 (\dd_+\Sigma)_s+(\lambda  u+1)^2\right)=0,\\
&\nonumber\\
u^6  \left(u^3 \Sigma_s+\lambda  u+1\right)^2A_s''&+6 u^5  \left(u^3 \Sigma_s+\lambda  u+1\right)^2A_s'+6 u^4  \left(u^3 \Sigma_s+\lambda  u+1\right)^2A_s\nonumber\\
-\frac{3}{2}  u^8 & \left(u B_s'+4 B_s\right) \left(u^3 \Sigma_s+\lambda  u+1\right)^2(\dd_+B)_s-\frac{1}{2} u^4  \left(u^3 \Sigma_s+\lambda  u+1\right)^2 \left(u \phi_s'+2 \phi_s\right)(\dd_+\phi)_s\nonumber\\
-\frac{1}{12}&\left(u^3 \Sigma_s+\lambda  u+1\right)^2\left(7  f(u^2 \phi_s)\mathcal{E}_s^2+2 V(u^2 \phi_s)\right) \nonumber\\
+3 &\left(2 u^4 (\dd_+\Sigma)_s+(\lambda  u+1)^2\right) \left(u^4 \Sigma_s'+2 u^3 \Sigma_s-1\right)+\left(u^3 \Sigma_s+\lambda  u+1\right)^2=0,\\
&\nonumber\\
u^2 \Sigma_s''+6 u \Sigma_s'+ &\frac{1}{6}\left(3 u^8 \left(u B_s'+4 B_s\right)^2+u^6 \left(\phi_s'\right)^2+4 u^5\phi_s \phi_s'+4 u^4\phi_s^2+36\right)\Sigma_s\nonumber\\
&+\frac{1}{6} u (\lambda  u+1) \left(3 u^4 \left(u B_s'+4 B_s\right)^2+u^2 \left(\phi_s'\right)^2+4 u\phi_s \phi_s'+4 \phi_s^2\right)=0,
\end{align}
\end{subequations}
with the prime now denoting the radial derivative $\partial_u$.

To integrate the equations of motion in the radial direction, we employ the same procedure detailed in section 5.4 of~\cite{Critelli:2017euk}, where an eigenvalue problem is formulated by discretizing the radial part of the differential equations with the help of the pseudospectral method~\cite{boyd01}. One therefore converts the original continuous differential equation problem into the discrete linear algebra problem of inverting a diagonal \( (N-1) \times (N-1) \) matrix for each bulk field, with \( N \) being the number of collocation points in the radial Chebyshev-Gauss-Lobatto grid. The larger the number of collocation points, the larger the order of truncation of the expansion of the bulk fields in the basis of Chebyshev polynomials used in the pseudospectral method, and the more precise the numerical results are. The matrices involved in this procedure encode the homogeneous contributions of the discretized radial equations, evaluated at all interior points of the grid, excluding the boundary point. To obtain the numerical profiles of the bulk fields, one solves the corresponding linear algebra system by applying the inverse of these matrices on the vectors containing the inhomogeneous terms. The point corresponding to the boundary in the radial grid is treated separately, with the UV values of the subtracted fields being imposed directly there. After determining the numerical solutions at the $N-1$ interior grid points, one appends the values at the boundary, which are computed via the imposed boundary conditions, thus reconstructing the full set of bulk field values across the entire radial domain. This yields the complete eigenvectors of size \( N \), where each component corresponds to the value of a given field at one of the \( N \) collocation points of the Chebyshev-Gauss-Lobatto grid in the radial direction. The above procedure is then implemented over the hypersurface defined on each time slice of the bulk manifold.

The radial integration of the equations of motion must cover the full region of the bulk geometry causally connected to boundary observers. For black hole geometries, this requires integrating the field equations at least from the event horizon up to the boundary (integrating from radial positions inside the event horizon adds no information to the boundary QFT observables, while integrating from radial positions starting outside the event horizon may lead to a loss of information). As mentioned before, in time-dependent backgrounds, as in the homogeneous isotropization dynamics, the event horizon is a global property whose location can only be determined by knowing first the entire future evolution of the bulk fields. In practice, one can more easily take as the starting point of the radial integration of the field equations the apparent horizon instead of the event horizon, since the radial location of the apparent horizon can be determined locally in time. Moreover, since it is always located within the event horizon, one guarantees that no information is lost.

\subsection{Apparent horizon and time evolution}
\label{sec:AH}

The apparent horizon is defined as the outermost trapped null surface lying within the event horizon. For the kind of metric given by Eq.~\eqref{eq:ansatzeds2}, the radial location of the apparent horizon, $u_\textrm{AH} = 1/r_\textrm{AH}$, can be obtained as the outermost solution of the equation $\left(d_+\Sigma\right)(v,r_\textrm{AH})=0$ \cite{Chesler:2013lia}.

In order to keep the radial position of the apparent horizon at some fixed value during the time evolution of the system, one may adequately rescale the radial coordinate on each time slice using the radial shift function $\lambda(v)$. Therefore, to obtain the equation of motion for $\lambda(v)$ under such condition, one requires that $\partial_v r_\textrm{AH} = 0$.  As a result, one finds that, $ d_+\left[d_+\Sigma\right](v,r_\textrm{AH}) = A(v,r_\textrm{AH}) \,\partial_r \left[d_+\Sigma\right](v,r_\textrm{AH})$, which leads to the following condition when substituted into the constraint~\eqref{eq:pdeh}, for both the 2RCBH and 1RCBH models,
\begin{align}
A(v,u_{\textrm{AH}}) = \frac{6([d_+B](v,u_{\textrm{AH}}))^2 + 2([d_+\phi](v,u_{\textrm{AH}}))^2}{2V(\phi)+f(\phi)\mathcal{E}^2}.
\label{eq:Astar}
\end{align}
By substituting~\eqref{eq:Astar} into the definition~\eqref{eq:Aa} of the subtracted field $A_s(v,u)$, one obtains the equation governing the time evolution of the radial shift function,
\begin{align}
\partial_v\lambda(v) &= u_{\textrm{AH}}^2 A_s(v,u_{\textrm{AH}})+\frac{1}{2u_{\textrm{AH}}^2}+\frac{\lambda(v)}{u_{\textrm{AH}}}+\frac{\lambda^2(v)}{2} -A(v,u_{\textrm{AH}})\nonumber\\
&=  u_{\textrm{AH}}^2 A_s(v,u_{\textrm{AH}})+\frac{1}{2u_{\textrm{AH}}^2}+\frac{\lambda(v)}{u_{\textrm{AH}}}+\frac{\lambda^2(v)}{2} - \frac{6 u_{\textrm{AH}}^6 (\dd_+ B)_s^2 + 2 u_{\textrm{AH}}^2 (\dd_+ \phi)_s^2}{2 V(u_{\textrm{AH}}^2 \phi_s) + f(u_{\textrm{AH}}^2 \phi_s) \mathcal{E}_s^2}.
\label{eq:dlambda}
\end{align}
Eq.~\eqref{eq:dlambda} evolves in time the initial condition $\lambda(v_0)$ by shifting the radial coordinate $u$ on each time slice such as to keep $u_\textrm{AH} = \textrm{constant}$.

To evolve the initial data $\{B_s(v_0,u),\phi_s(v_0,u)\}$ in time, one requires the specification of the time derivatives $\partial_v B_s$ and $\partial_v \phi_s$. These can be obtained from the expressions for $d_+B=\partial_v B+A\partial_r B$ and $d_+\phi=\partial_v \phi+A\partial_r \phi$ rewritten in terms of the subtracted bulk fields and the compact radial coordinate $u=1/r$. The resulting expressions, also valid for both models, are given below,
\begin{subequations}
\begin{align}
\partial_v B_s(v,u) &= \frac{[d_+B]_s}{u} + \frac{2B_s}{u} + \frac{B_s'}{2} + 4u^3A_sB_s + u^4A_sB_s' + \left(4B_s+uB_s'\right) \lambda + \left(2uB_s +\frac{u^2B_s'}{2} \right)\lambda^2 \nonumber\\
& \,\,\,\,\,\,\, - \left( 4uB_s + u^2B_s' \right)\partial_v\lambda, \label{eq:dtBsa}\\
\partial_v B_s(v,0)&=\partial_vB_4(v), \label{eq:dtBsb}
\end{align}
\end{subequations}

\begin{subequations}
\begin{align}
\partial_v\phi_s(v,u) & = \frac{[d_+\phi]_s}{u} + \frac{\phi_s}{u} + \frac{\phi_s'}{2} +2u^3A_s\phi_s +u^4A_s\phi_s' + \left(u\phi_s'+2\phi_s\right)\lambda +\left(u\phi_s+\frac{u^2\phi_s'}{2}\right)\lambda^2 \nonumber\\
& \,\,\,\,\,\,\, -\left(2u\phi_s+u^2\phi_s'\right)\partial_v\lambda, \label{eq:dtphisa}\\
\partial_v\phi_s(v,0)&=\partial_v \phi_2(v),\label{eq:dtphisb}
\end{align}
\end{subequations}
with the derivatives $Z_s'(v,u)\equiv \partial_u Z_s(v,u)$ being calculated at any fixed time slice by applying the pseudospectral finite differentiation matrix~\cite{boyd01} on the numerical solution $Z_s(v,u)$.

\subsection{Initial data and energy conditions}
\label{sec:initialdata}

As discussed in section \ref{sec:1stedoSystem}, the set of initial data needed to be chosen to evolve in time the bulk fields in the homogeneous isotropization dynamics correspond to, $\left\{\mu/T,B_s(v_0,u),\phi_s(v_0,u),\lambda(v_0)\right\}$, where we choose here $v_0=0$ as the initial time slice, and we take $\lambda(v_0)=0$ as an initial condition. Therefore, for each value of $\mu/T$, one still needs to choose the initial radial profiles for the subtracted metric anisotropy function, $B_s(v_0,u)$, and the subtracted  dilaton field, $\phi_s(v_0,u)$. In the present work, we consider the following functional forms,
\begin{align}
B_s(v_0=0,u) &= \mathcal{B}\, e^{-\mathcal{S}_B (u-u_B)^2}, \label{eq:Bs0}\\
\phi_s(v_0=0,u) &= \mathcal{F}\, e^{-\mathcal{S}_\phi (u-u_\phi)^2}, \label{eq:phis0}
\end{align}
where, for different values of $\mu/T$, we chose to analyze here four different initial conditions (ICs) as indicated in Table~\ref{tabICs}.

\begin{table}[h!]
\centering
\begin{tabular}{|c||c|c|c||c|c|c|}
\hline
IC $\#$ & $\mathcal{B}$ & $\mathcal{S}_B$ & $u_B$ & $\mathcal{F}$ & $\mathcal{S}_\phi$ & $u_\phi$ \\
\hline
\hline
1 & 0.1 & $10^2$ & 0.4 & $-0.01$ & $10^2$ & 0.3 \\
\hline
2 & 0.1 & 0 & 0 & $-0.1$ & 1 & 0.3 \\
\hline
3 & 0.5 & $10^2$ & 0.4 & $-\sqrt{2/3}\, Q^2$ & 0 & 0 \\
\hline
4 & 0.5 & 10 & 0.4 & $-\sqrt{2/3}\, Q^2$ & 0 & 0 \\
\hline
\end{tabular}
\caption{Parameter sets chosen for the initial profiles of the subtracted metric anisotropy function \eqref{eq:Bs0} and the subtracted dilaton field \eqref{eq:phis0}.}
\label{tabICs}
\end{table}

As in~\cite{Rougemont:2021qyk,Rougemont:2021gjm} for the purely thermal SYM plasma, and in~\cite{Rougemont:2022piu,Rougemont:2024hpf} for the 1RCBH plasma, both in the case of homogeneous isotropization and Bjorken flow dynamics, also for the 2RCBH plasma, the weak energy condition (WEC), which is equivalent to the strong energy condition (SEC) in the case of conformal field theories as the 2RCBH and 1RCBH models, reads as follows,
\begin{align}
\textrm{WEC / SEC:}\qquad \hat{\varepsilon}\ge0 ,\quad -4 \le \frac{\Delta\hat{p}}{\hat{\varepsilon}} \le 2,\label{eq:WEC}
\end{align}
while the dominant energy condition (DEC) is given by,
\begin{align}
\textrm{DEC:}\qquad \hat{\varepsilon}\ge0 ,\quad -1 \le \frac{\Delta\hat{p}}{\hat{\varepsilon}} \le 2.\label{eq:DEC}
\end{align}
As we shall see, some initial data preserving all the energy conditions on the initial time slice may transiently violate these energy conditions when the 2RCBH plasma is still far-from-equilibrium, similarly to what was reported in previous works for the purely SYM plasma and the 1RCBH plasma~\cite{Rougemont:2021qyk,Rougemont:2021gjm,Rougemont:2022piu,Rougemont:2024hpf}. These transient violations are related to violations of the inequalities in~\eqref{eq:WEC} and~\eqref{eq:DEC} involving the pressure anisotropy, while the energy density is always non-negative. Out-of-equilibrium transient violations of the classical energy conditions are possible to happen due to the quantum nature of the holographic fluids considered here. In fact, some quantum effects are well-known to violate classical energy conditions~\cite{Visser:1999de,Costa:2021hpu}.

\begin{figure}[h!]
\centering  
\subfigure[Scalar Condensate]{\includegraphics[width=0.425\linewidth]{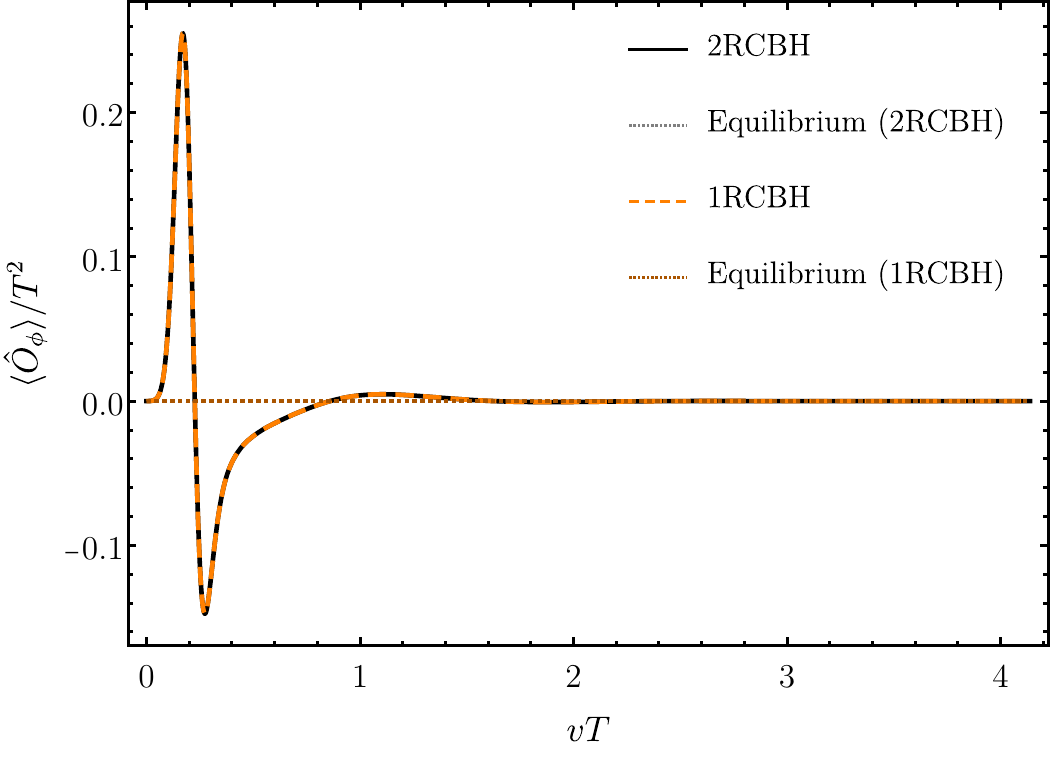}\label{fig:ModelmuT0a}}
\subfigure[Logarithmic Scalar Condensate]{\includegraphics[width=0.425\linewidth]{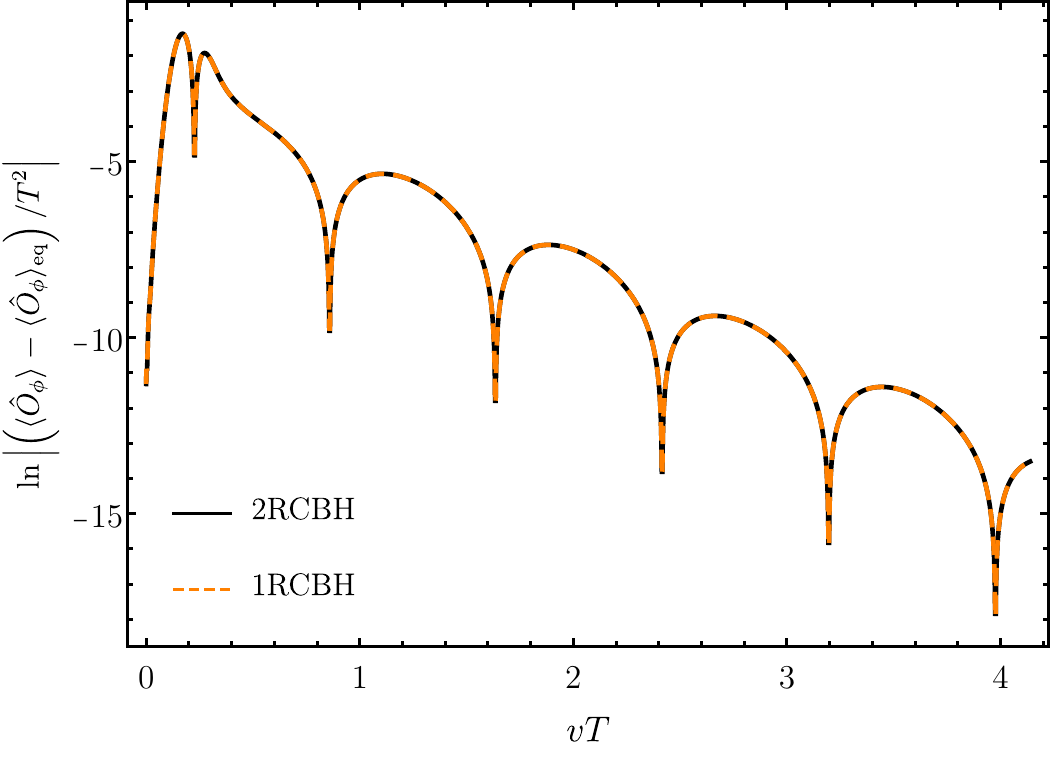}\label{fig:ModelmuT0b}}
\subfigure[Pressure anisotropy]{\includegraphics[width=0.425\linewidth]{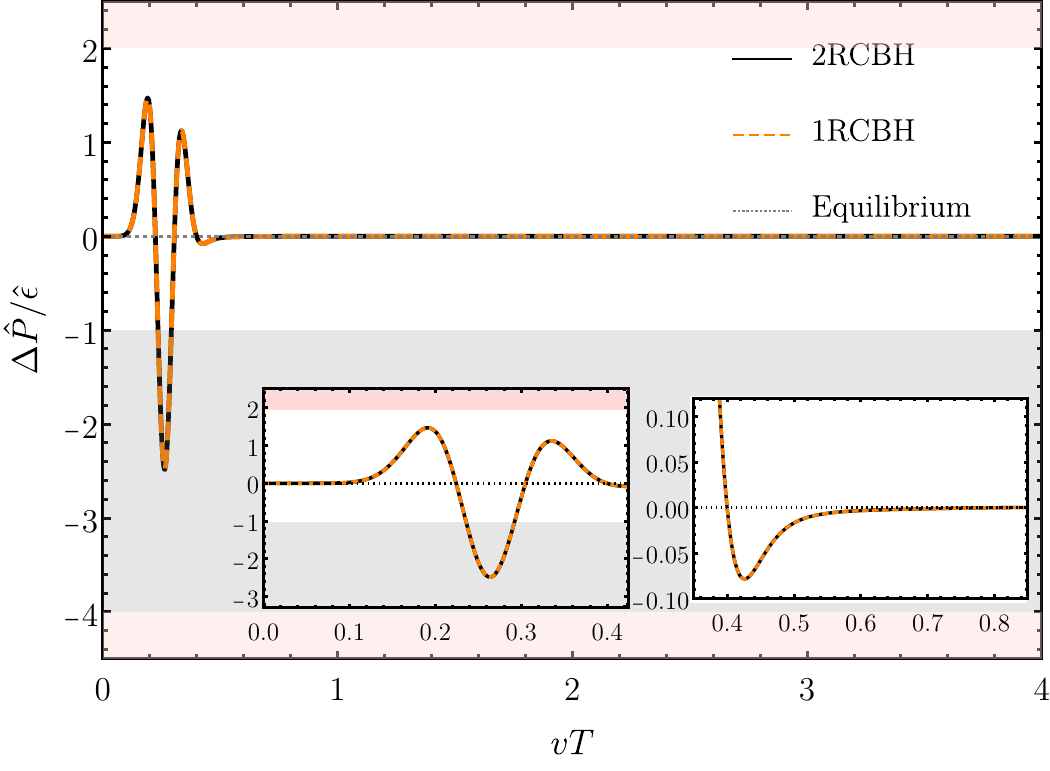}\label{fig:ModelmuT0c}}
\subfigure[Logarithmic pressure anisotropy]{\includegraphics[width=0.425\linewidth]{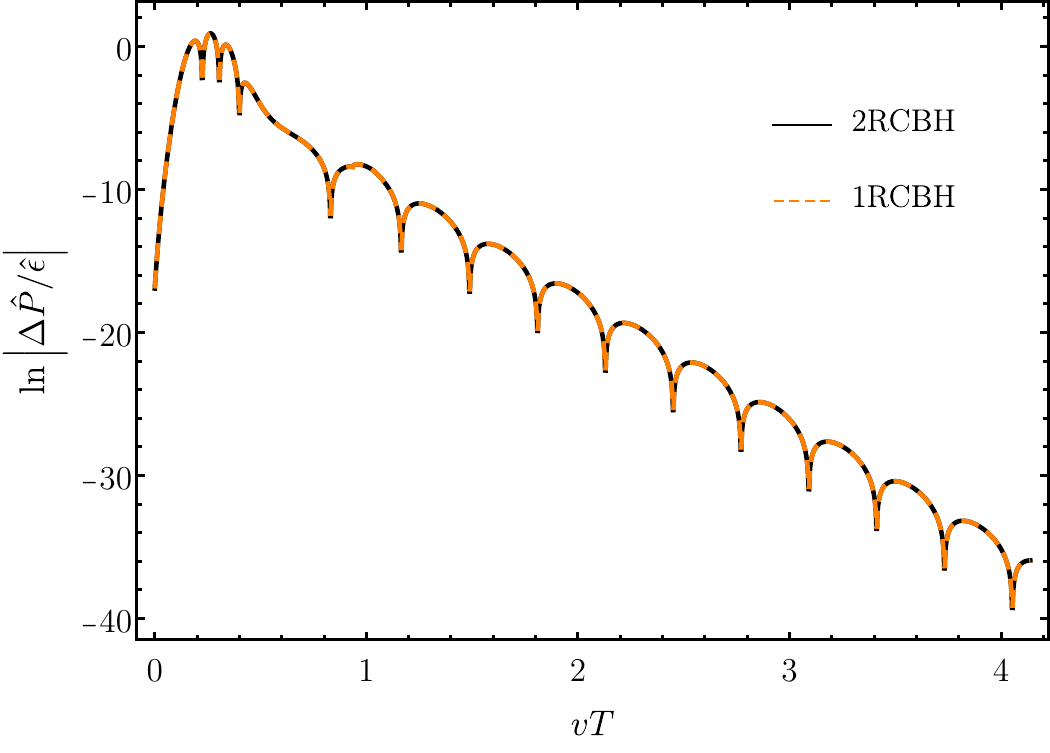}\label{fig:ModelmuT0d}}
\subfigure[Entropy Density]{\includegraphics[width=0.425\linewidth]{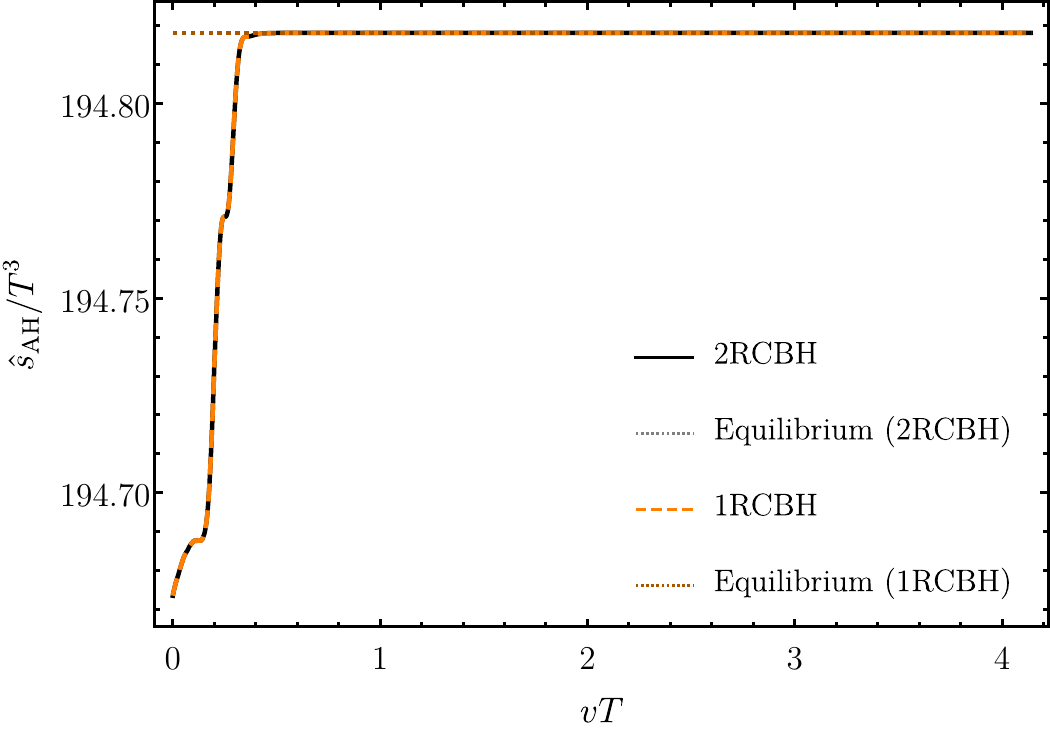}\label{fig:ModelmuT0e}}
\subfigure[Logarithmic Entropy Density]{\includegraphics[width=0.425\linewidth]{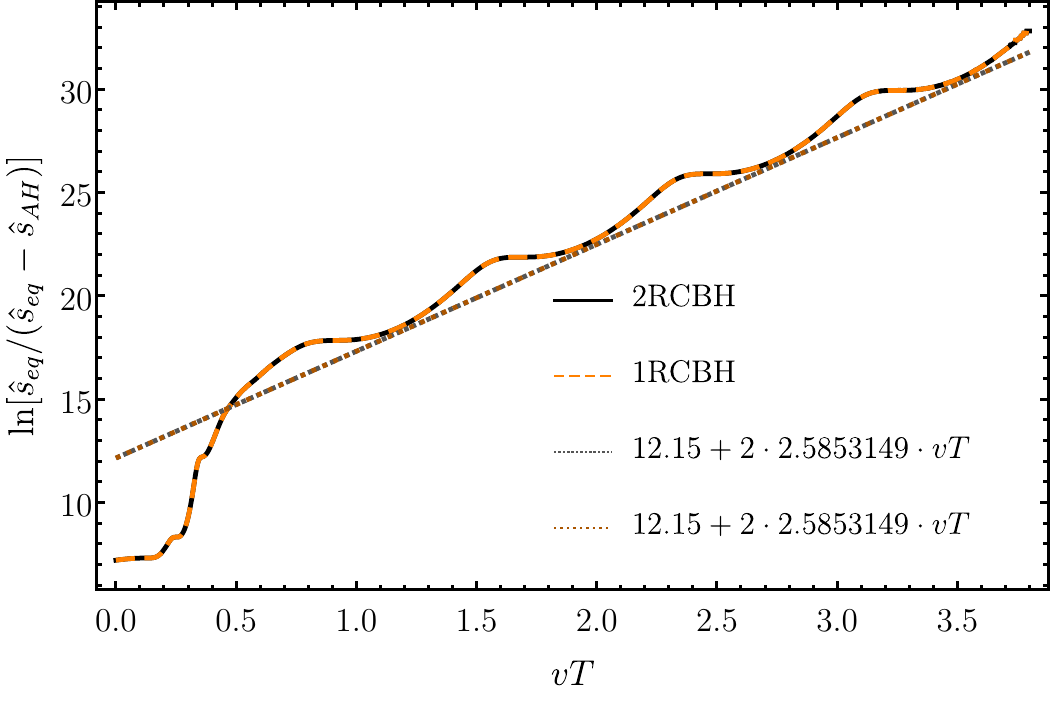}\label{fig:ModelmuT0f}}
\caption{Comparison between the 2RCBH and 1RCBH models for different observables at $\mu/T=0$ considering IC1 in Table~\ref{tabICs}. In Fig.~\ref{fig:ModelmuT0c}, the gray shaded band demarcates the region where the DEC is violated, while the pink shaded band delimits the region where both the WEC and DEC are violated. In Fig.~\ref{fig:ModelmuT0f} the parametrization of the support slopes is also shown for the entropy stairways.}
\label{fig:ModelmuT0}
\end{figure}

\begin{figure}[h!]
\centering  
\subfigure[Scalar Condensate]{\includegraphics[width=0.425\linewidth]{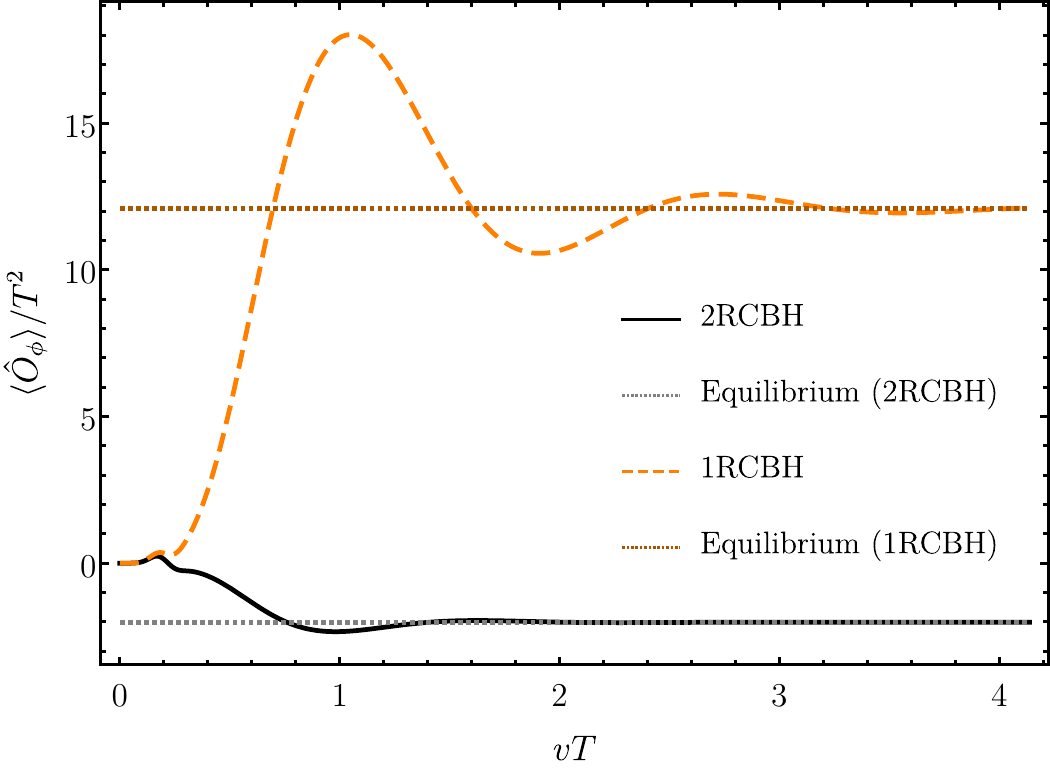}\label{fig:ModelmuT22a}}
\subfigure[Logarithmic Scalar Condensate]{\includegraphics[width=0.425\linewidth]{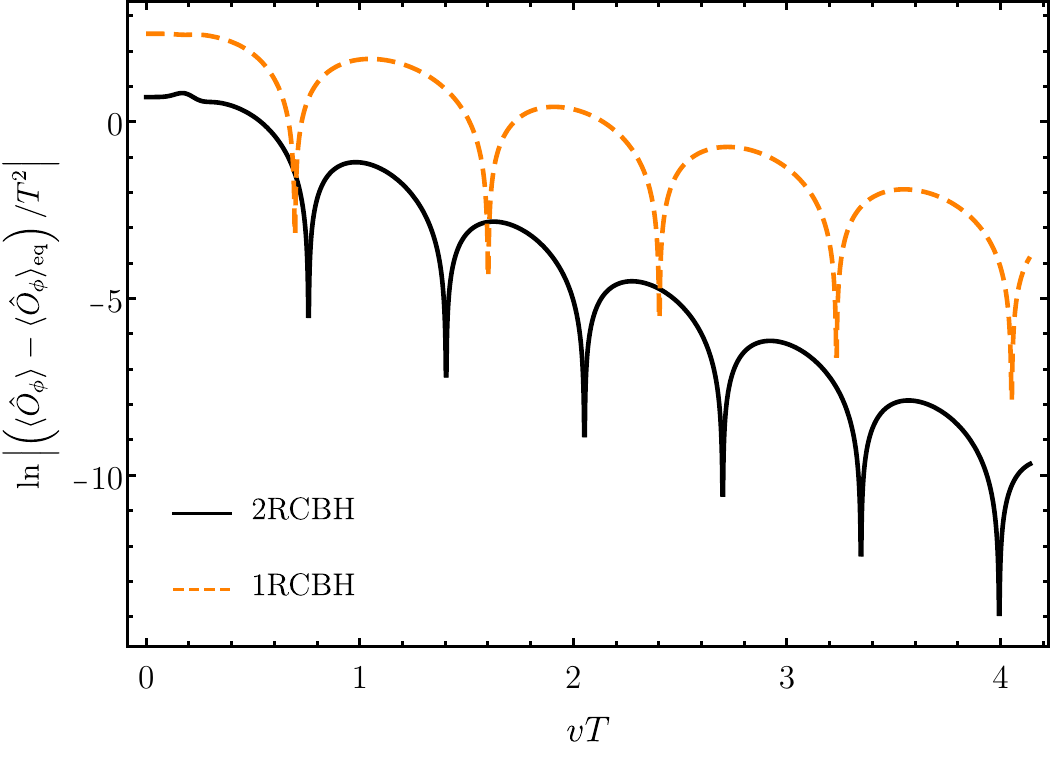}\label{fig:ModelmuT22b}}
\subfigure[Pressure anisotropy]{\includegraphics[width=0.425\linewidth]{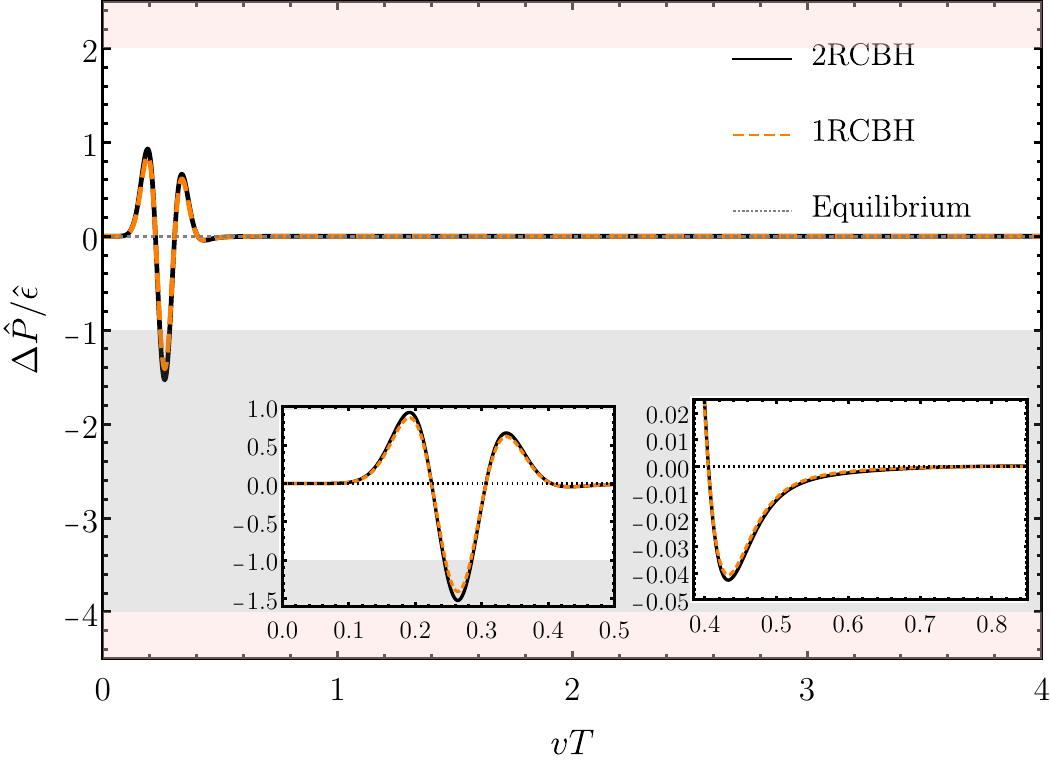}\label{fig:ModelmuT22c}}
\subfigure[Logarithmic pressure anisotropy]{\includegraphics[width=0.425\linewidth]{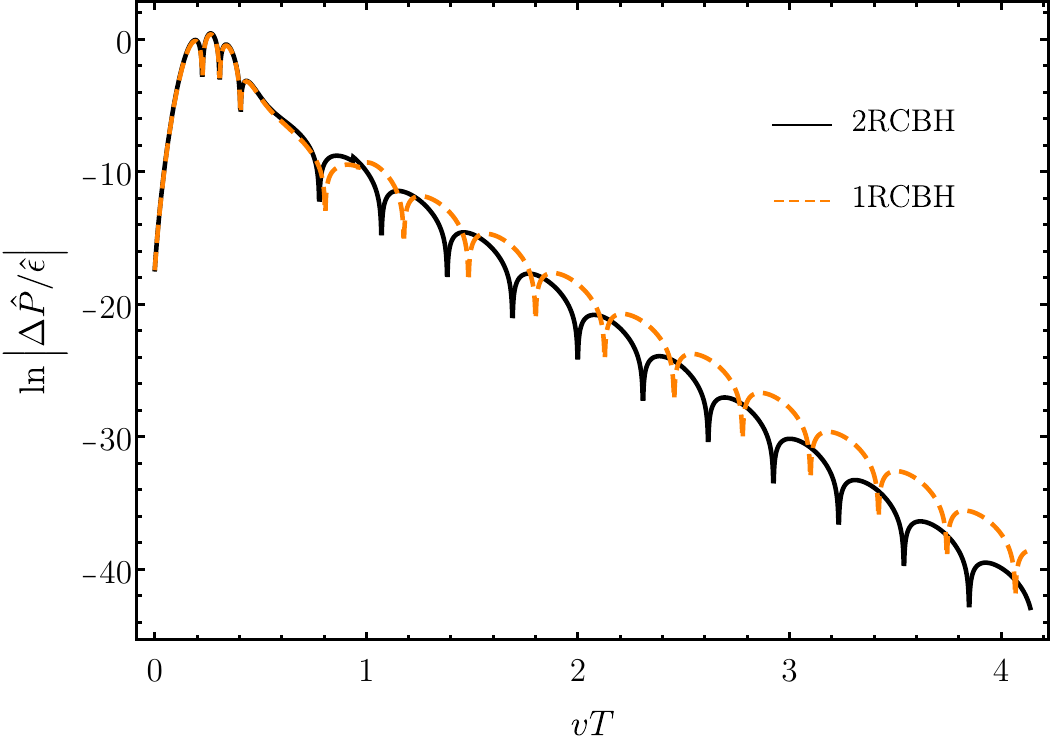}\label{fig:ModelmuT22d}}
\subfigure[Entropy Density]{\includegraphics[width=0.425\linewidth]{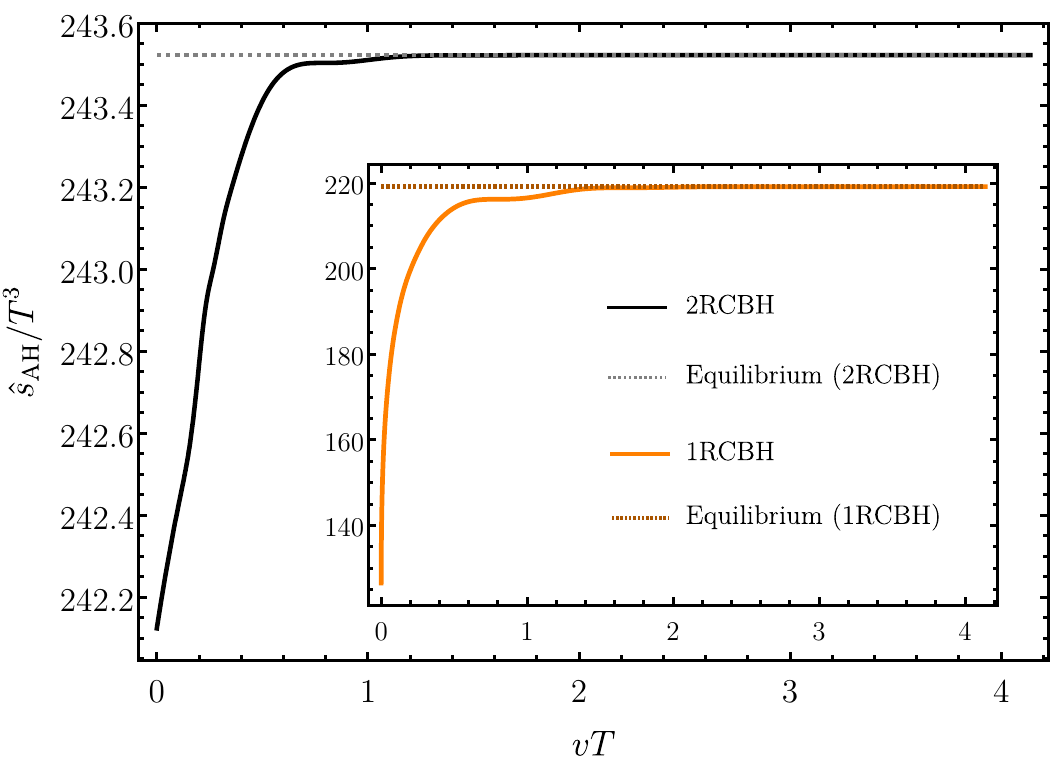}\label{fig:ModelmuT22e}}
\subfigure[Logarithmic Entropy Density]{\includegraphics[width=0.425\linewidth]{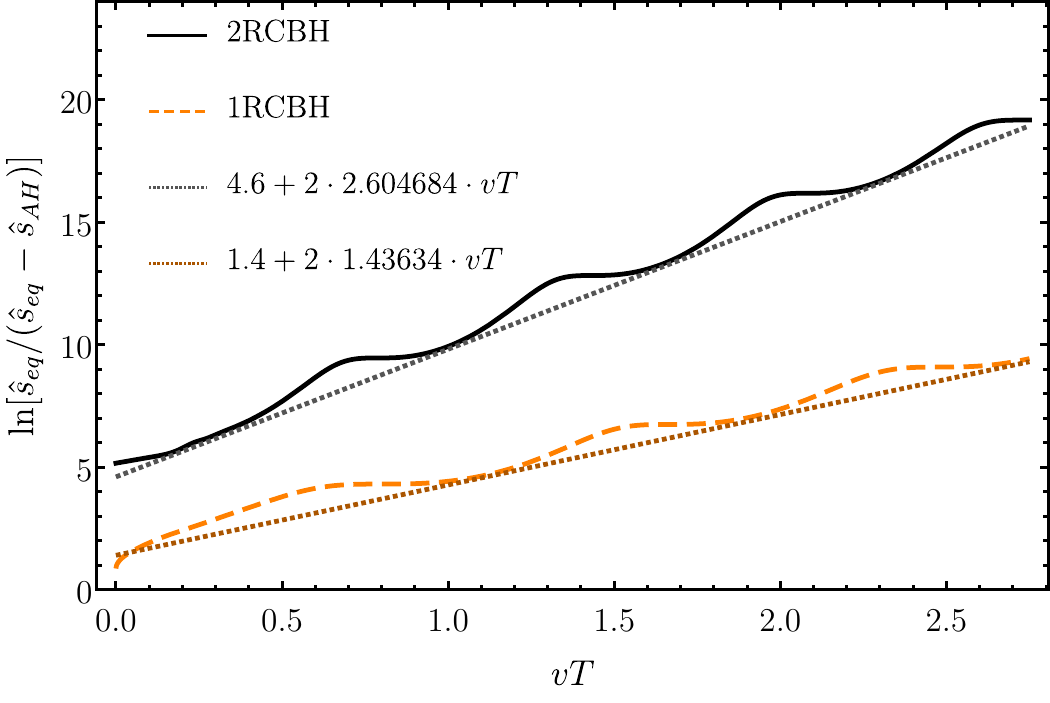}\label{fig:ModelmuT22f}}
\caption{Comparison between the 2RCBH and 1RCBH models for different observables at $\mu/T=\pi/\sqrt{2}$ considering IC1 in Table~\ref{tabICs}. In Fig.~\ref{fig:ModelmuT22c}, the gray shaded band demarcates the region where the DEC is violated, while the pink shaded band delimits the region where both the WEC and DEC are violated. In Fig.~\ref{fig:ModelmuT22f} the parametrization of the support slopes is also shown for the entropy stairways.}
\label{fig:ModelmuT22}
\end{figure}

\begin{figure}[h!]
\centering  
\subfigure[Condensate Scalar]{\includegraphics[width=0.425\linewidth]{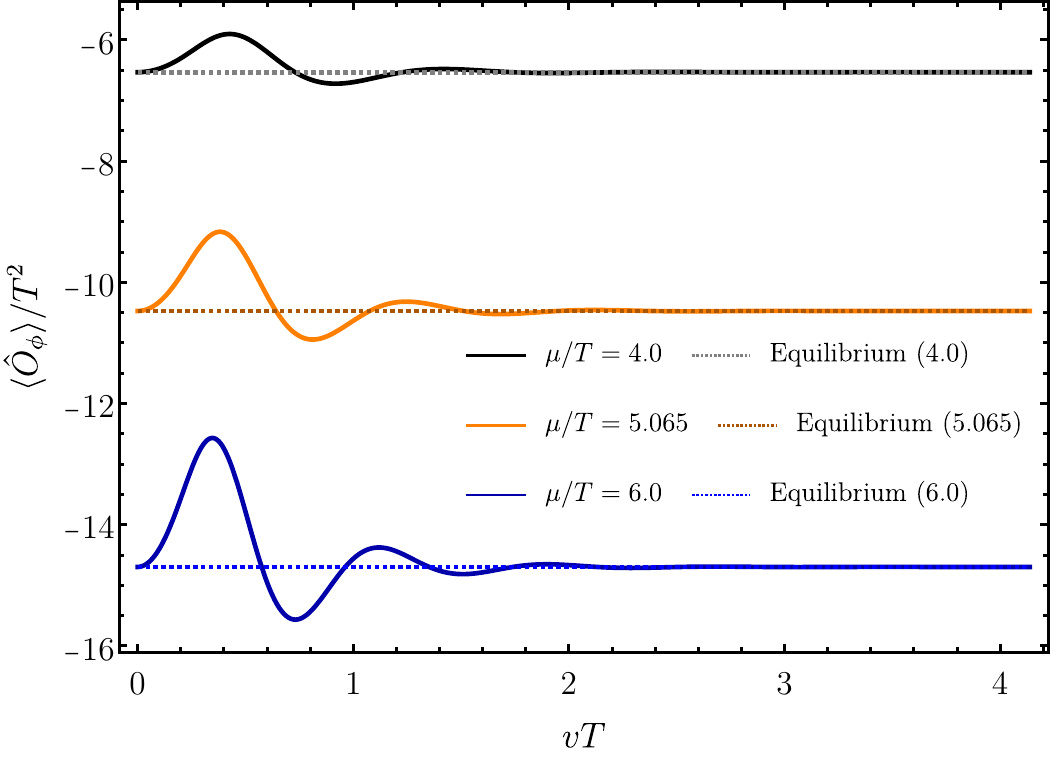}\label{fig:muTIC3a}}
\subfigure[Logarithmic Condensate Scalar]{\includegraphics[width=0.425\linewidth]{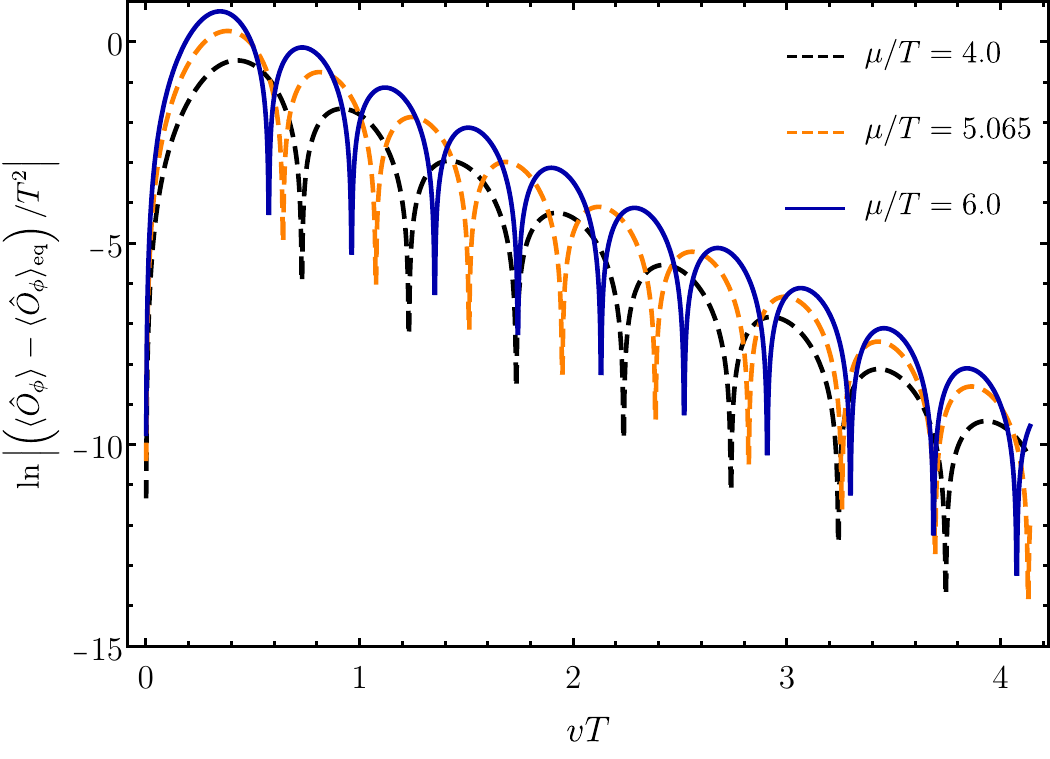}\label{fig:muTIC3b}}
\subfigure[Pressure anisotropy]{\includegraphics[width=0.425\linewidth]{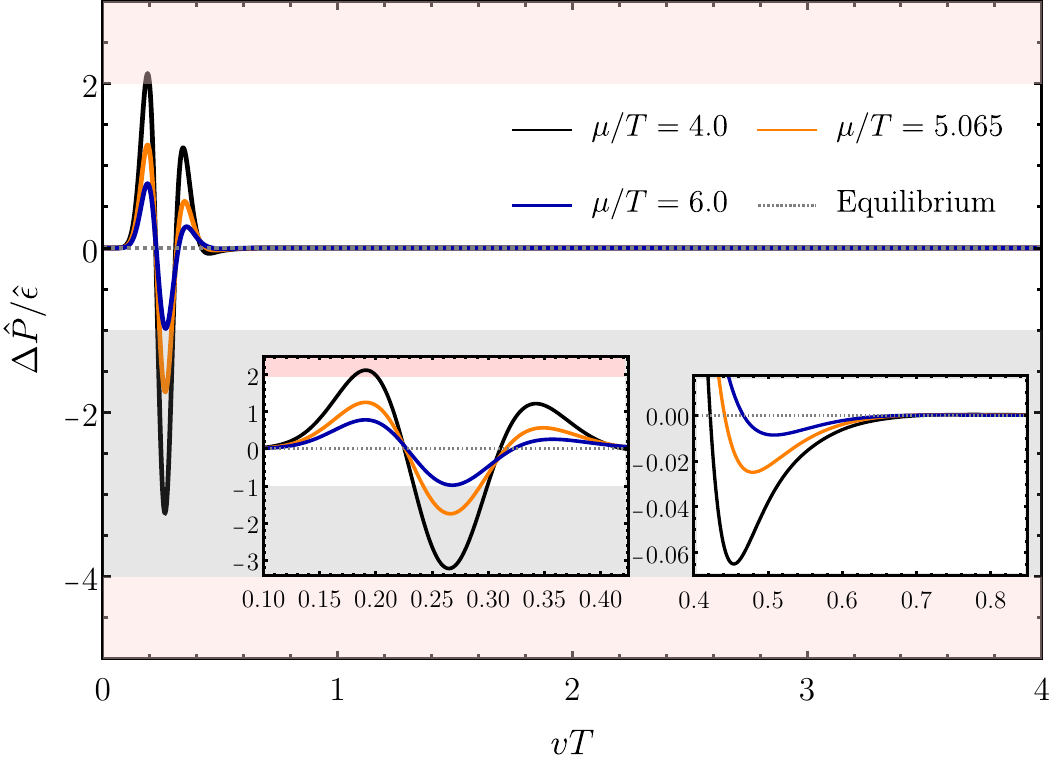}\label{fig:muTIC3c}}
\subfigure[Logarithmic pressure anisotropy]{\includegraphics[width=0.425\linewidth]{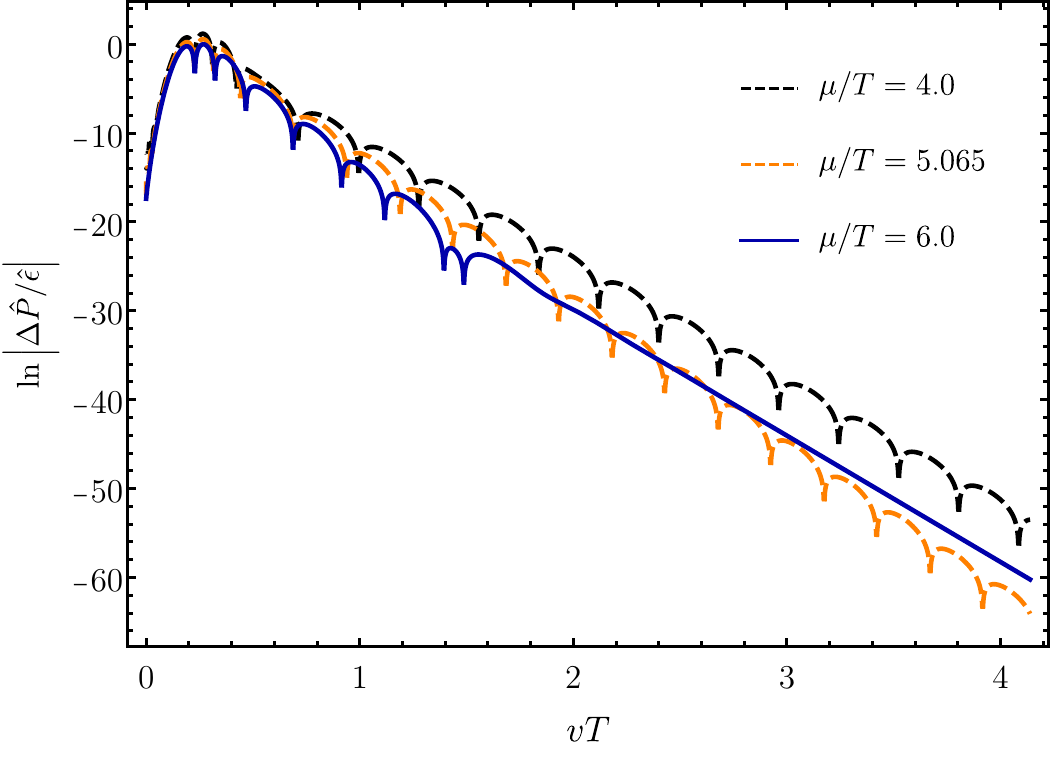}\label{fig:muTIC3d}}
\subfigure[Entropy Density]{\includegraphics[width=0.425\linewidth]{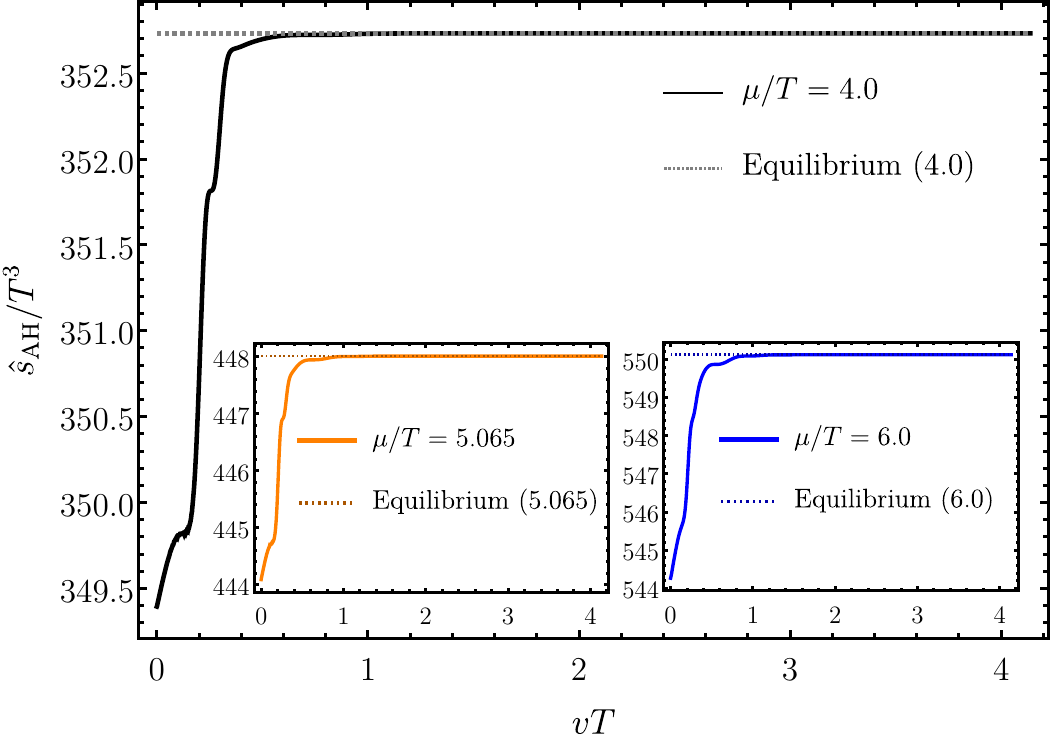}\label{fig:muTIC3e}}
\subfigure[Logarithmic Entropy Density]{\includegraphics[width=0.425\linewidth]{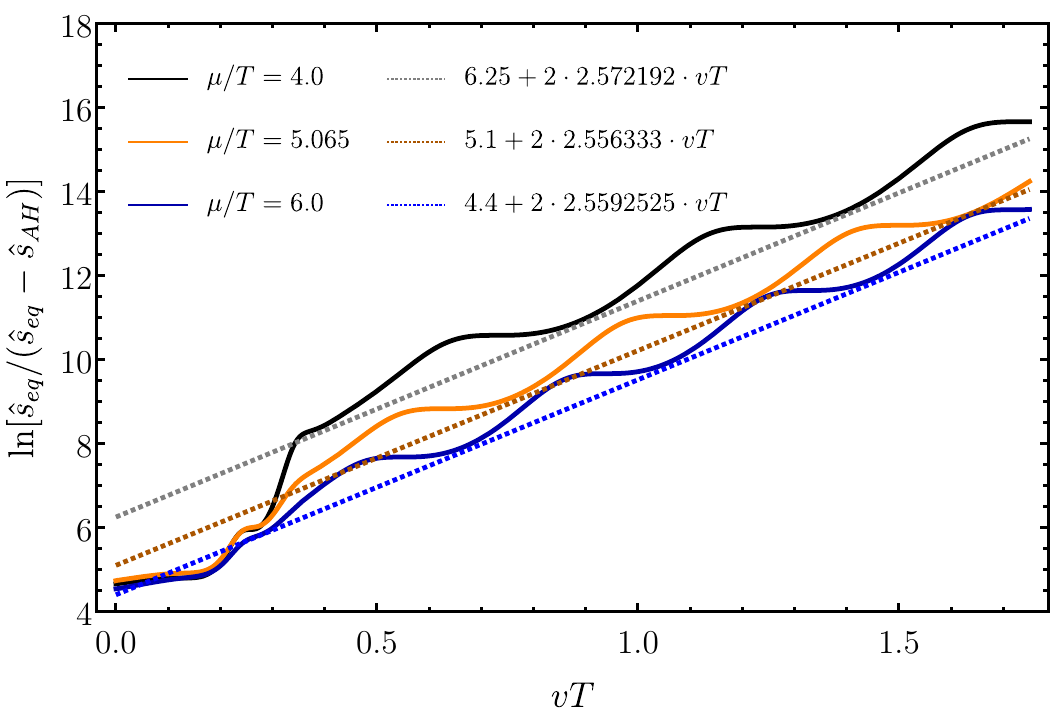}\label{fig:muTIC3f}}
\caption{Comparison for different observables evaluated at $\mu/T=4$, $\mu/T=5.065$, and $\mu/T=6$ in the case of the initial condition IC3 in Table~\ref{tabICs} for the 2RCBH model. In Fig.~\ref{fig:muTIC3c}, the gray shaded band demarcates the region where the DEC is violated, while the pink shaded band delimits the region where both the WEC and DEC are violated. In Fig.~\ref{fig:muTIC3f} the parametrization of the support slopes is also shown for the entropy stairways.}
\label{fig:muTIC3}
\end{figure}

\begin{figure}[h!]
\centering  
\subfigure[Condensate Scalar]{\includegraphics[width=0.425\linewidth]{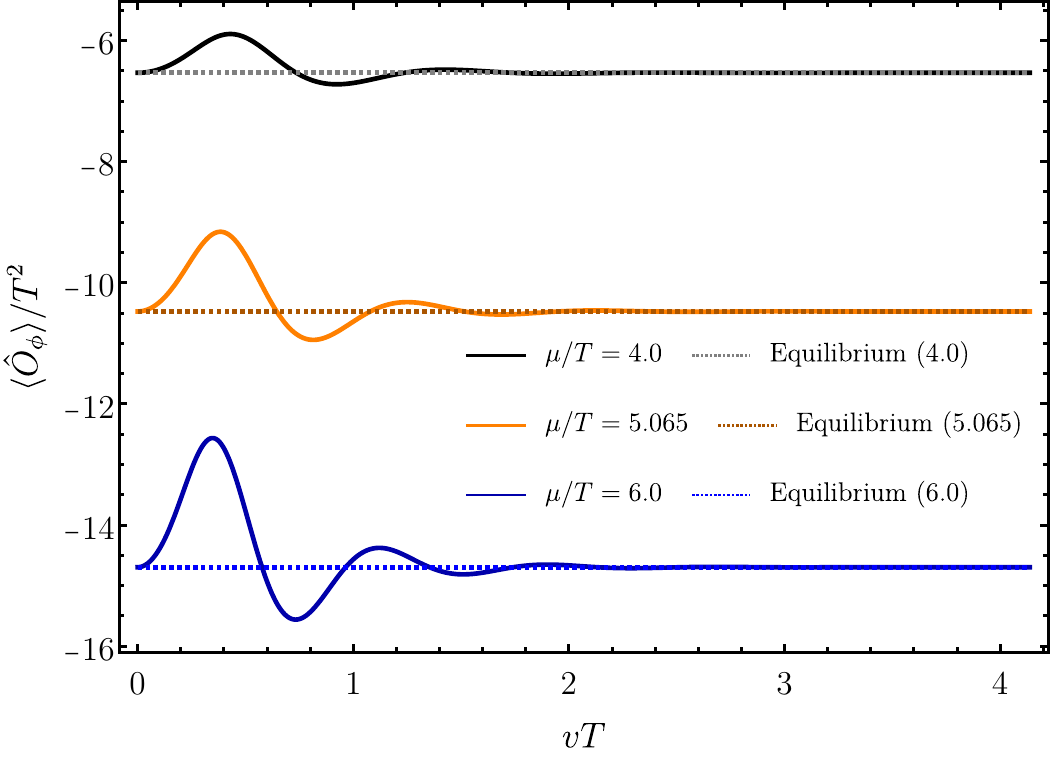}\label{fig:muTIC4a}}
\subfigure[Logarithmic Condensate Scalar]{\includegraphics[width=0.425\linewidth]{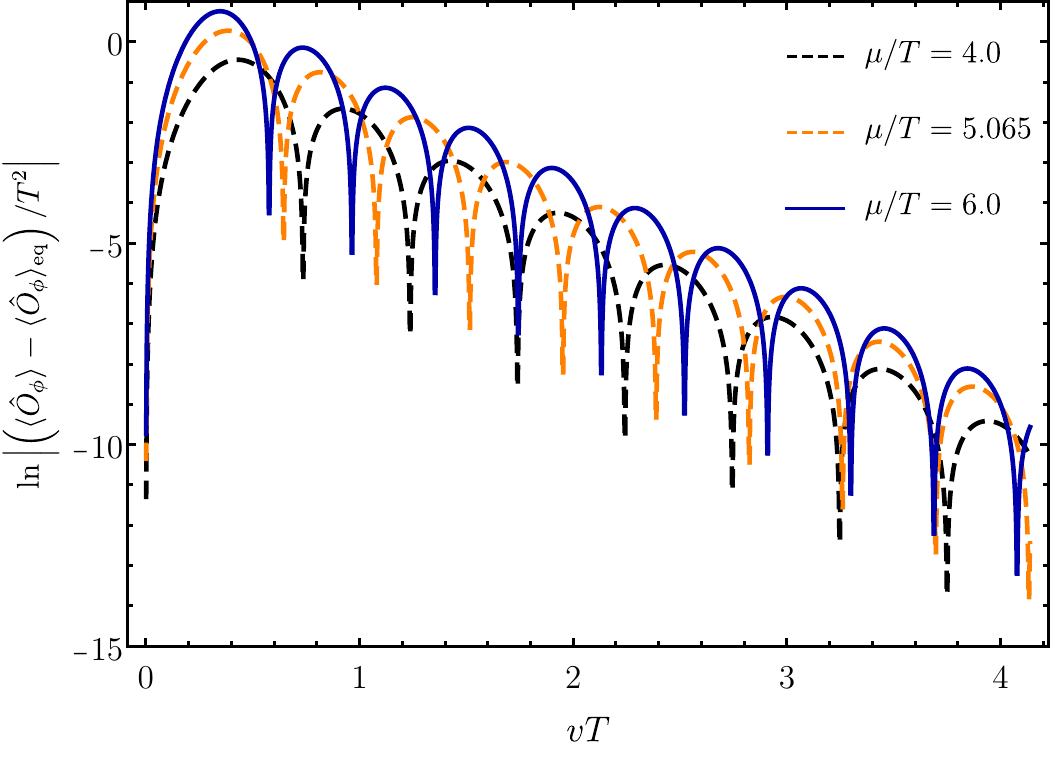}\label{fig:muTIC4b}}
\subfigure[Pressure anisotropy]{\includegraphics[width=0.425\linewidth]{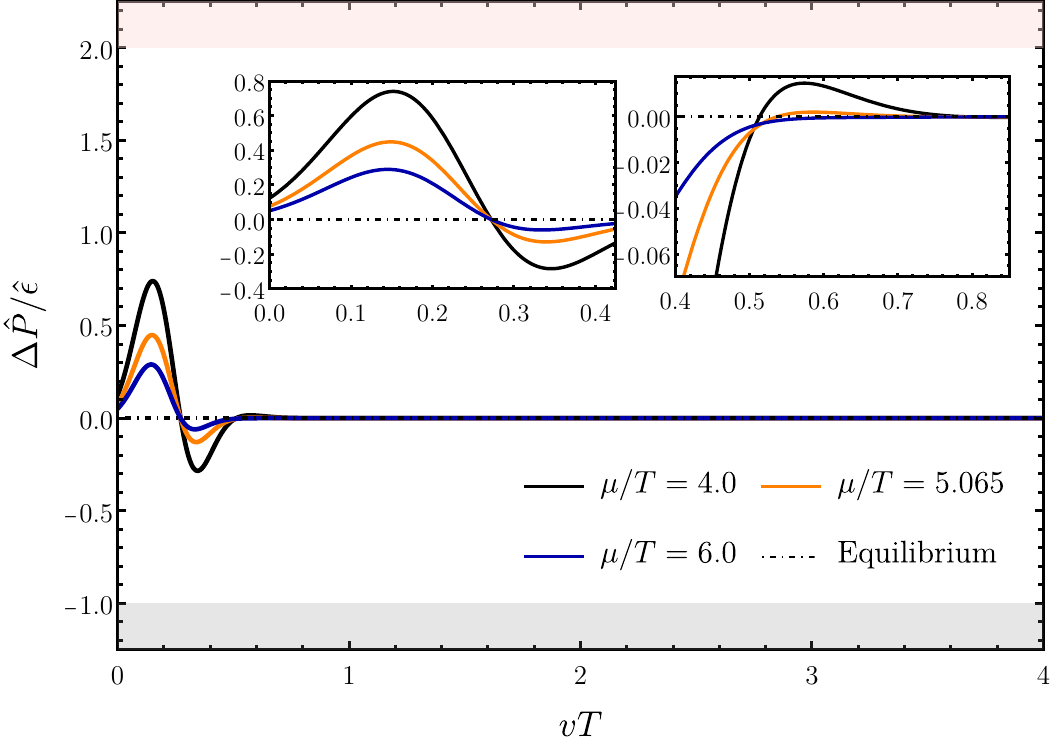}\label{fig:muTIC4c}}
\subfigure[Logarithmic pressure anisotropy]{\includegraphics[width=0.425\linewidth]{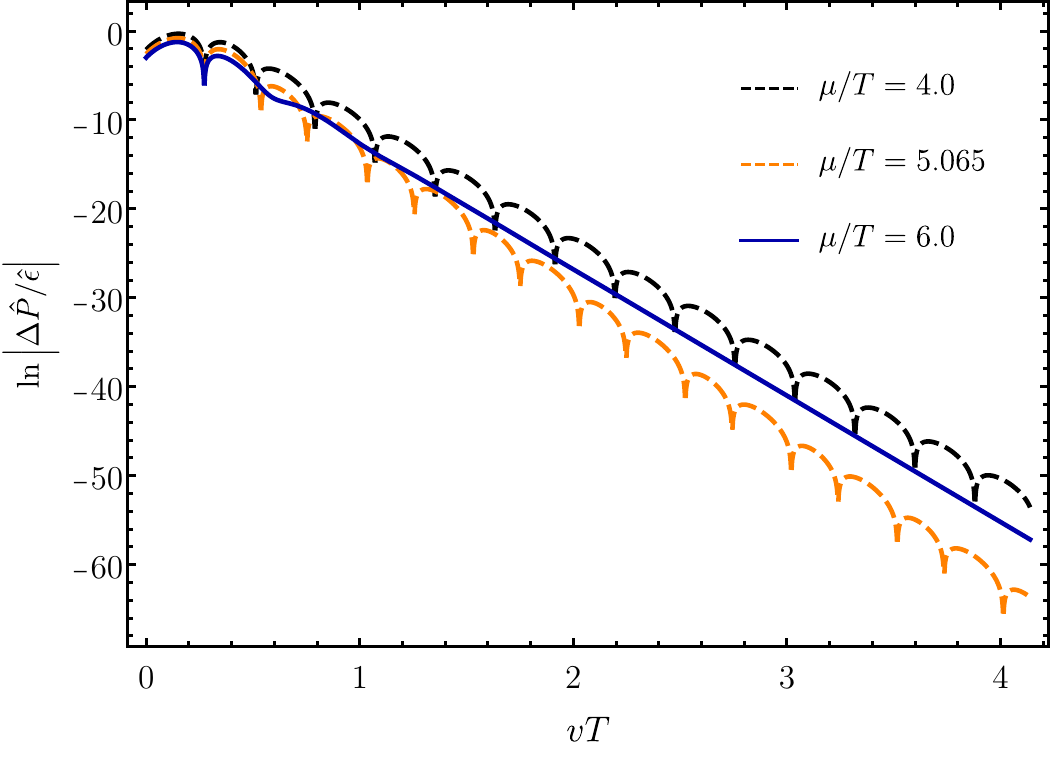}\label{fig:muTIC4d}}
\subfigure[Entropy Density]{\includegraphics[width=0.425\linewidth]{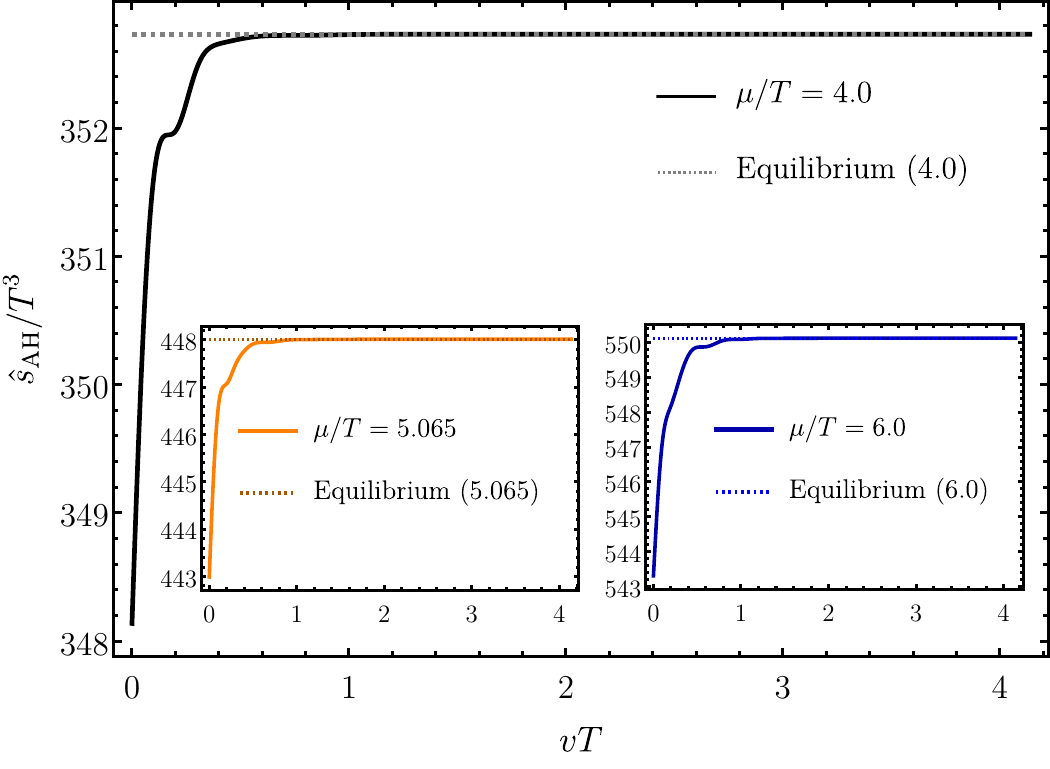}\label{fig:muTIC4e}}
\subfigure[Logarithmic Entropy Density]{\includegraphics[width=0.425\linewidth]{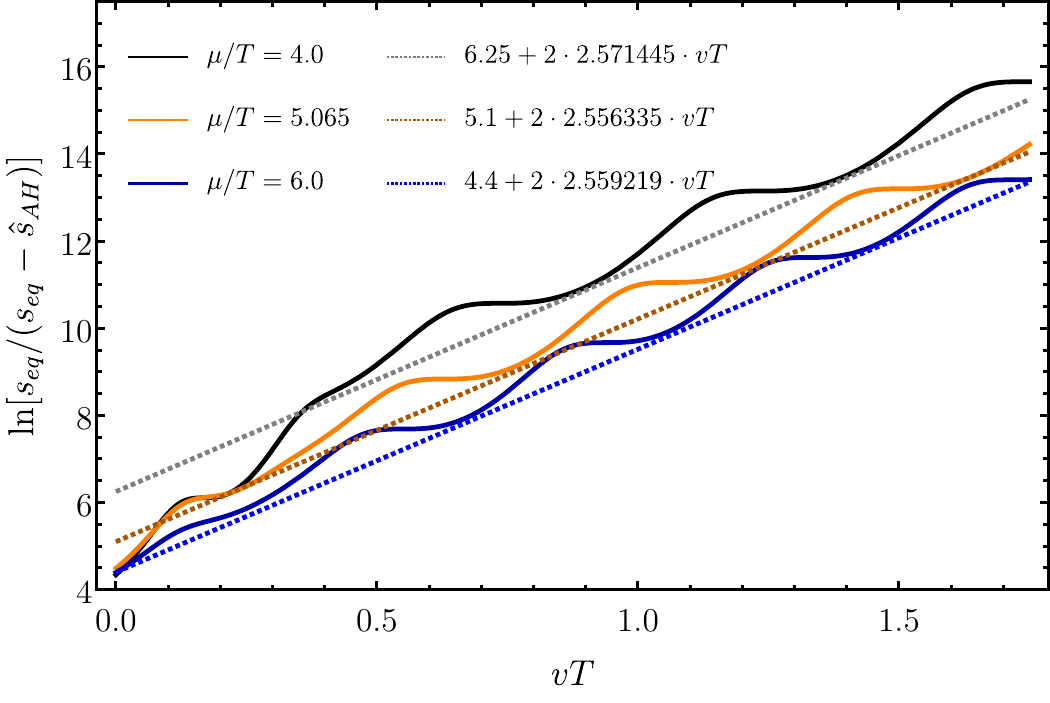}\label{fig:muTIC4f}}
\caption{Comparison for different observables evaluated at $\mu/T=4$, $\mu/T=5.065$, and $\mu/T=6$ in the case of the initial condition IC4 in Table~\ref{tabICs} for the 2RCBH model. In Fig.~\ref{fig:muTIC4c}, the gray shaded band demarcates the region where the DEC is violated, while the pink shaded band delimits the region where both the WEC and DEC are violated. In Fig.~\ref{fig:muTIC4f}, the parametrization of the support slopes is also shown for the entropy stairways.}
\label{fig:muTIC4}
\end{figure}

\section{Numerical results for the homogeneous isotropization dynamics}
\label{sec:results}

\begin{figure}[h!]
\centering  
\subfigure[Condensate Scalar]{\includegraphics[width=0.425\linewidth]{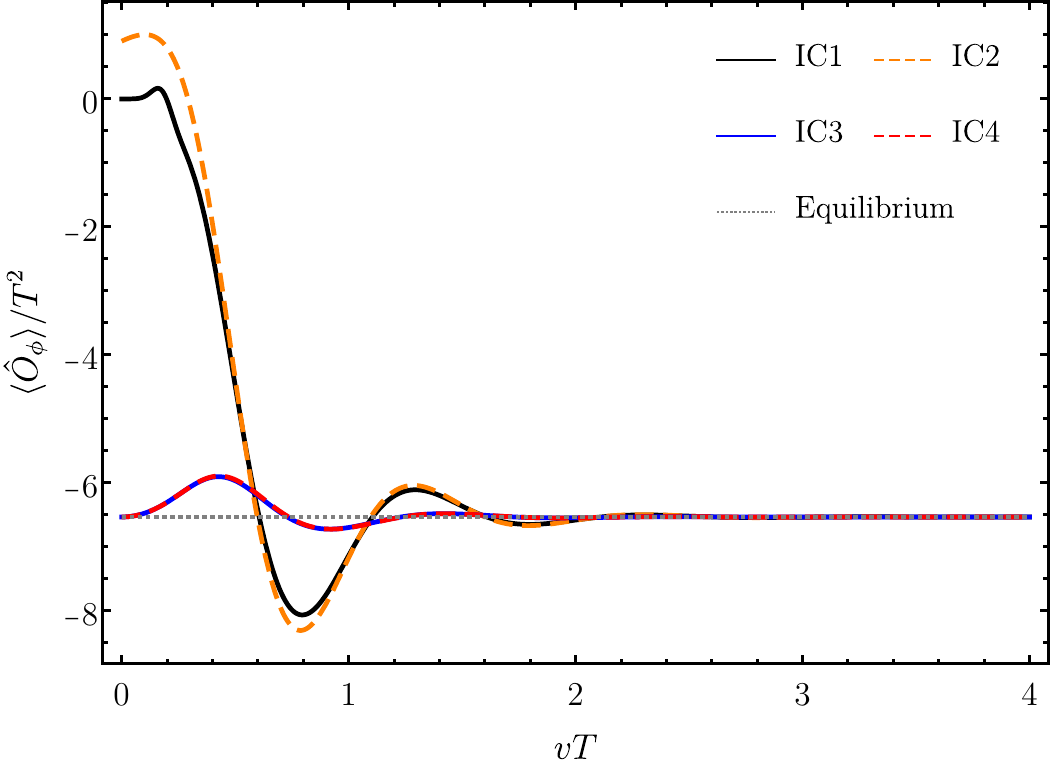}\label{fig:ICmuT4a}}
\subfigure[Logarithmic Condensate Scalar]{\includegraphics[width=0.425\linewidth]{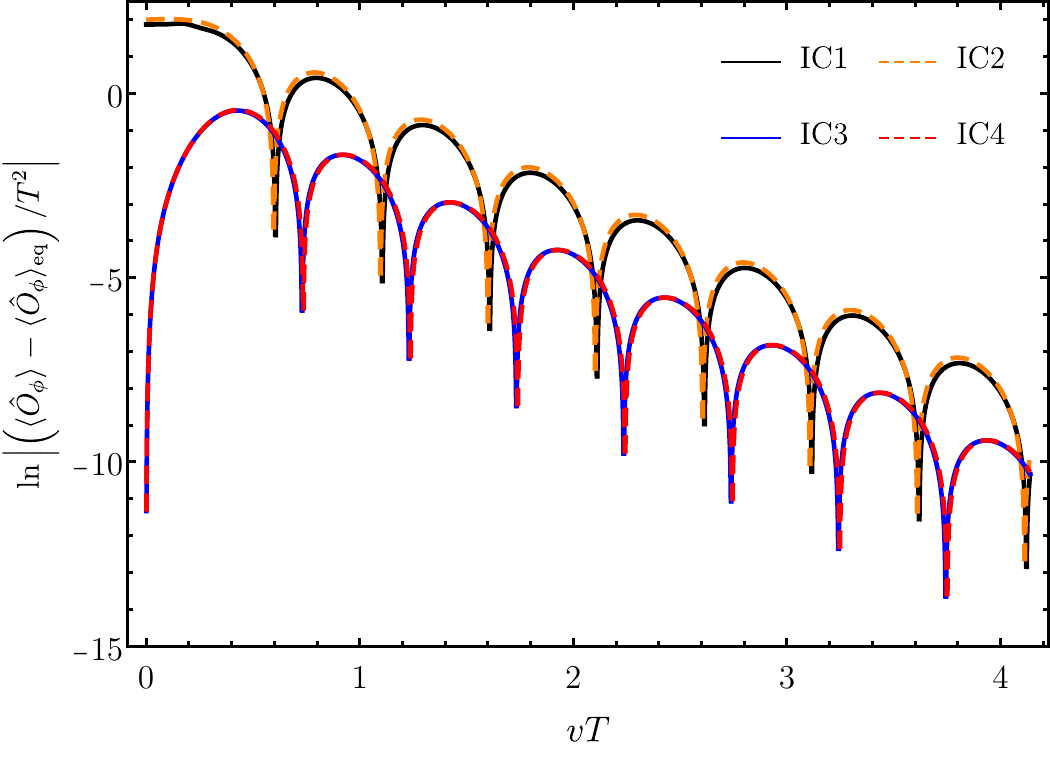}\label{fig:ICmuT4b}}
\subfigure[Pressure anisotropy]{\includegraphics[width=0.425\linewidth]{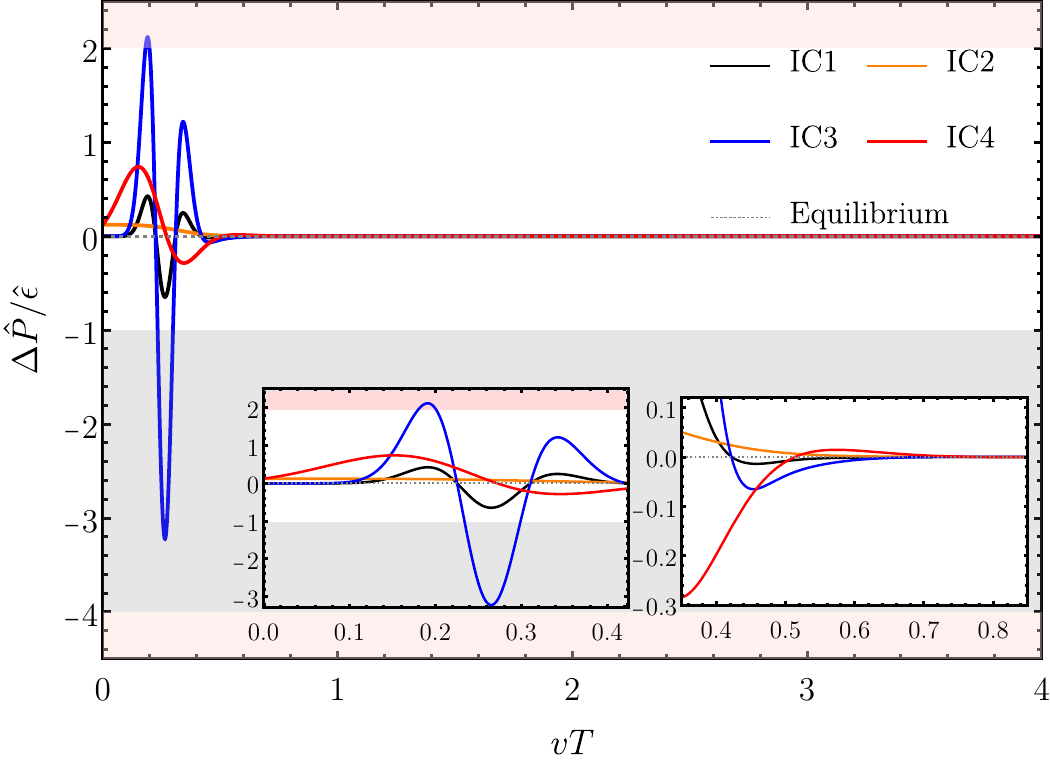}\label{fig:ICmuT4c}}
\subfigure[Logarithmic pressure anisotropy]{\includegraphics[width=0.425\linewidth]{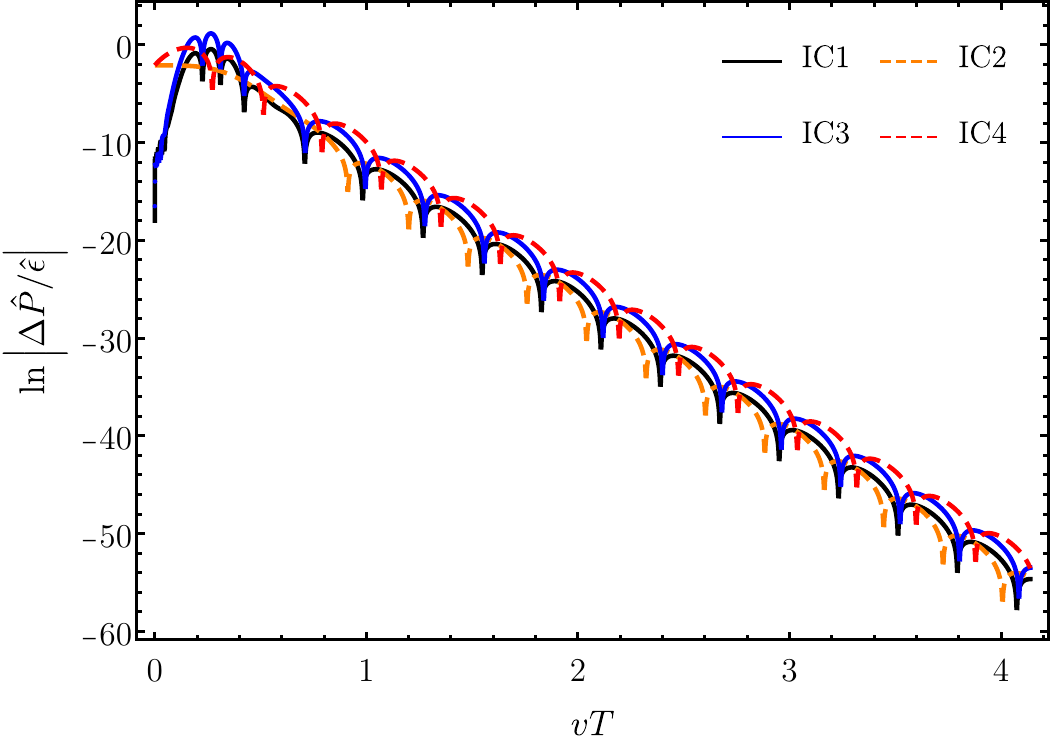}\label{fig:ICmuT4d}}
\subfigure[Entropy Density]{\includegraphics[width=0.425\linewidth]{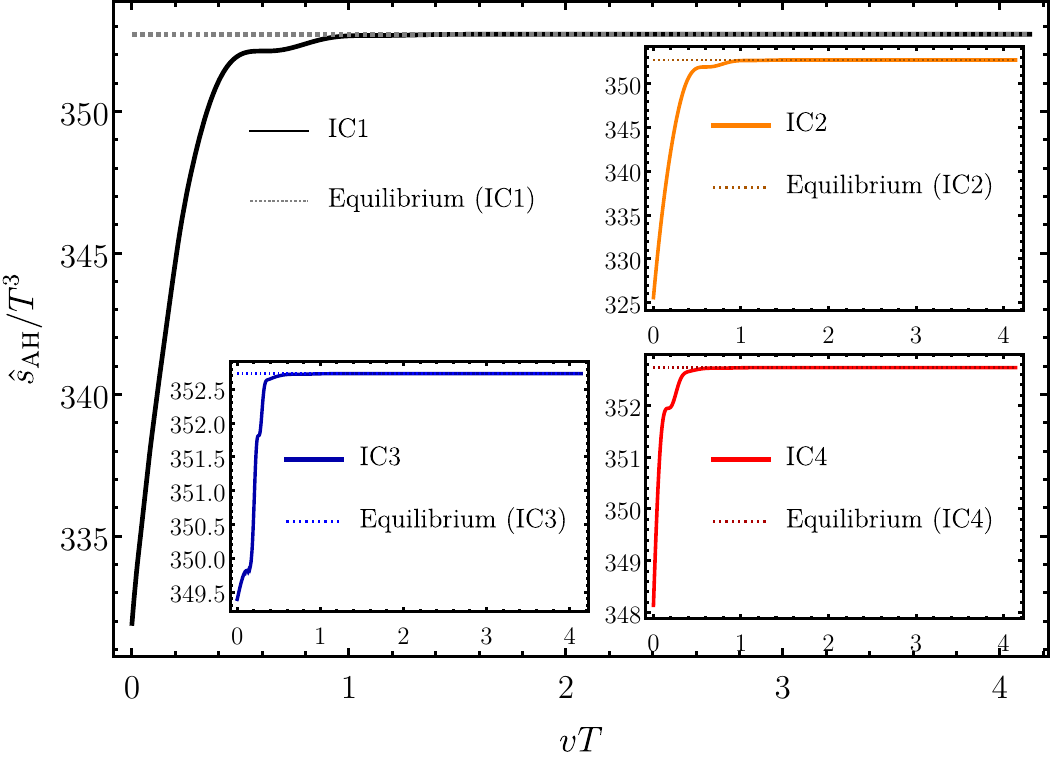}\label{fig:ICmuT4e}}
\subfigure[Logarithmic Entropy Density]{\includegraphics[width=0.425\linewidth]{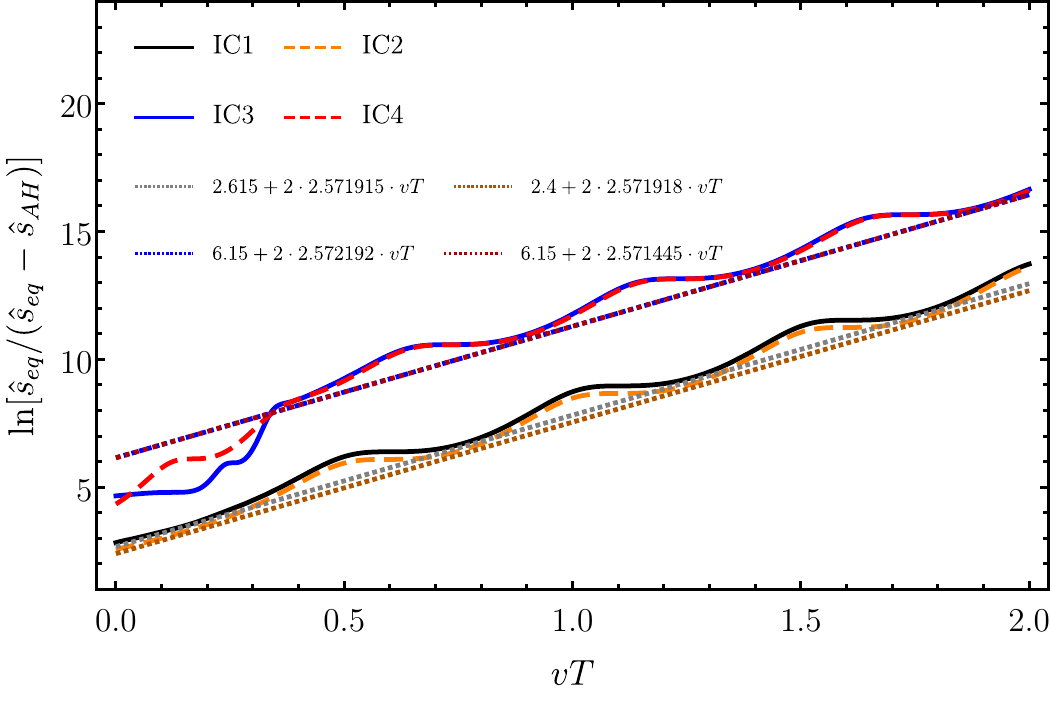}\label{fig:ICmuT4f}}
\caption{Comparison between the initial conditions IC1 --- IC4 in Table~\ref{tabICs} for different observables evaluated at $\mu/T=4$ for the 2RCBH model. In Fig.~\ref{fig:ICmuT4c}, the gray shaded band demarcates the region where the DEC is violated, while the pink shaded band delimits the region where both the WEC and DEC are violated. In Fig.~\ref{fig:ICmuT4f} the parametrization of the support slopes is also shown for the entropy stairways.}
\label{fig:ICmuT4}
\end{figure}

\begin{figure}[h!]
\centering  
\subfigure[Condensate Scalar]{\includegraphics[width=0.425\linewidth]{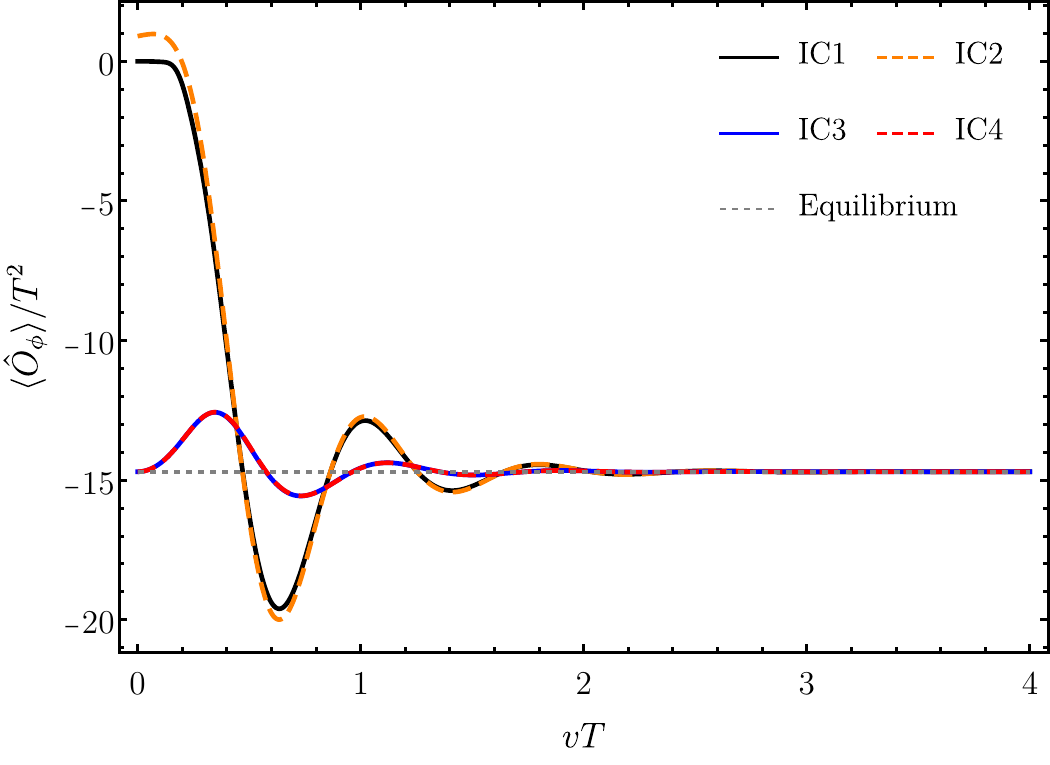}\label{fig:ICmuT6a}}
\subfigure[Logarithmic Condensate Scalar]{\includegraphics[width=0.425\linewidth]{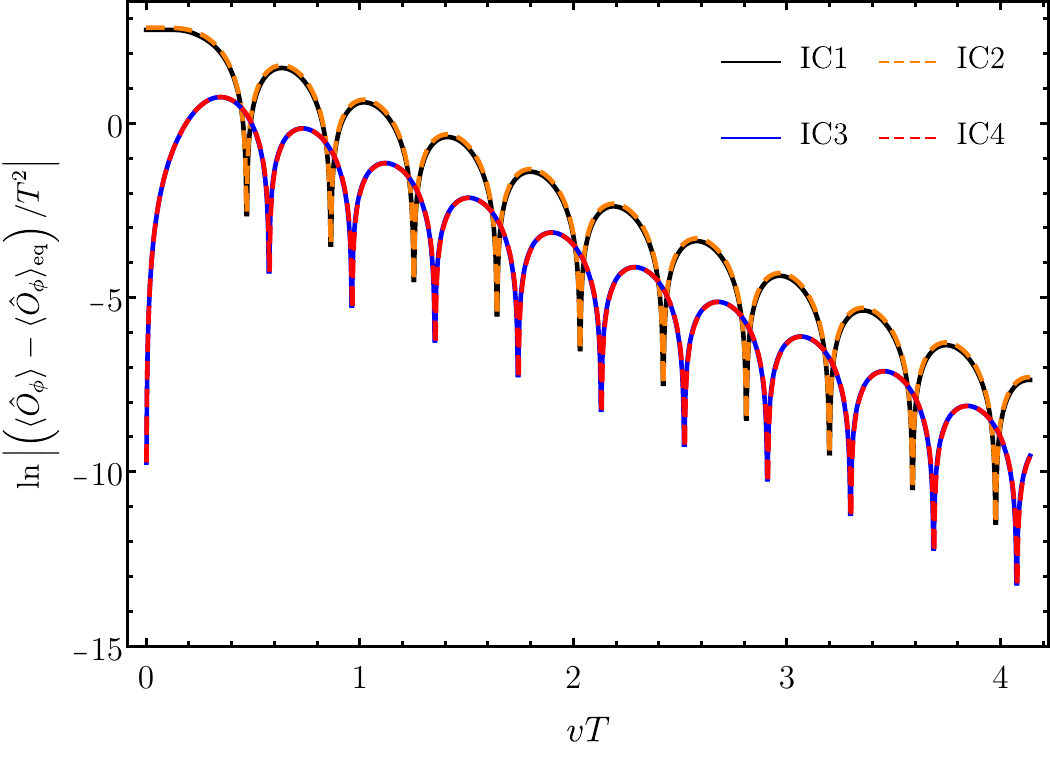}\label{fig:ICmuT6b}}
\subfigure[Pressure anisotropy]{\includegraphics[width=0.425\linewidth]{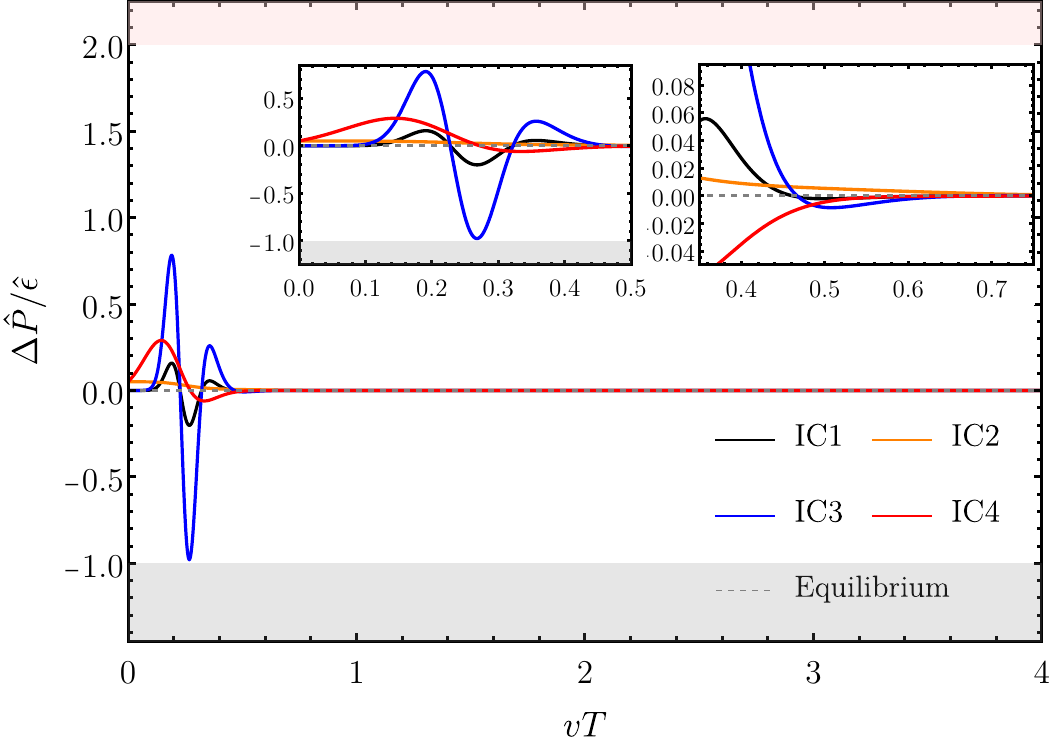}\label{fig:ICmuT6c}}
\subfigure[Logarithmic pressure anisotropy]{\includegraphics[width=0.425\linewidth]{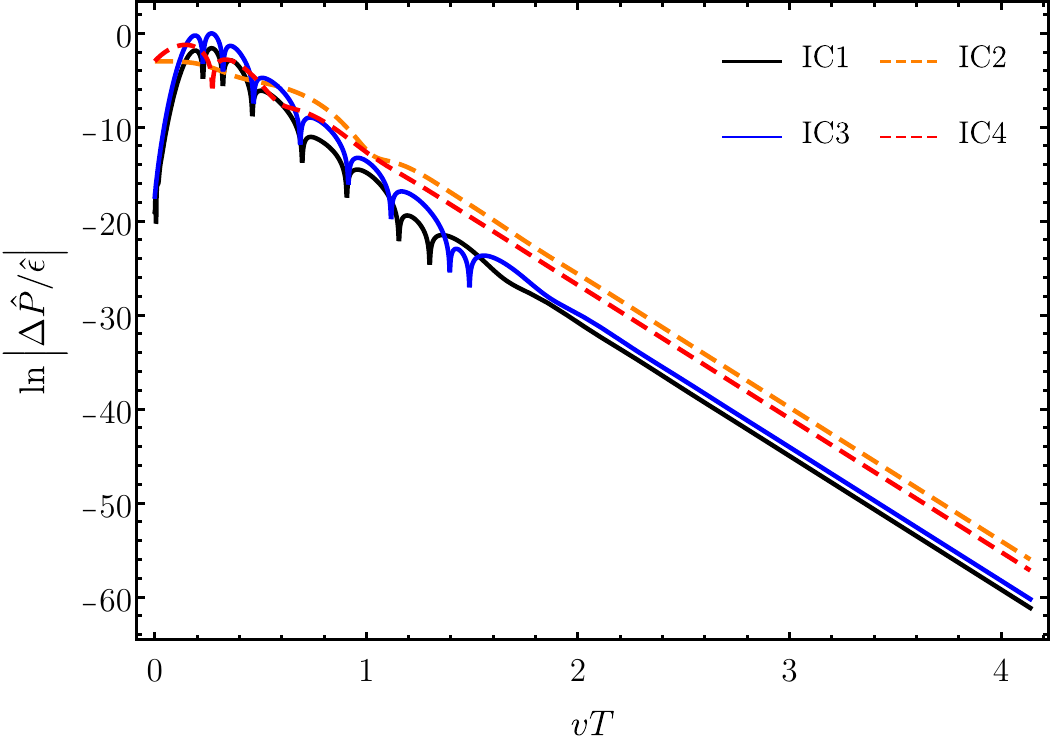}\label{fig:ICmuT6d}}
\subfigure[Entropy Density]{\includegraphics[width=0.425\linewidth]{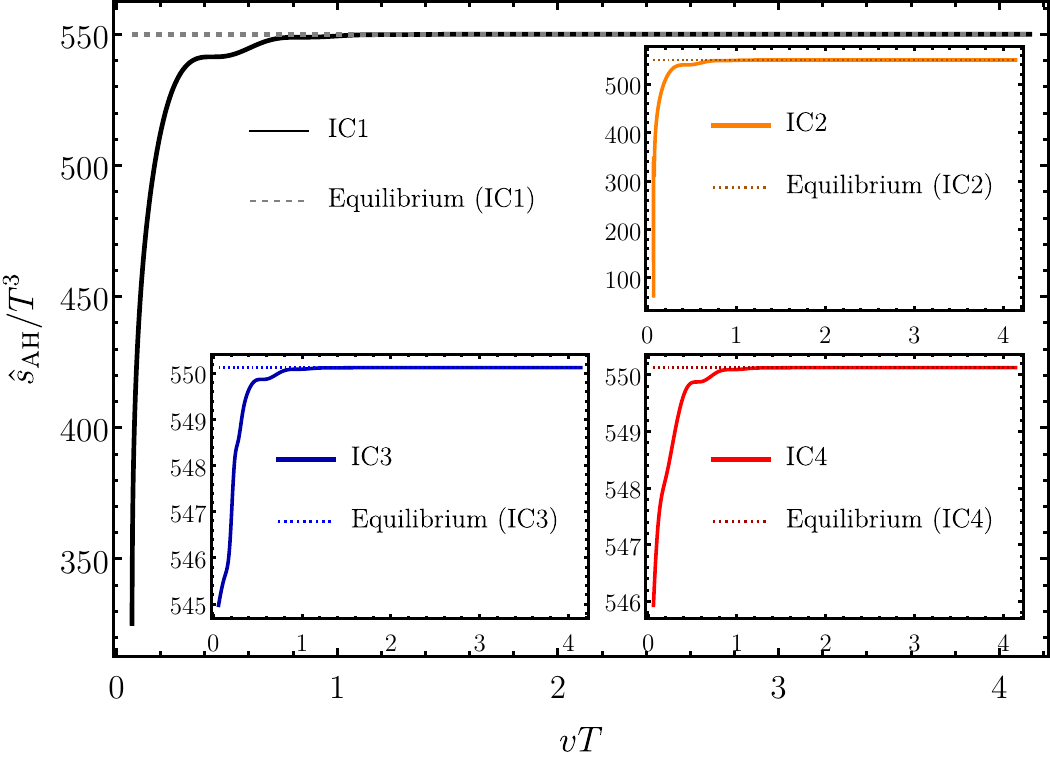}\label{fig:ICmuT6e}}
\subfigure[Logarithmic Entropy Density]{\includegraphics[width=0.425\linewidth]{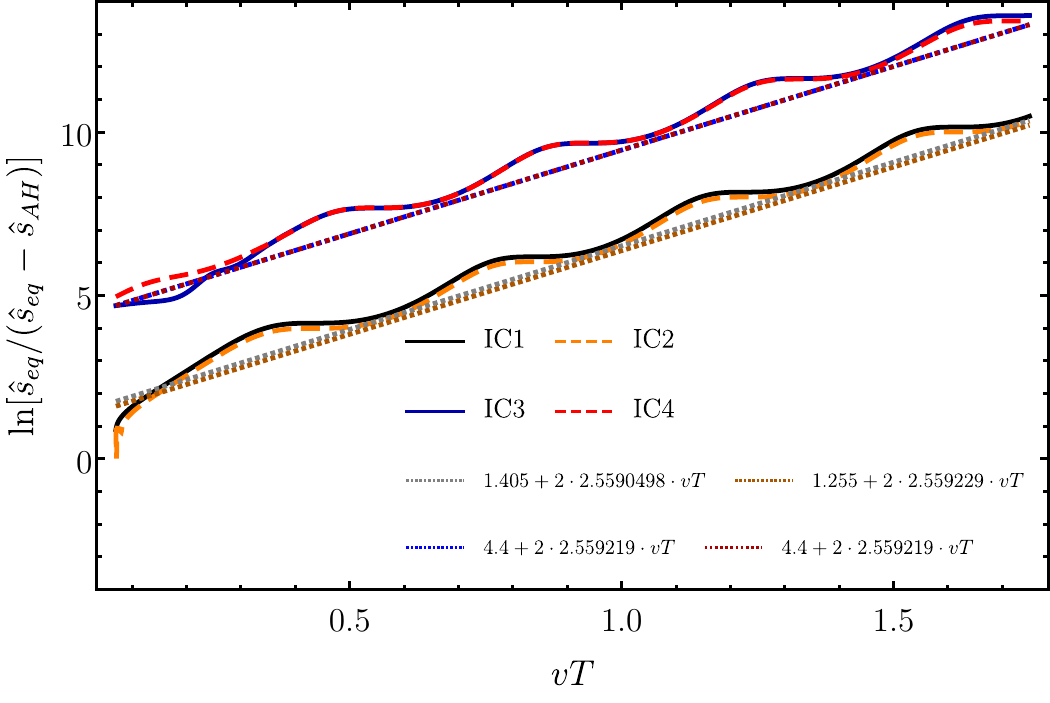}\label{fig:ICmuT6f}}
\caption{Comparison between the initial conditions IC1 --- IC4 in Table~\ref{tabICs} for different observables evaluated at $\mu/T=6$ for the 2RCBH model. In Fig.~\ref{fig:ICmuT6c}, the gray shaded band demarcates the region where the DEC is violated, while the pink shaded band delimits the region where both the WEC and DEC are violated. In Fig.~\ref{fig:ICmuT6f}, the parametrization of the support slopes is also shown for the entropy stairways.}
\label{fig:ICmuT6}
\end{figure}

We show in Figs.~\ref{fig:ModelmuT0} ---~\ref{fig:ICmuT6} our numerical results for the time evolution of the scalar condensate, the pressure anisotropy, and the non-equilibrium entropy density of the strongly coupled fluid at the boundary QFT, considering different initial conditions which are representative of the homogeneous isotropization dynamics of the medium.  We also display the corresponding logarithmic versions of the plots, built with the purpose of highlighting the exponentially small variations of the physical observables as they approach thermodynamic equilibrium in the long-time regime. Information about the initial conditions and the chosen values of $\mu/T$ considered in each figure is summarized in Table~\ref{tab:blocks}. Figs.~\ref{fig:ModelmuT0} and~\ref{fig:ModelmuT22} compare the 2RCBH and 1RCBH models by displaying the behavior of the different physical observables for IC1 in Table~\ref{tabICs} evaluated, respectively, at $\mu/T=0$ and $\mu/T=\pi/\sqrt{2}$ (corresponding to the critical point of the 1RCBH model, which is also the end point of its phase diagram). Figs.~\ref{fig:muTIC3} ---~\ref{fig:ICmuT6} display results for the 2RCBH model at larger values of $\mu/T$, considering the different initial conditions in Table~\ref{tabICs}.

\begin{table}[h!]
\centering
\begin{tabular}{cccc}
\hline
Figure & IC & $\mu/T$ & Holographic Model  \\
\hline
2 & 1 & $0$ &1RCBH/2RCBH \\
3 & 1 & $\pi/\sqrt{2}$ &1RCBH/2RCBH \\
4 & 3 & $4.000;\ 5.065;\ 6.000$ &2RCBH \\
5 & 4 & $4.000;\ 5.065;\ 6.000$ &2RCBH \\
6 & 1-4 & $4.000;\ 6.000$ & 2RCBH  \\
7 & 1-4 & $4.000;\ 6.000$ & 2RCBH \\
\hline
\end{tabular}
\caption{Initial conditions in Table~\ref{tabICs}, $\mu/T$ values, and the holographic models considered in the analysis of Figs.~\ref{fig:ModelmuT0} ---~\ref{fig:ICmuT6}.}
\label{tab:blocks}
\end{table}

As illustrated in Fig.~\ref{fig:ModelmuT0}, at $\mu/T=0$ one finds that the curves for the 2RCBH and 1RCBH models coincides in all the plots. This constitutes the first consistency check of our numerical results because both models should indeed coincide at $\mu/T= 0$, since the differences between the 2RCBH and 1RCBH models concern their R-charges, which are turned off at $\mu/T=0$. Notice, however, that in Fig.~\ref{fig:ModelmuT0} we are not considering the evolution of a purely thermal SYM plasma, even though we are taking $\mu/T=0$, because the initial condition chosen for the subtracted dilaton field in IC1 is non-trivial (if one otherwise sets $\phi_s(v_0,u)=0$, then one has a purely thermal SYM evolution). That means that we are evolving the fluid with a scalar perturbation (which is not present in a purely thermal SYM plasma) --- nonetheless, since each thermodynamic equilibrium state which is approached in the long-time regime of both models only depends on the corresponding value of $\mu/T$, the equilibrium states of both models coincide with the equilibrium state of a purely thermal SYM plasma when the fluid is taken at zero R-charge density, as expected.

\vspace{2pt}

\noindent In a transient period where the fluid is still far-from-equilibrium, one can see from Fig.~\ref{fig:ModelmuT0e} the formation of short-lived plateaus with no entropy production. These far-from-equilibrium plateaus in the entropy density are different from the near-equilibrium plateaus revealed in Fig.~\ref{fig:ModelmuT0f}, corresponding to the so-called entropy stairway (in fact, the stairway in this case involves near-equilibrium plateaus which are so close to each other that only in the logarithmic plot it is possible to resolve its structure). Indeed, far-from-equilibrium plateaus for the entropy density may or may not be present depending on the initial data considered, however, the near-equilibrium entropy stairway structure is universal in the homogeneous isotropization dynamics, being present for all initial data in the different models.

\vspace{3pt}

At $\mu/T=\pi/\sqrt{2}$, corresponding to the critical point of the 1RCBH model and also to the end point of its phase diagram, both models display clear differences, as shown in Fig.~\ref{fig:ModelmuT22}. This is best illustrated by the scalar condensate in Fig.~\ref{fig:ModelmuT22a}, where the curves for both models, although coinciding at early times for IC1, very quickly bifurcate into distinct curves, with the 2RCBH model equilibrating into a negative asymptotic value, while the 1RCBH model equilibrates into a positive value, which precisely coincide with the analytical equilibrium results given by the formulas~\eqref{eq:O12}, which are illustrated in Fig.~\ref{fig:OThermo}. In fact, for all the observables and initial data considered here, the long-time regime of the numerical solutions of the homogeneous isotropization dynamics converges to the corresponding equilibrium results displayed in Fig.~\ref{fig:Thermo}, which is another check of the consistency of our numerical results.

\vspace{2pt}

\noindent In the case of the pressure anisotropy in Fig.~\ref{fig:ModelmuT22c}, both models present almost the same behavior for IC1 at $\mu/T=\pi/\sqrt{2}$, although very small differences in the corresponding isotropization times can be seen in the logarithmic plot of Fig.~\ref{fig:ModelmuT22d}, which highlights the oscillating behavior of the pressure anisotropy around zero in finer details.
In Fig.~\ref{fig:ModelmuT22c}, the gray and pink shaded bands represent, respectively, the regions with violations of the DEC and WEC, as given by Eqs.~\eqref{eq:DEC} and~\eqref{eq:WEC}. One also notices by comparing Figs.~\ref{fig:ModelmuT0c} and~\ref{fig:ModelmuT22c} that the transient violation of energy conditions gets reduced by increasing the value of $\mu/T$ in both the 2RCBH and 1RCBH models, which is tied to the overall suppression in the magnitude of the oscillations of the pressure anisotropy at higher R-charge densities.

\vspace{2pt}

\noindent In the model comparison of the non-equilibrium entropy density displayed in Fig.~\ref{fig:ModelmuT22e}, both models exhibit qualitatively similar curves, with the 2RCBH model equilibrating into a higher value of entropy. One can also see in Fig.~\ref{fig:ModelmuT22f} the entropy stairway for each model at $\mu/T=\pi/\sqrt{2}$.

\vspace{3pt}

The entropy stairways illustrated in Figs.~\ref{fig:ModelmuT0f} and~\ref{fig:ModelmuT22f} exhibit the periodic formation of ever-increasing plateaus asymptotically approaching their corresponding equilibrium ceiling values.\footnote{Notice that contrary to Figs.~\ref{fig:ModelmuT0e} and~\ref{fig:ModelmuT22e} for $\hat{s}_\textrm{AH}(v)/T^3$, which asymptote to fixed values in the long-time regime, Figs.~\ref{fig:ModelmuT0f} and~\ref{fig:ModelmuT22f} for $\ln\left[\hat{s}_\textrm{eq}/\left(\hat{s}_\textrm{eq}-\hat{s}_\textrm{AH}(v)\right)\right]$ indefinitely increase in magnitude as time passes. Of course, since the plateaus are progressively closer to each other, resolving the structure of the stairway at longer times is a task which requires increasing the precision of the numerical calculations, demanding higher computational power.} A remarkable feature of this behavior is that it encodes a periodic structure without producing oscillations in the near-equilibrium entropy (such oscillations would lead to violations of the second law of thermodynamics). In fact, a possible way for the dynamical entropy to encode a periodicity while simultaneously respecting the second law of thermodynamics is by means of periodically producing ever-increasing and closer plateaus as the system approaches thermodynamic equilibrium.

The term \emph{entropy stairway} was coined in Ref.~\cite{Rougemont:2024hpf}, where its observation together with a connection with the lowest QNM of the system was reported in the numerical analysis of the time evolution of the purely thermal SYM and 1RCBH plasmas undergoing homogeneous isotropization dynamics. This is a particular instance of the quite general connection between entropy production and QNMs first discovered in Refs.~\cite{Jansen:2016zai,Jansen:2020ign}. In fact, it was considered in~\cite{Jansen:2016zai} a model of gravity coupled to a massive quadratic scalar field in time-dependent asymptotically AdS black hole geometries taken to be homogeneous and isotropic at all times (which is different from homogeneous isotropization dynamics), where it was shown that as the system evolves in time approaching thermodynamic equilibrium, the non-equilibrium entropy measured by the area of the apparent horizon displays periodic plateaus, corresponding to what we call the entropy stairway, modulated by twice the lowest complex QNM frequency of the system. Furthermore, it was also shown that the non-equilibrium entropy is extensive and that its time derivative is positive semi-definite, in agreement with the second law of black hole thermodynamics coming from the area theorem for the apparent horizon. Such conclusions were numerically checked in a number of different spacetime dimensions and different scaling dimensions for the boundary QFT operator dual to the bulk scalar field. Moreover, Eq.~(2.9) of~\cite{Jansen:2016zai} was first proposed through numerical analysis as an ansatz quantitatively describing the connection between near-equilibrium entropy production and the lowest QNM of the system. This connection was later proved in an analytical way in~\cite{Jansen:2020ign}, under much more general situations. Indeed, in~\cite{Jansen:2020ign} the connection between near-equilibrium entropy production and QNMs was considered for general EMD models, with ansatze for the bulk fields which encompass the particular kind of dynamics considered in~\cite{Jansen:2016zai}, and also the homogeneous isotropization dynamics considered in~\cite{Rougemont:2024hpf} and in the present work, besides other kinds of more general out-of-equilibrium dynamics. We shall also numerically confirm the validity of Eq.~(2.9) of~\cite{Jansen:2016zai} for the late-time stages of the homogeneous isotropization dynamics of the 2RCBH model later on in the present work.

In what concerns the homogeneous isotropization dynamics of the 1RCBH plasma at finite temperature and R-charge chemical potential, the period of plateau production in the entropy stairway was numerically reported in~\cite{Rougemont:2024hpf} to be half the period of oscillation associated with the inverse of the real part of the complex frequency of the lowest QNM (lQNM) of the system for the considered value of $\mu/T$,

\begin{align}
\tau_{\mathcal{S}}(\mu/T) = \frac{1}{2\,\textrm{Re}\left[\nu_\textrm{lQNM}(\mu/T)\right]} = \frac{\pi}{\textrm{Re}\left[\omega_\textrm{lQNM}(\mu/T)\right]} \quad\Rightarrow\quad \omega_{\mathcal{S}}(\mu/T) = 2 \,\textrm{Re}\left[\omega_\textrm{lQNM}(\mu/T)\right].
\label{eq:tauS}
\end{align}
Being a near-equilibrium fingerprint of the system, as is also the case with the QNMs, the entropy stairway does not depend on the particular initial data considered in the time evolution of the medium, being determined by the value of $\mu/T$ in thermodynamic equilibrium. Eq.~\eqref{eq:tauS}, numerically checked in~\cite{Rougemont:2024hpf} to hold for the purely thermal SYM and the 1RCBH plasmas, is confirmed here to be valid also for the 2RCBH plasma undergoing homogeneous isotropization dynamics, as anticipated above.

The calculation involves two complementary procedures employed to numerically extract the characteristic complex (angular) frequencies $\omega$ encoded in the time series of the physical observables. They are precisely stated as follows (we recall that the QNM spectra of homogeneous EMD models are classified under different irreducible representation of the $SO(3)$ rotation symmetry group, which define different channels related to gauge and diffeomorphism invariant perturbations of the bulk EMD fields that do not mix at the linear level~\cite{DeWolfe:2010he,Finazzo:2016psx,Critelli:2017euk,Rougemont:2018ivt,Rougemont:2024hpf,deOliveira:2024bgh}): \\
\newline
\noindent{\bf Procedure A (analysis of the pressure anisotropy and the scalar condensate)}
\begin{enumerate}
\item We isolate the periodic part of the late-time behavior of the pressure anisotropy (whose equilibration is associated with the lowest QNM of the $SO(3)$ quintuplet channel) by multiplying the respective numerical result by $e^{+\tau b_{\Delta P}}$, where $-b_{\Delta P}$ is a dimensionless number corresponding to the numerical value of the expected decay rate of the pressure anisotropy and $\tau\equiv vT$ is a dimensionless time measure;
\item We isolate the periodic part of the late-time behavior of the scalar condensate (whose equilibration is associated with the lowest QNM of the $SO(3)$ singlet channel) minus its equilibrium value, by multiplying the respective numerical result by $e^{+\tau b_{\langle\mathcal{O}\rangle}}$, where $-b_{\langle\mathcal{O}\rangle}$ is a dimensionless number corresponding to the numerical value of the expected decay rate of the aforementioned scalar condensate difference. The dominant dissipative channel in the long-time regime of the system, corresponding to the less damped channel which takes longer to equilibrate, is then identified by comparing the results in the $SO(3)$ quintuplet and singlet channels;
\item We select a time window in which the numerical results being analyzed are clearly periodic and employ the {\sc Harminv} software~\cite{harminv} to extract the complex frequency associated with the late-time behavior of the observable under consideration, which is written in the form $\omega=a-(b+\delta b)i$, along with the associated error, where $\delta b$ accounts for corrections due to the real data.
\end{enumerate}

\vspace{3pt}

\noindent{\bf Procedure B (analysis of the entropy stairway)}
\begin{enumerate}
\item By defining the entropy stairway function, $\mathcal{S}(vT,\mu/T)\equiv\ln\left[\hat{s}_\textrm{eq}(\mu/T)/\left(\hat{s}_\textrm{eq}(\mu/T)-\hat{s}_\textrm{AH}(vT,\mu/T)\right)\right]$, one computes its first finite difference, allowing for the identification of the locations of the corresponding minima and maxima, as well as the time window where the corresponding numerical results are clearly periodic;
\item We then compute the second finite difference of the entropy stairway to estimate, through the use of {\sc Harminv}, the frequency of the corresponding numerical results, including an assessment of the associated error.
\end{enumerate}

\vspace{3pt}

The complex (angular) frequencies extracted from the implementation of the above procedures are listed in Tables~\ref{tab:comparison} ---~\ref{tab:stairways2}. One notices excellent agreement with the independent reference results for the QNM spectra of the 2RCBH~\cite{deOliveira:2024bgh} and 1RCBH~\cite{Finazzo:2016psx,Critelli:2017euk} models obtained by numerically solving the linearized equations of motion for the bulk EMD field perturbations in the appropriate $SO(3)$ channels. Such an agreement provides another highly non-trivial check of the consistency of our numerical results.

Moreover, except for purely SYM evolutions (where the scalar condensate identically vanishes and the stairway structure is governed by the lowest QNM of the quintuplet channel~\cite{Rougemont:2024hpf}), for all considered values of $\mu/T$, the entropy stairway is universally governed by the lowest QNM of the singlet channel, independent of the specific initial condition being evolved in time. This is in full agreement with Eq.~\eqref{eq:tauS}, according to which the frequency of plateau formation in the entropy stairway is twice the real part of the lowest QNM of the system, which is associated with the equilibration of the scalar condensate, which is the observable taking longer to equilibrate in the strongly coupled medium. Notice also that the support slope lines for the entropy stairways shown in Figs.~\ref{fig:ModelmuT0} ---~\ref{fig:ICmuT6} are given by minus twice the imaginary part of the lowest QNM of the system, which is another feature of the relation between the entropy stairway structure and the lowest QNM of the medium undergoing homogeneous isotropization dynamics.
 
\begin{table}[h!]
\centering
\begin{tabular}{ccccc}
\hline
$\mu/T$ & Holographic Model & $SO(3)$ Channel & Extracted & Reference \\
\hline
0.00 & 1RCBH & singlet & $4.0260578(0)-2.5853149(7)i$ & $4.0260-2.5853i$ \\
0.00 & 1RCBH & quintuplet & $9.799884(1)-8.628808(4)i$ & $9.8000-8.6289i$ \\
0.00 & 2RCBH & singlet & $4.0260578(0)-2.5853149(7)i$ & $4.0260-2.5853i$ \\
0.00 & 2RCBH & quintuplet & $9.799884(1)-8.628808(4)i$ & $9.8000-8.6289i$ \\
2.22 & 1RCBH & singlet & $3.81696(6)-1.43634(3)i$ & $3.8212-1.4402i$ \\
2.22 & 1RCBH & quintuplet & $9.71279(9)-9.18361(3)i$ & $9.7295-9.1923i$ \\
2.22 & 2RCBH & singlet & $4.853911(4)-2.604684(3)i$ & $4.8538-2.6047i$ \\
2.22 & 2RCBH & quintuplet & $10.1882478(1)-10.1014461(1)i$ & $10.1883-10.1015i$ \\
\hline
\end{tabular}
\caption{Numerically extracted lowest QNM complex frequencies compared with the reference values from~\cite{deOliveira:2024bgh}.}
\label{tab:comparison}
\end{table}

\begin{table}[h!]
\centering
\begin{tabular}{cccc}
\hline
$\mu/T$ & Holographic Model & Stairway Frequency & Support Slope\\
\hline
0    & 1RCBH/2RCBH   &  $2 \times 4.039(1)$ & $\;\;\;\,2\times 2.5853149(7)$\\
2.22 & 1RCBH &  $\;\,2 \times 3.8452(2)$ & $2\times 1.43634(3)$\\
2.22 & 2RCBH  &  $2 \times 4.857(9)$ &  $\;\;2\times 2.604684(3)$ \\
\hline
\end{tabular}
\caption{Numerically extracted frequencies from the entropy stairways and the associated support slopes.}
\label{tab:stairways}
\end{table}

\begin{table}[h!]
\centering
\begin{tabular}{ccccc}
\hline
IC & $\mu/T$ & Singlet 1st OQNM & Stairway Frequency & Support Slope\\
\hline
3 & 4.000 & $6.251342(1)-2.572192(1)i$  & $2\times 6.30(4)$  & $2\times 2.572192(1)$ \\
3 & 5.065 & $7.203357(8)-2.556333(6)i$ & $2\times 7.194(0)$ &  $2\times 2.556333(6)$\\
3 & 6.000 &$8.0747727(7)-2.5592525(4)i$  & $2\times 8.0787(9)$ &   $2\times 2.5592525(4)$\\
4 & 4.000 &$6.251480(4)-2.571445(7)i$  & $2\times 6.2558(2)$ &  $2\times 2.571445(7)$\\
4 & 5.065 & $7.203546(3)- 2.556335(4)i$ & --\footnote{The stairway in this run is numerically too noisy to reliably extract its frequency; however, the support line provides a good fit for the associated slope in the interval $vT\in[0.4,1.4]$.} & $2\times 2.556335(4)$ \\
4 & $6.000$ & $8.074772(8)-2.559219(1)i$ & $2\times 8.116(8)$ & $2\times 2.559219(1)$ \\
1 & $4.000$ & $6.251455(2)-2.571915(7)i$ & $2\times 6.2676(3)$ & $2\times 2.571915(7)$\\
1 & $6.000$ & $8.0745842(7)-2.5590498(5)i$ &  $2 \times 8.0943(8)$ & $2\times 2.5590498(5)$\\
2 & $4.000$ & $6.251392(4)-2.571918(9)i$ & $2\times 6.2431(0)$ & $2\times 2.571918(9)$\\
2 & $6.000$ & $8.074772(8)-2.559229(7)i$ & $2\times 8.0680(2)$ & $2\times 2.559229(7)$\\
\hline
\end{tabular}
\caption{Numerically extracted frequencies for the 1st ordinary QNM (OQNM) in the $SO(3)$ singlet channel and the entropy stairways with the associated support slopes for the 2RCBH model.}
\label{tab:stairways2}
\end{table}

Having discussed above the illustrations in Figs.~\ref{fig:ModelmuT0} and~\ref{fig:ModelmuT22} concerning general comparisons between the 2RCBH and 1RCBH models, in Figs.~\ref{fig:muTIC3} ---~\ref{fig:ICmuT6} we illustrate the behavior of the 2RCBH model for higher values of $\mu/T$. As discussed in detail in~\cite{deOliveira:2024bgh}, all the $SO(3)$ channels of the 2RCBH model, besides the ordinary QNMs (OQNMs) corresponding to complex eigenfrequencies with symmetric nonzero real parts, also display purely imaginary quasinormal modes (PIQNMs) corresponding to purely damped modes with no oscillatory behavior. At low values of $\mu/T$, the lowest QNMs of each $SO(3)$ channel correspond to OQNMs. However, above certain thresholds, the lowest QNMs of each channel become dominated by PIQNMs. For the quintuplet channel the threshold value of dominance of PIQNMs is $(\mu/T)_\textrm{5et}\approx 5.065$, while in the singlet channel the threshold value is $(\mu/T)_\textrm{1et}\approx 133.68$, which is much higher than the values of $\mu/T$ reachable by our current numerical codes for the homogeneous isotropization dynamics of the 2RCBH model. However, we can probe with our current numerical codes the threshold value $(\mu/T)_\textrm{5et}\approx 5.065$ for the dominance of PIQNMs in the quintuplet channel, and such an analysis explicitly confirms a prediction we put forward in~\cite{deOliveira:2024bgh}. In fact, since the PIQNMs have no real/oscillatory part, as the R-charge chemical potential is increased beyond the threshold value $(\mu/T)_\textrm{5et}\approx 5.065$, the late-time behavior of the pressure anisotropy gets increasingly deformed, eventually losing the oscillatory behavior observed at lower values of chemical potential. This becomes clear by comparing e.g.~the logarithmic plots for the pressure anisotropy in Fig.~\ref{fig:ICmuT4d} (where $\mu/T=4<5.065$ and the pressure anisotropy equilibrates oscillating around zero) and in Fig.~\ref{fig:ICmuT6d} (where $\mu/T=6>5.065$ and the pressure anisotropy equilibrates without oscillating around zero, by following a purely damped pattern at late-times). This behavior is also explicitly manifest in Figs.~\ref{fig:muTIC3d} and~\ref{fig:muTIC4d}.

Finally, we follow below the analysis performed in~\cite{Jansen:2016zai}, which was later analytically explained in~\cite{Jansen:2020ign}, where the numerical results for the difference between the dynamical entropy measured by the area of the apparent horizon and its asymptotic equilibrium value, $\delta s(vT)\equiv s_{\text{eq}}-s_{\text{AH}}(vT)$, were compared to the following analytical formula corresponding to Eq.~(2.9) of~\cite{Jansen:2016zai},\footnote{Remember that we fixed here the temperature scale $T=1/\pi$, so that $\pi T=1$. The factor of $\pi T$ is explicitly shown in Eq.~\eqref{eq:AnalyticEntropyDifference} just for the purpose of dimensional analysis.}
\begin{equation}
\label{eq:AnalyticEntropyDifference}
    \delta s(vT)\equiv s_{\text{eq}}-s_{\text{AH}}(vT)\equiv A\exp{2\,\frac{\textrm{Im}[\omega_{l\text{QNM}}]}{\pi T}\, vT}\left[\cos(2\,\frac{\textrm{Re}[\omega_{l\text{QNM}}]}{\pi T}\, v T+\delta)+B\right],
\end{equation}
where $\omega_{l\text{QNM}}$ corresponds to the complex frequency of the lowest quasinormal mode of the system. The parameters $A$ and $\delta$ depend on the initial conditions and can be found by fitting the numerical data within an adequate time window using the analytical formula~\eqref{eq:AnalyticEntropyDifference}, while the parameter $B$ is fixed in terms of the lowest quasinormal mode of the system as follows,
\begin{equation}
    B=B_\textrm{min}= -\frac{|\omega_{l\text{QNM}}|}{\textrm{Im}[\omega_{l\text{QNM}}]},
    \label{eq:Bmin}
\end{equation}
where $B_\textrm{min}$ is the minimum value of the parameter $B$ in~\eqref{eq:AnalyticEntropyDifference} compatible with the second law of black hole thermodynamics, corresponding to the area theorem for the apparent horizon, $s'_\textrm{AH}(vT)\ge 0$. In order to show Eq.~\eqref{eq:Bmin}, notice that by applying the second law of thermodynamics to Eq.~\eqref{eq:AnalyticEntropyDifference} it follows that,
\begin{align}
B\ge -\frac{\textrm{Im}[\omega_{l\text{QNM}}] \cos\left(\Delta\phi\right) - \textrm{Re}[\omega_{l\text{QNM}}] \sin\left(\Delta\phi\right)}{\textrm{Im}[\omega_{l\text{QNM}}]}\equiv f(vT),
\label{ineq}
\end{align}
where $\Delta\phi\equiv 2\,(\textrm{Re}[\omega_{l\text{QNM}}]/\pi T)\,vT+\delta$. The requirement that $B\ge f(vT)$ implies that the minimum value allowed for $B$ is the maximum value of $f(vT)$. Therefore, one evaluates the condition $f'(vT)=0$, which implies that,
\begin{align}
\sin(\Delta\phi)=-\frac{\textrm{Re}[\omega_{l\text{QNM}}]}{\textrm{Im}[\omega_{l\text{QNM}}]} \cos(\Delta\phi)\,\,\,\,\,\Rightarrow\,\,\,\,\, \left(\frac{\textrm{Re}[\omega_{l\text{QNM}}]}{\textrm{Im}[\omega_{l\text{QNM}}]}\right)^2=\frac{\sin^2(\Delta\phi)}{\cos^2(\Delta\phi)}.
\label{eq:fmax}
\end{align}
By substituting into~\eqref{ineq} both relations displayed in~\eqref{eq:fmax}, it follows the result given in~\eqref{eq:Bmin},
\begin{align}
B_\textrm{min}=f_\textrm{max}=-\frac{|\omega_{l\text{QNM}}|}{\textrm{Im}[\omega_{l\text{QNM}}]}\times \frac{|\omega_{l\text{QNM}}|}{\textrm{Im}[\omega_{l\text{QNM}}]}\cos(\Delta\phi)=-\frac{|\omega_{l\text{QNM}}|}{\textrm{Im}[\omega_{l\text{QNM}}]}\times \sqrt{1+\left(\frac{\textrm{Re}[\omega_{l\text{QNM}}]}{\textrm{Im}[\omega_{l\text{QNM}}]}\right)^2}\cos(\Delta\phi)=-\frac{|\omega_{l\text{QNM}}|}{\textrm{Im}[\omega_{l\text{QNM}}]}.
\label{eq:Bminproof}
\end{align}

\begin{figure}[h!]
\centering  
\subfigure[IC1: 1RCBH vs 2RCBH models at $\mu/T=\pi/\sqrt{2}$]{\includegraphics[width=0.425\linewidth]{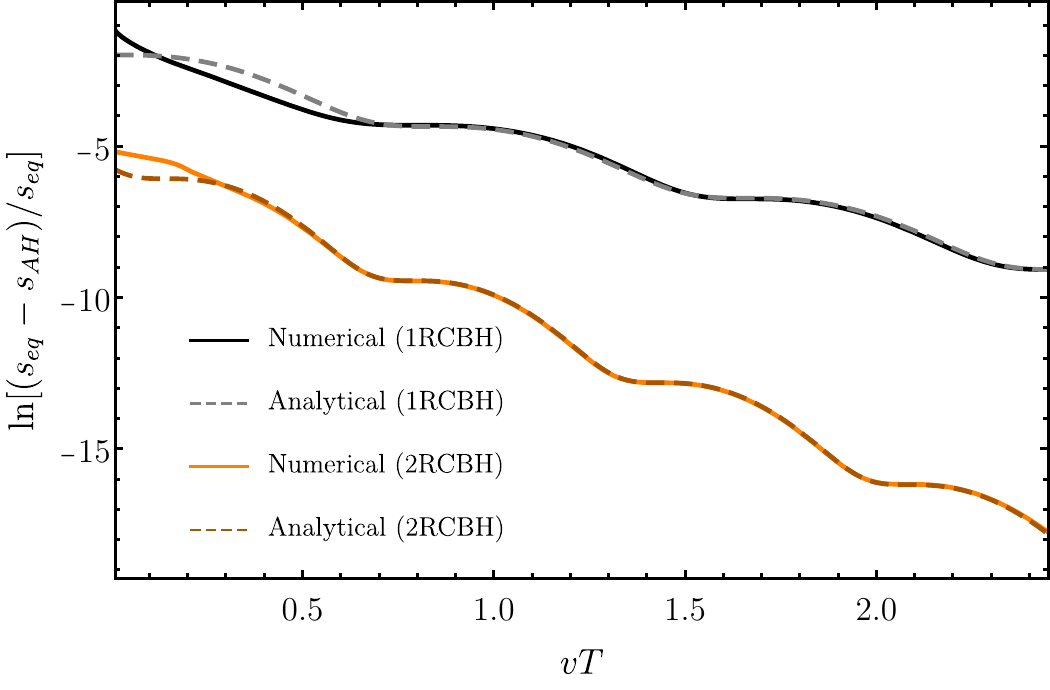}\label{fig:NumvsFita}}
\subfigure[IC2: $\mu/T=4.0$ vs $\mu/T=6.0$ (2RCBH model)]{\includegraphics[width=0.425\linewidth]{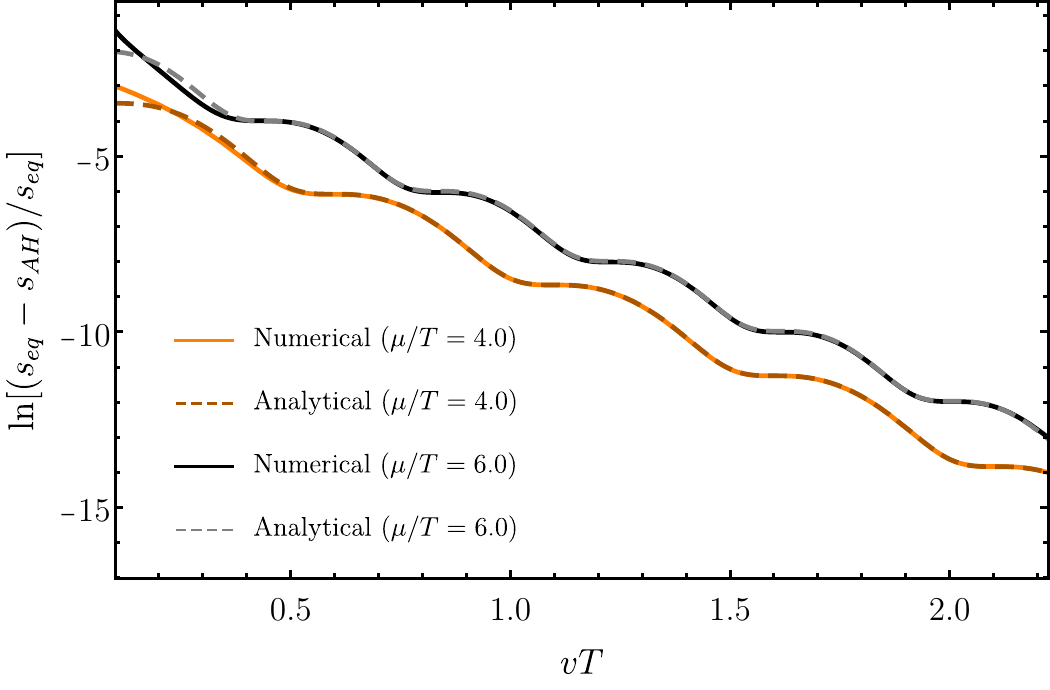}\label{fig:NumvsFitb}}
\subfigure[IC3: $\mu/T=4.0$ vs $\mu/T=6.0$ (2RCBH model)]{\includegraphics[width=0.425\linewidth]{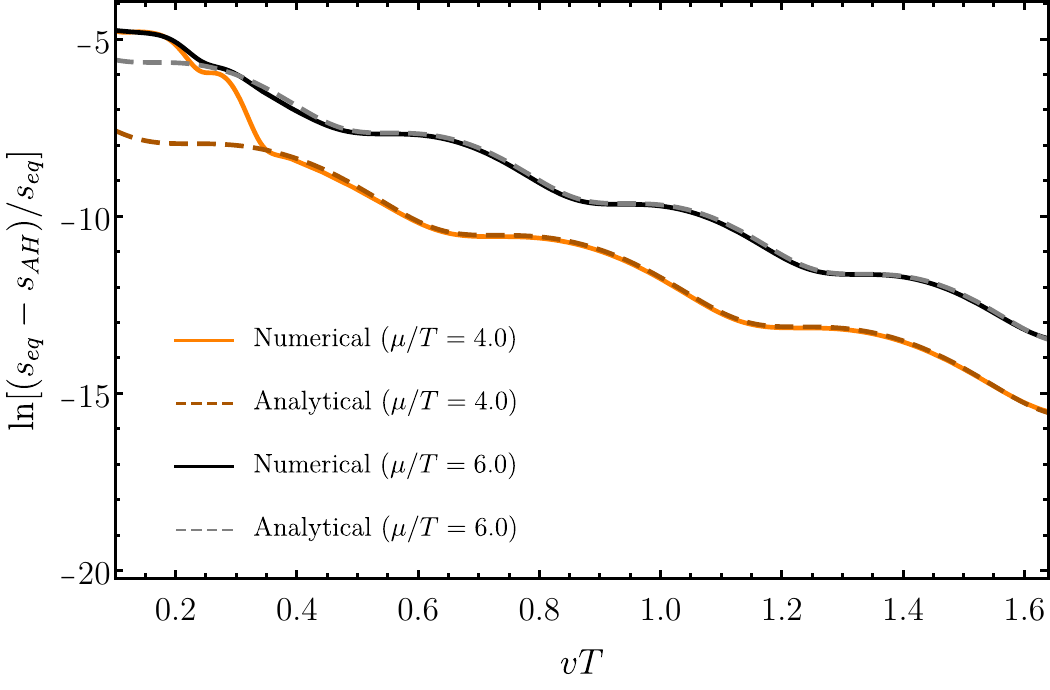}\label{fig:NumvsFitc}}
\subfigure[IC4: $\mu/T=4.0$ vs $\mu/T=6.0$ (2RCBH model)]{\includegraphics[width=0.425\linewidth]{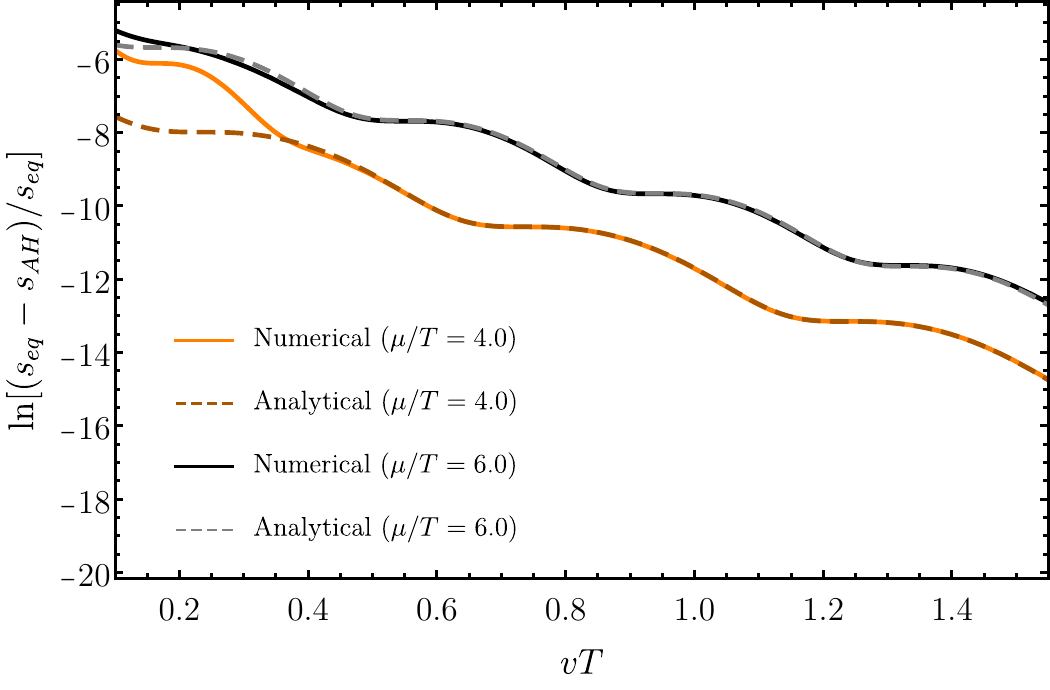}\label{fig:NumvsFitd}}
\caption{Comparison between full numerical data and the analytical fit from formula~\eqref{eq:AnalyticEntropyDifference} for the entropy production. Except for the initial stages when the fluid is very far-from-equilibrium, the analytical formula~\eqref{eq:AnalyticEntropyDifference} works perfectly well to describe the entropy production in connection with the lowest QNM of the system.}
\label{fig:NumericvsFit}
\end{figure}

Fig.~\ref{fig:NumericvsFit} displays comparisons between the full numerical data (solid lines) and the analytical results obtained by using the formula~\eqref{eq:AnalyticEntropyDifference} (dashed lines). In Fig.~\ref{fig:NumvsFita}, the comparison is shown for the 1RCBH and 2RCBH models at $\mu/T=\pi/\sqrt{2}$ with the initial condition IC1. Moreover, Figs.~\ref{fig:NumvsFitb},~\ref{fig:NumvsFitc}, and~\ref{fig:NumvsFitd} display comparisons for different values of $\mu/T$ in the 2RCBH model with the initial conditions IC2, IC3, and IC4. The corresponding fitted values for the free parameters $A$ and $\delta$ in~\eqref{eq:AnalyticEntropyDifference}, the fitting windows, and quasinormal mode frequencies used in the fitting procedure are summarized in Table~\ref{tab:comparison2}.

\begin{table}[h!]
\centering
\begin{tabular}{ccccccccc}
\hline
Model & IC & $\mu/T$ & A & $\delta$ & B & $\textrm{Re}[\omega_{l\text{QNM}}]$ & $\textrm{Im}[\omega_{l\text{QNM}}]$ & Fitting Window ($\mu/T$)\\
\hline
1RCBH & 1 & $\pi/\sqrt{2}$ & 0.6380 & 2.8576 & 2.8394 & 3.8169 & -1.4363 & [0.75,2.5] \\
2RCBH & 1 & $\pi/\sqrt{2}$ & 12.5927 & -2.0399 & 2.1149 & 4.8169 & -2.6046 & [0.75,2.5]\\
2RCBH & 2 & 4.0 & 7.6767 & 3.1588 & 2.6286 & 6.2518 & -2.5717 & [0.75,2.5]\\
2RCBH & 2 & 6.0 & 32.8249 & 3.4684 & 3.3098 & 8.0746 & -2.5592 & [0.75,2.5]\\
2RCBH & 3 & 4.0 & 0.1765 & 1.4953 & 2.6286 & 6.2518 & -2.5717 & [0.75,2.0]\\
2RCBH & 3 & 6.0 & 1.4752 & 1.7595 & 3.3098 & 8.0746 & -2.5592 & [0.75,2.0]\\
2RCBH & 4 & 4.0 & 0.1769 & 1.4119 & 2.6286 & 6.2518 & -2.5717 & [0.75,1.6]\\
2RCBH & 4 & 6.0 & 1.4472 & 1.7716 & 3.3098 & 8.0746 & -2.5592 & [0.75,1.6]\\
\hline
\end{tabular}
\caption{Fitted parameters $A$ and $\delta$ in Eq.~\eqref{eq:AnalyticEntropyDifference} used to plot the analytical curves in Fig.~\ref{fig:NumericvsFit}.}
\label{tab:comparison2}
\end{table}


\section{Conclusions}
\label{sec:conc}

In the present work, we numerically investigated the behavior of the top-down 2RCBH plasma undergoing homogeneous isotropization dynamics, and also compared it with previous results obtained for the 1RCBH plasma. We list below the overall results found here for the 2RCBH model:
\begin{enumerate}
\item There are some initial data which, although respecting all the energy conditions at the initial time slice, evolve in time such as to transiently violate the dominant and even the weak energy condition, when the fluid is still far-from-equilibrium. In the cases where transient violations of energy conditions are present, they get reduced and eventually disappear by increasing the value of the R-charge chemical potential, which is tied to the fact that the magnitude of oscillations of the pressure anisotropy is diminished at higher R-charge densities in the homogeneous isotropization dynamics. These transient violations of classical energy conditions are possible due to the quantum nature of the strongly interacting fluids considered here;

\item The scalar condensate gets more negative (augmenting its absolute value) by increasing the value of the R-charge chemical potential of the medium, while the magnitude of its oscillations is increased, leading to a delay in the equilibration of the scalar condensate at higher R-charge densities;

\item The entropy density of the medium is larger at higher R-charge densities and two different classes of transient plateaus for the entropy can be formed during the time evolution of the system (the transient plateaus correspond to finite time windows where there is no entropy production in the fluid). One class of plateaus may or may not be present, depending on the initial data being evolved in time, and refers to far-from-equilibrium plateaus. The other class of plateaus is universal and it is present in the long-time regime of all the initial data, corresponding to the near-equilibrium plateaus which comprise the entropy stairway structure.
\end{enumerate}

In fact, the universality of the entropy stairway in the homogeneous isotropization dynamics refers not only to its presence in the long-time regime of all initial data for the 2RCBH plasma, but it is much more general than that. Indeed, the entropy stairway was previously observed in~\cite{Rougemont:2024hpf} to be also present in the homogeneous isotropization dynamics of the purely thermal SYM and the 1RCBH plasmas. These are all particular realizations of the quite general connection originally discovered in~\cite{Jansen:2016zai,Jansen:2020ign} relating entropy production and the lowest dominant quasinormal mode of strongly interacting media with holographic duals.

In order to resolve the structure of the entropy stairway, a high numerical precision is needed due to the increasingly close proximity of the entropy plateaus near the corresponding equilibrium ceiling value. In~\cite{Rougemont:2024hpf}, it was numerically checked for the purely thermal SYM and the 1RCBH plasmas the relation given in Eq.~\eqref{eq:tauS} between the frequency of plateau formation in the entropy stairway and the real part of the complex eigenfrequency corresponding to the lowest QNM of the system. This is the QNM with lowest imaginary part in modulus, which is the dominant mode near-equilibrium, since this is the less damped non-hydrodynamic mode of the system. Here we confirmed that relation~\eqref{eq:tauS} also holds for the 2RCBH model. Moreover, we also numerically checked here a more complete picture of the quantitative behavior of the entropy stairway, valid for all the holographic models considered. Besides Eq.~\eqref{eq:tauS} involving the real part of the lowest QNM of the system, which is associated with the frequency of oscillation of different physical observables near thermodynamic equilibrium, also the imaginary part of the lowest QNM, which is associated with the decay of different physical observables towards their equilibrium values, becomes manifest in the behavior of the entropy stairway through their support slope lines depicted in Figs.~\ref{fig:ModelmuT0} ---~\ref{fig:ICmuT6}. As described in Tables~\ref{tab:comparison} ---~\ref{tab:stairways2}, the slope of the support lines is precisely given by minus twice the imaginary part of the lowest QNM of the system. More generally, as explicitly checked in Fig.~\ref{fig:NumericvsFit} and in Table~\ref{tab:comparison2}, the general formula originally discovered in~\cite{Jansen:2016zai}, given by Eq.~\eqref{eq:AnalyticEntropyDifference} with $B$ universally corresponding to~\eqref{eq:Bmin} and the parameters $A$ and $\delta$ fitted to the numerically computed time evolution of the dynamical entropy for a given set of initial data, provides the complete quantitative connection between the entropy stairway and the lowest QNM of the system.

The complete picture for the quantitative relation between the entropy stairway and the lowest QNM of the system discussed above provides a powerful tool for predicting the behavior of one from the behavior of the other. Indeed, the observed fact that the entropy near thermodynamic equilibrium in homogeneous isotropization dynamics is tied to the lowest QNM of the system implies that when this mode has a nonzero real part, the entropy density must somehow embody a periodic behavior. This periodic behavior cannot become manifest in the form of oscillations, since oscillations in the entropy density near thermodynamic equilibrium would violate the second law of thermodynamics. Therefore, the way found by the entropy density to embody a periodic behavior is through the production of the entropy stairway, corresponding to ever-increasing periodic plateaus with transient zero entropy production, where the consecutive plateaus are progressively closer to each other due to the corresponding ceiling thermodynamic equilibrium value.

In the case of the 2RCBH model, the homogeneous QNM spectra investigated in~\cite{deOliveira:2024bgh} revealed that all the $SO(3)$ channels, besides the OQNMs corresponding to ordinary quasinormal frequencies with symmetric nonzero real part, also have PIQNMs with zero real part, describing purely damped modes with no oscillatory behavior. Moreover, the PIQNMs were found to become the lowest QNMs of the system for $\mu/T$ above a certain threshold value. Specifically, in the quintuplet channel associated with the equilibration of the pressure anisotropy, the threshold value for dominance of the PIQNMs is $(\mu/T)_\textrm{5et}\approx 5.065$, while in the singlet channel, associated with the equilibration of the scalar condensate, the threshold value is $(\mu/T)_\textrm{1et}\approx 133.68$. Due to this, three predictions were put forward in~\cite{deOliveira:2024bgh}: (1) for $\mu/T>5.065$ the pressure anisotropy should eventually lose its oscillatory behavior near thermodynamic equilibrium, (2) for $\mu/T>133.68$ the scalar condensate should eventually lose its oscillatory behavior near thermodynamic equilibrium, and (3) since the entropy stairway is tied to the lowest QNM of the system, which is generally the lowest QNM of the $SO(3)$ singlet channel associated with the equilibration of the scalar condensate, when the lowest QNM of the system corresponds to a PIQNM, the entropy stairway should progressively evanesce as $\mu/T$ is further increased, until the stairway eventually disappears. In the present work we were able to explicitly confirm the validity of prediction (1). However, predictions (2) and (3) require simulating the time evolution of the 2RCBH plasma for much higher values of $\mu/T$ than those which can be achieved with our current numerical codes and are left to be tested in the future.

\acknowledgments
G.O. acknowledges financial support from the Coordination for the Improvement of Higher Education Personnel (CAPES). W.B. and R.R. acknowledge financial support by National Council for Scientific and Technological Development (CNPq) under grant number~407162/2023-2. W.B. is grateful to S\~{a}o Paulo Research Foundation (FAPESP) for the received support under grant number~2022/02503-9. R.R. acknowledges financial support by CNPq under grant number~305466/2024-0.




\bibliographystyle{apsrev4-2}
\bibliography{bibliography,extrabiblio} 

\end{document}